\documentclass[aps,rmp,preprint,groupedaddress,floatfix]{revtex4}
\usepackage{epsfig}
\usepackage{natbib}
\usepackage{genetics}

\begin{document}

\newcommand{\mon}{\begin{displaymath}}
\newcommand{\moff}{\end{displaymath}}
\newcommand{\sumi}[1]{\sum_{{#1}=-\infty}^{\infty}}
\renewcommand{\b}[1]{\mbox{\boldmath ${#1}$}}
\newcommand{\sumy}{\sum_{\b{y}}}
\newcommand{\sumz}{\sum_{\b{z}}}
\renewcommand{\u}{U_b}

\newcommand{\pd}[2]{\frac{\partial {#1}}{\partial {#2}}}
\newcommand{\od}[2]{\frac{d {#1}}{d {#2}}}
\newcommand{\inti}{\int_{-\infty}^{\infty}}
\newcommand{\eon}{\begin{equation}}
\newcommand{\eoff}{\end{equation}}
\newcommand{\e}[1]{\times 10^{#1}}

\newcommand{\chem}[2]{{}^{#2} \mathrm{#1}}
\renewcommand{\sb}{s}
\newcommand{\s}{s}
\newcommand{\zetaexp}{\left( \zeta e^{q \s t} \right)}
\newcommand{\taunuc}{\tau_{est}}
\newcommand{\eq}[1]{Eq. (\ref{#1})}
\newcommand{\ev}[1]{\langle #1 \rangle}
\newcommand{\mat}[1]{\bf{\mathcal{#1}}}
\newcommand{\fig}[1]{Fig. \ref{#1}}
\newcommand{\degrees}{\,^{\circ}\mathrm{C}}
\newcommand{\p}{r}
\renewcommand{\log}{\ln}
\newcommand{\fit}{s}

\renewcommand{\sec}[1]{section \ref{#1}}
\newcommand{\pr}[1]{\mathrm{Prob}\left[{#1}\right]}
\newcommand{\dpr}[1]{\mathrm{prob}\left[{#1}\right]}

\renewcommand{\ni}{\noindent}
\newcommand{\din}{\indent \indent}
\newcommand{\mni}{\medskip \noindent}
\newcommand{\bni}{\bigskip \noindent}
\newcommand{\sni}{\smallskip \noindent}

\newcommand{\be}{\begin{equation}}
\newcommand{\ee}{\end{equation}}

\newcommand{\sci}{s_{ci}}
\newcommand{\CI}{clonal interference }
\newcommand{\La}{\Lambda}
\newcommand{\la}{\lambda}
\newcommand{\st}{\tilde{s}}
\renewcommand{\th}{\beta}
\newcommand{\si}{\sigma}
\renewcommand{\u}{U_b}
\newcommand{\ut}{\tilde{U}_b}
\renewcommand{\L}{L}

\newcommand{\Q}{\hat{q}}

\title{Beneficial Mutation-Selection Balance and the Effect of Linkage on Positive Selection}
\author{Michael M. Desai$^{*, \dagger, 1}$}
\author{Daniel S. Fisher$^{*, \dagger \dagger}$}
\affiliation{${}^*$Department of Physics, ${}^\dagger$Department of
Molecular and Cell Biology, ${}^{\dagger \dagger}$Division of
Engineering and Applied Sciences, Harvard University, Cambridge MA
02138 \\ ${}^{1}$ Present Address:  Lewis-Sigler Institute for
Integrative Genomics, Princeton University, Princeton NJ 08544,
mmdesai@princeton.edu \\ \\}
\date{\today ) \newpage Running Head:  Linkage and Positive Selection 
Corresponding Author: \\  Michael Desai \\ Lewis-Sigler Institute for Integrative Genomics \\ Carl Icahn Laboratory \\ Princeton University \\ Princeton, NJ 08544 \\
617-710-1318 or 609-258-8327 \\ mmdesai@princeton.edu \\ \newpage}

\begin{abstract}

When beneficial mutations are rare, they accumulate by a series of
selective sweeps.  But when they are common, many beneficial
mutations will occur before any can fix, so there will be many
different mutant lineages in the population concurrently. In an
asexual population, these different mutant lineages interfere and
not all can fix simultaneously.  In addition, further beneficial
mutations can accumulate in mutant lineages while these are still a
minority of the population. In this paper, we analyze the dynamics
of such multiple mutations and the interplay between multiple
mutations and interference between clones. These result in
substantial variation in fitness accumulating within a single
asexual population.  The amount of variation is determined by a
balance between selection, which destroys variation, and beneficial
mutations, which create more. The behavior depends in a subtle way
on the population parameters: the population size, the beneficial
mutation rate, and the distribution of the fitness increments of the
potential beneficial mutations.  The mutation-selection balance
leads to a continually evolving population with a steady-state
fitness variation. This variation increases logarithmically with
both population size and mutation rate and sets the rate at which
the population accumulates beneficial mutations, which thus also
grows only logarithmically with population size and mutation rate.
These results imply that mutator phenotypes are less effective  in
larger asexual populations. They also have consequences for the
advantages (or disadvantages) of sex via the Fisher-Muller effect;
these are discussed briefly.

\end{abstract}

\maketitle

\section{Introduction}

The vast majority of mutations are neutral or deleterious. Extensive
study of such mutations has explained the genetic diversity in many
populations, and been useful for inferring population parameters and
histories from data.  Yet beneficial mutations, despite their
rarity, are what causes long-term adaptation. Unfortunately, our
understanding of their dynamics remains poor by comparison.
Beneficial mutations also alter the distributions of neutral and
deleterious mutations: in asexual populations, as well as  in
regions of a sexual genome that remain linked on the relevant
timescales, positive selection can dramatically affect the genetic
diversity.

When beneficial mutations are rare and selection is strong, positive
selection results in a succession of selective sweeps.  A mutation
occurs, spreads through the population due to selection, and soon
fixes.  Some time later, another such event may occur. This
situation is sometimes called the strong selection weak mutation
regime --- to make its character clear, we will refer to it as the
\emph{successional-mutations regime}: between sweeps, there is a
single ``ruling" population. In this regime, the effect of positive
selection on patterns of genetic variation is reasonably well
understood. A selective sweep reduces the genetic variation in
regions of the genome linked, over the timescale of the sweep, to
the site at which a beneficial mutation occurs: other mutations in
these regions hitchhike to fixation.

Successional-mutations behavior typically occurs in small to
moderate sized populations in which beneficial mutations are
sufficiently rare. However, a different regime occurs in larger
populations, in which  beneficial mutations occur frequently. When
beneficial mutations are common enough that many mutant lineages can
be simultaneously present in the population, selective sweeps will
overlap and interfere with one another (i.e. different beneficial
mutations will grow in the population concurrently).  If, in
addition, selection is strong enough that it is not dominated by
random drift (except while mutants are very rare), we have a
``strong selection strong mutation''  regime. For clarity, we will
refer to this as the \emph{concurrent-mutations regime}. The effects
of concurrent mutations in asexual populations are the focus of the
present paper. As we will see, the concurrent beneficial mutations
regime is not an unusual special case: many viral, bacterial and
simple eukaryotic populations likely experience evolution via
multiple concurrent mutations.

In populations which contain many different beneficial mutants,
there will be substantial variation in the fitness within the
population.  This variation will be acted on by selection. But in
the absence of new mutations, the variation will soon disappear.
Thus the traditional approach to evolution of quantitative traits
--- to assume that there always exists genetic variation (as for
traits not subject to selection) --- fails badly. New mutations are
crucially needed to maintain the variation on which further
selection can act. Thus to understand adaptation when multiple
mutations are involved, it is essential to analyze the interplay
between selection and new beneficial mutations, especially how the
latter maintains the former. Understanding this beneficial
mutation-selection balance and the resulting dynamics is the primary
goal of this paper.

Both the successional- and concurrent-mutations regimes require that
selection dominates drift except while mutants are very rare.  A
qualitatively different regime occurs with weakly beneficial
mutations: these do not sweep in the traditional sense because drift
dominates their dynamics. This weakly beneficial regime most readily
occurs in small populations, where selective forces cannot overcome
drift, or when considering mutations of very small effect, such as
those that affect synonymous codon usage \citep{570, 572, 569, 232}.
But when populations are large, selection of beneficial mutations of
fitness increment $s$ will tend to dominate over drift (unless the
beneficial mutations are extremely rare).  In this paper we are
interested in beneficial mutations in large populations, thus we
focus exclusively on the {\it strong selection} regimes for which
drift is only important for beneficial  mutant lineages while they
are a tiny minority of the population.

The essential difference between the successional-mutations and
concurrent-mutations regimes is presented in Fig. 1, which depicts
beneficial mutations in an asexual population. In a small enough
population, or one whose beneficial mutation rate ($\u$) is low,
beneficial mutations occur rarely enough that they are well
separated in time and one can sweep before another arises (Fig. 1a).
This is the successional-mutations regime, in which the beneficial
mutations all behave independently. However, in a larger population
or at higher $\u$, multiple mutant populations exist concurrently
and they are no longer independent  (Fig. 1b). Mutations that occur
in different lineages cannot both fix in the absence of
recombination:  at least one of them must be ``wasted".  In this
paper, we focus on understanding how an asexual population in the
concurrent-mutation regime  accumulates beneficial mutations.

In the concurrent mutations regime, two important effects occur. The
first is when a moderately beneficial mutation occurs and begins to
sweep, only to be outcompeted by a later, more strongly beneficial
mutation that occurs in a wild-type individual (i.e. an individual
of the majority population).  The first mutation is then wasted as
it is eliminated along with the then-majority type by the sweep of
the stronger mutation.  This effect is referred to as {\it clonal
interference}; it is illustrated in Fig. 1c.  Note that (despite
earlier broader definitions) we will use the term ``clonal
interference'' to refer to only this first effect, consistent with
the focus of recent work on the subject.  The second effect is when
{\it multiple mutations} occur in the {\it same} lineage before the
first beneficial mutation fixes.  For example, a second
(moderate-effect) beneficial mutation can occur in an individual
that already has the one beneficial mutation. The double mutant can
then benefit from the combined effect of the two mutations and
outcompete the single mutant as well as some other stronger
single-mutants that  arise in the majority population.  This process
is illustrated in Fig. 1d.

The dynamics of evolution in the concurrent-mutations regime is
important to understand.  At the very least, this is  essential for
forming sensible null expectations about experimental, observational
and genomic data from large populations, which --- after all
--- are generally evolving.  Knowing how the effects of beneficial
mutations depend on mutation rate and population size is crucial for
making meaningful comparisons between different populations.  Most
important, in our view, is developing an intuition for how large
populations evolve. The simple picture of successive selective
sweeps in the successional-mutations regime is a valuable guide to
thinking about positive selection. Yet we have little intuitive
guidance when the successional-mutations approximation does not
apply.  This is a serious shortcoming in our understanding of the
evolution of a wide array of populations, including viruses and most
unicellular organisms.

Although it is not as well understood as the successional-mutations
regime, the concurrent-mutations regime has been the subject of
substantial interest since the 1930s.  \citet{567} and \citet{571}
first noted the potential importance of interference between
beneficial mutations (Muller drew diagrams very similar to our
Fig. 1).  They proposed what has come to be known as the
Fisher-Muller hypothesis for the advantage of sex: sexual
populations can recombine beneficial mutations in competing
lineages into the same individual. This prevents mutational events
from being wasted, as they often are in asexual populations.

Much subsequent work on positive selection in the
concurrent-mutations regime has focused on the implications for the
evolution of sex. \citet{297}, \citet{309}, and \citet{564}
attempted to quantify the Fisher-Muller effect in the late 1960s and
early 1970s. However, their analysis was incomplete --- it did not
fully account for stochastic behavior, ignored triple and higher
mutations, and did not correctly account for the effects of sex.
Contemporaneously, \citet{573} looked at this problem from the
perspective of the linkage disequilibrium generated by multiple
linked beneficial mutations segregating simultaneously. This has
become known as the Hill-Robertson effect.  It is essentially
equivalent to the Fisher-Muller effect and the analysis of
\citet{297} and \citet{309} (see \citet{12} for a detailed
discussion). In recent years, \citet{568, 276} and \citet{234, 263}
have analyzed the Fisher-Muller effect from the Hill-Robertson
perspective. Their work focuses on the buildup of linkage
disequilibrium due to mutations and selection, and the average
effect of recombination on the variance in fitness and the
destruction of disequilibrium. This provides useful insight into the
effects of sex, but does not explain the full evolutionary dynamics
or population genetic structure created by this type of positive
selection.

In this paper, we step back from the long tradition of studying the
implications of concurrent mutations for the evolution of sex and
focus instead on the basic dynamics shown schematically
 in Fig. 1b. We show, both heuristically and
quantitatively, how an asexual population in the
concurrent-mutations regime accumulates many beneficial
mutations, what the fitness distribution looks like, how it
develops, and how quickly selected substitutions occur via
collective sweeps. We develop a framework for thinking more
generally about positive selection and its effects that is
applicable to large populations of asexuals or any other case
where linkage between mutations is important.

We do not analyze the questions about sex or patterns of diversity
in this paper. However,  these questions should  be informed by our
results; some can be studied within the framework we present in this
paper. For example, when recombination is uncommon, the average
effects of sex may be irrelevant --- instead all that matters is
whether or not it creates rare individuals that are much more fit
than the majority of the population.   To study this, we must first
understand the full distribution of genetic diversity within the
population. Similarly, before analyzing the patterns of genetic
variation exhibited by populations in which multiple linked
beneficial mutations have occurred --- or are occurring --- one must
understand the rate of beneficial substitutions and typical
interference patterns between these within the linked regions.

To understand the concurrent-mutations dynamics in detail, it is
essential to start with a specific model that focuses on some subset
of the important effects.  Features can then be added after enough
understanding has been gleaned to enable predictions of which
effects are model-specific and which are more general.  Positive
selection can involve various complications, including epistasis
(interactions between effects of mutations), conditionally
beneficial mutations, frequency-dependent benefits, and changing
environments, among others.  Many different scenarios are possible.
At present we have little understanding of which, if any, of these
situations are biologically ``typical,'' and which are unusual. In
this paper, we do not attempt to catalogue all possible
complications; this is an impossibly broad subject. Instead we look
at the simplest possible situation involving positive selection of
concurrent mutations. We suppose that a variety of beneficial
mutations are available to a population, and ask how the population
acquires them.  We assume these mutations interact in a simple
multiplicative way (additive for the growth rates) with no
epistasis, frequency dependence, or changing environment of any
kind.  In short, we ask how the population climbs a single smoothly
sloped ``hill'' in fitness space (Fig. 2).

This simple scenario is probably common. Populations often find
themselves in an environment where they can accumulate quite a few
different beneficial mutations which each independently (or at least
roughly so) help them adapt.  Even when this simple hill-climbing
scenario does not apply, it is an important null model.  Some more
complex forms of positive selection may also prove tractable within
the framework we describe, while others will not; these leave open
many avenues for future work.

Various other authors have studied the dynamics of multiple
concurrent beneficial mutations under the simple assumptions
outlined above. \citet{1} analyzed ``clonal interference'' between
mutations of different strengths; this has since been extended by
various authors \citep{410, 417, 351, 331, 292, 294}. This work
focuses on the interference between mutations of different strengths
that occur in the \emph{same} lineage, while neglecting the
competition between mutations that arise in different lineages ---
in particular multiple mutants. Our analysis in the present paper
starts instead with the other concurrent-mutation effect, multiple
mutants, initially in a model in which clonal interference is
absent. In any real situation,  the two effects will both occur.  We
thus discuss the interplay between clonal interference and multiple
mutations in a later section.  The detailed analysis  will be
presented elsewhere:  in the present paper we summarize a few of the
salient results. \citet{562} have also recently analyzed a model
which combines some aspects of clonal interference and multiple
mutations. At this point, it is important to note that if population
parameters are such that clonal interference is important, the
effects of multiple mutants are usually at least of comparable
importance.  Thus there is some inconsistency in focusing on clonal
interference alone.

To focus on the effects of multiple mutants without clonal
interference, two additional simplifying approximations are useful.
For most of this paper, we study a model in which each beneficial
mutation has the {\it same} effect $s$ on fitness (i.e. each step
uphill is of the same size).  Furthermore, to focus  on the effects
of positive selection, we neglect deleterious mutations in the
primary analysis. Even though neither assumption will typically be
true, these turn out to be reasonable approximations in many
circumstances. Situations in which they are not appropriate are more
complicated scenarios for positive selection, some of which,
especially the effects of a distribution of fitness increments,  we
discuss briefly.

Remarkably, even the simplest possible model with many
equal-strength beneficial mutations available is only partially
understood.  \citet{21} and \citet{20} analyzed a similar simple
model, but their initial work did not handle random drift correctly.
More recently, they have developed a sophisticated although somewhat
unwieldy moment-based approach \citep{kesslerlevineunpub} from which
it is unfortunately hard to understand the essential aspects of the
dynamics. \citet{313} also studied a model similar in its essential
aspects to our simplest model (although also including deleterious
mutations of the same magnitude).  They were concerned with viral
evolution, and their results are primarily valid for very large
mutation rates appropriate for many viruses. Nevertheless, if worked
out more fully from Rouzine et. al.'s  analysis, several results can
be obtained that are closely related to ours. But our analysis
involves a less mathematically formal approach --- we believe it is
both clearer and a better basis for further development (some of
which is included herein).   We focus in this paper on a different
regime from \cite{313}: it should obtain in single celled organisms
(and some viruses).  We discuss in detail, below, the relationship
between our analysis and that of \citet{313}.

The outline of this paper is as follows.  We begin by describing in
the next section a heuristic approach to the dynamics. This analysis
gets the behavior roughly correct, and illustrates the ideas
underlying our approach. We then describe the simplest model more
precisely, and analyze it the following section.  We next discuss
transient behavior before the population has reached its steady
state fitness distribution, and address the effects of deleterious
mutations. In the next section, we make comparisons between our
analytic results and simulations. We then relax our assumption that
all mutations have the same effect, and discuss the relationship
between our theory and clonal interference analysis. Finally, we
summarize our results and discuss future directions.

\section{Heuristic Analysis and Intuition \label{intuitive}}

In the simplest situation with multiple concurrent beneficial
mutations available, there are three important parameters:  the
population size, $N$, the beneficial mutation rate per individual
per generation, $\u$, and the fitness increase $s$ provided by each
mutation.  We will refer to the basic exponential growth rate, $r$,
of a population as its fitness (rather than its growth factor per
generation $R = e^r \approx 1+r$).  Thus two mutations of magnitude
$s_1$ and $s_2$ increase fitness by $s_1+s_2$ (in the absence of
epistasis, which we will generally assume).  We call the rate of
increase, $d\ev{r}/dt$,  of the average fitness of a population the
\emph{speed of evolution} and denote it $v$.

To focus on the effects of multiple mutants in a situation in
which clonal interference does not occur, we initially restrict
consideration to the approximation that all beneficial mutations
have the same effect.  A $k$-tuple mutant thus has fitness $ks$
greater than the original wild-type.  The speed of evolution is
then simply $v = s \langle \frac{dk}{dt} \rangle$.

We begin by reviewing the successional-mutations regime where
beneficial mutations are sufficiently separated in time for them
to sweep independently, as in Fig. 1a.  Although this is exactly
analyzable and well known, it is instructive to consider it from a
heuristic perspective.  We then turn to a heuristic analysis of
the more complex concurrent-mutations dynamics illustrated in
Fig. 1d.

\subsection{Successional-Mutations Regime and the Establishment of
Mutants}

Small asexual populations evolve by accumulating beneficial
mutations sequentially.  Beneficial mutations occur in the
population at a total rate $N \u$.  If its selective advantage $s$
is in the range $\frac{1}{N} \ll s \ll 1$, the probability that a
particular mutant will survive random drift is proportional to $s$
(the constant of proportionality depends on the specific model for
the stochastic dynamics; for our model it is $1$ and we discuss in
the Model section below the minor modifications of our results that
are needed for other stochastic dynamics).  We call the process by
which the lineage of a beneficial mutant that survives drift becomes
large enough for the population of its descendants to grow
deterministically the \emph{establishment} of the mutant clone. Thus
new beneficial mutations are established at a rate $N \u s$ per
generation (other mutant populations die out due to random drift) so
that a new mutation will become established about once every
$\tau_{establish} = \frac{1}{N \u s}$ generations. As we will show
later, a mutant population typically becomes established when its
size is of order $\frac{1}{s}$.  Thus once it has become
established, the mutant takes of order $\tau_{fix} = \frac{1}{s} \ln
[Ns]$ generations to fix (we will loosely call ``fixed'' a mutant
lineage that has grown to represent a large fraction of the
population; the conventional definition corresponds to fully fixed,
which takes about twice as long).

When the population size or mutation rate is small enough, fixation
will happen more quickly than establishment.  This occurs when \eon
\tau_{fix} \approx \frac{\log[N s]}{s} \ll \tau_{establish} \sim
\frac{1}{N \u s}, \label{t-est-fix-comp} \eoff which corresponds to
$N \u \ll \frac{1}{\ln [N s]}$.  When this condition holds, we are
in the successional mutations regime, in which the establishment
rate is limiting:  a mutation A that arises and fixes will do so
long before the next mutation destined to survive drift, B, is
established.  Thus a relevant mutation B occurs in a population that
has already fixed A, yielding AB, and B fixes well before mutation C
is established as mutant ABC. Beneficial mutations continue to
accumulate in this simple way. New mutations arise and fix at
average rate $N \u s$, each one increasing the fitness by $s$. Thus
fitness increases at a speed \be v = N \u s^2 \ , \ee linear in the
product $N \u$.  This linear mutation-limited behavior characterizes
the successional-mutations regime of successional selective sweeps.

\subsection{Concurrent-Mutations Regime}

In larger populations, the behavior is more complex, as illustrated
by Fig. 1d.  In this case, the establishment times of new mutants
are shorter than their fixation times, corresponding to \eon N \u
\gtrsim \frac{1}{\ln [N s]}. \eoff  Thus new beneficial mutations
arise and become established before earlier ones can sweep, causing
them to interfere with one another.

Two types of interference are important. Recent clonal interference
analyses have focused on one: the competition that occurs when two
mutations which have different strengths occur independently in
individuals with similar initial fitness \citep{1, 410, 417, 351,
331, 292, 294, 562}.  We focus in the bulk of this paper on the
other type of interference: a mutation that arises in a fitter
background --- e.g. one with an earlier beneficial mutation --- will
outcompete another mutation of similar effect that occurs in a less
fit background.  In the constant-$s$ model clonal interference is
explicitly absent, and we thus initially focus exclusively on this
latter effect. In this constant $s$ approximation, two different
mutants that occur among those with the same fitness (in particular
members of the same clone) will compete equally and sweep together,
each becoming only partially fixed.  Unless we are interested in the
neutral genetic variability of the population, all subpopulations
with the same fitness can be considered as a single subpopulation:
we will do this except in the discussion at the end of this paper.
Also, we postpone discussion of the interplay between clonal
interference and multiple mutants (i.e. going beyond the constant
$s$ model) to a later section below.

First consider starting from a monoclonal population. Mutations
initially give rise to a subpopulation with  fitness increased by
$s$ (Fig. 3a). The size of this mutant subpopulation drifts
stochastically, but eventually becomes large enough, roughly
$\frac{1}{s}$ individuals, to become deterministic. This takes a
(stochastic) establishment time, $\tau_1$. After its establishment
but before its fixation, mutations can occur in the still-small
mutant subpopulation to create double mutants with fitness $2s$
(Fig. 3b). This typically happens well before the single mutants
have fixed (else we are by \eq{t-est-fix-comp} in the
successional-mutations regime). A double-mutant population thereby
becomes established a time $\tau_2$ after the establishment of the
single-mutant population. Triple mutants then begin to arise, and
become established after an additional time $\tau_3$. This interval
is typically \emph{shorter} than $\tau_2$ primarily because
double-mutants grow faster than single-mutants and hence generate
more mutations, and, in addition, because  the triple-mutants are
more fit than double-mutants and hence survive drift more easily
(with probability $3s$ rather than $2s$).

This process continues, accelerating at each step. Eventually,
however, enough time passes that the single-mutant subpopulation (or
one of the multiple-mutant subpopulations) becomes larger than the
original wild-type. This near-fixation of the single-mutants
increases the mean fitness by $s$, which balances the accelerating
front and creates a moving fitness distribution which will attain a
(roughly) steady state width with the mean fitness increasing with a
steady state average speed, $v$. This is a form of
mutation-selection balance: as each new beneficial mutation becomes
established, the mean fitness increases by $s$ and the fitness
distribution moves to higher fitness while maintaining the same
shape.

It is useful to consider this process in more general terms. The key
to the behavior is the balance between mutation, which increases the
variation in fitness within the population, and selection, which
decreases the variation by eliminating all but the fittest
individuals.  If we were discussing deleterious mutations, mutation
would also oppose the tendency of selection to increase the mean
fitness, leading to a steady-state distribution of fitness (ignoring
Muller's Ratchet, which for large populations only matters on
extremely long timescales). This deleterious mutation-selection
balance, which is independent of population size for large $N$, has
long been understood \citep{565}. In our case, the dynamics are more
subtle because the important mutations are beneficial. The basic
idea of mutation-selection balance, however, is unchanged. Mutations
broaden the fitness distribution while selection narrows it,
creating a steady state variance around an increasing mean fitness.
But unlike the deleterious case, the dynamics of the rare
individuals near the most-fit tail of the fitness distribution (the
``nose'') control the behavior. Selection moves the distribution
towards higher fitness at a rate very close to the steady state
variance in fitness --- the classic result in the absence of
mutations (the ``fundamental theorem of natural selection'')
\citep{567}. But new beneficial mutations at the nose are essential
to {\it maintain} this variance: in their absence the fitness
distribution would collapse to a narrow peak near the most-fit
individual and evolution would grind to a halt.

The crucial dependence on new mutations in the nose makes the
analysis of the beneficial mutation-selection balance more complex
than in the deleterious case.  It is now essential to account
properly for random drift in the small populations  near the nose.
In the case of deleterious mutation-selection balance, rare new
mutants are less fit than the rest of the population. They will die
out soon anyway, so failing to account properly for the stochastic
dynamics by which they do so has no serious consequences.  Random
drift is only important with solely deleterious mutations if
Muller's ratchet is operating, i.e. if the most-fit individuals are
rare enough that they can die out due to random drift. The
beneficial mutation-selection balance is quite analogous to this
Muller's ratchet case. Here too the subpopulations that are more fit
than average control the long-term behavior of the population, and
these are small enough that  correct stochastic treatment is
essential. As is the case with Muller's ratchet, infinite-$N$
deterministic approximations are not even qualitatively correct.
Indeed, with a large supply of  beneficial mutations, deterministic
analysis incorrectly predicts a rapid acceleration of the nose
towards an infinite speed of evolution. This nonsense result is
because of  the creation in the deterministic approximation of (what
are effectively) fractional numbers of new much fitter mutants which
then grow exponentially, unhampered by drift, and dominate the
behavior soon after (we describe this in more detail in Appendix A).

There are two factors that determine the dependence of the speed of
evolution on the population size. The first is the dynamics of
already established populations, which is dominated by selection.
The second is the new mutations that occur in the fittest
subpopulation. We define the {\it lead} of the distribution, $Q$, as
the difference between the fitness of the most-fit individual and
the mean fitness of the population (more precisely, $Q-s$ is the
difference between the mean fitness and that of the most-fit
established mutant class).  We define $q$ by $Q = qs$, so that if
the lead is $Q$, the most-fit individuals have $q$ more beneficial
mutations than the average individual: they have a ``lead'' $Q$ in
the race to higher fitness. Once it is established, the fittest
population grows exponentially, first at rate $qs$ but more slowly
as the mean advances.  Growing from its establishment upon reaching
size $\frac{1}{qs}$ until it reaches a large fraction of $N$ will
thus take time $\log (Nqs) /(\frac{qs}{2})$, since $\frac{qs}{2}$ is
the average growth rate. In this time the mean fitness will increase
by $qs$. Therefore $v \approx (qs)^2/[2 \log (Nqs)]$. One can show
that this $v$ is equal to the variance in fitness, as expected if
mutation is indeed negligible compared to selection in the bulk
(i.e. away from the nose) of the distribution, so that the
fundamental theorem of natural selection applies.

The other factor is the dynamics of the nose, where mutations are
essential.  A more-fit mutant that moves the nose forward by $s$
will be established some time $\tau_q$ after the previous most-fit
mutant. Thus the nose advances at a speed $v = s/ \ev{\tau_q}$,
where $\ev{\tau_q}$ is the average $\tau_q$. After it is
established, the fittest population $n_q$ will grow exponentially at
rate $qs$ and produce mutants at a rate $\u n_q \approx \u
e^{qst}/qs$. Many new mutants will occur soon after the time at
which $\u \int n_q(t) dt$ becomes equal to one, so the time it takes
a new mutant to establish is $\tau_q \sim \frac{1}{qs} \log (s/\u)$.
This means the nose advances at rate $v\approx s/\ev{\tau_q} \approx
qs^2/\log(s/\u)$. Significantly, the behavior of the nose depends
only on mutations from the most-fit subpopulation; it is almost
independent of the less fit populations and thus can depend on $N$
only via the lead, $qs$. As far as the nose is concerned, the
majority of the population --- destined to die out shortly --- is
important only to ease the competition for the fittest few. Yet we
argued above that the bulk of the population fixes the speed of the
mean via the selection pressure: $v \approx (qs)^2 /[2\log(Nqs)]$.
In steady state, the speed of the mean must equal the speed of the
nose --- the mutation-selection balance.  This implies that \be q
\sim \frac{2 \ln [Ns]}{\ln \left[ s/\u \right]} \ee and \be v \sim
\frac{2 s^2 \ln [Ns]}{\ln^2 \left[ s/\u \right]} \ . \ee These
results are very close to the more careful calculations below.  All
the basic qualitative behavior follows from this intuitive
reasoning.

For large $N\u$, we have found that $v$ depends
\emph{logarithmically} on $N$ and $\u$, much slower than the
linear dependence on $N\u$ which holds for smaller populations.
This reduction occurs because at large $N\u$, almost all
beneficial mutations occur in individuals far from the nose of the
fitness distribution (i.e. in a bad genetic background) and are
therefore ``wasted,'' since these subpopulations are doomed to
extinction. Thus increasing $N$ does not directly increase the
supply of {\it important} mutations, as these occur in the
relatively few individuals at the nose. Rather, the effect of
increasing $N$ is to increase the time required for selection to
move the mean fitness, which increases the lead, which makes
individuals at the nose more fit relative to the mean fitness,
which speeds the establishments at the nose. Similarly, increasing
$\u$ does not directly affect the dynamics of most of the fitness
distribution.  Rather, it decreases the time for new mutations to
occur at the nose, which means that more mutations can occur
before the mean moves, which increases the lead and speeds the
evolution.

This also explains why $v$ is \emph{not} a function of $N \u$: $N$
directly affects only selection timescales, while $\u$ directly
affects only the mutation supply rate, so $v$ depends on $N$ and
$\u$ \emph{separately}.  It is {\it not} a function of the commonly
used parameter $\theta = 2 N \u$.  Instead, it is a function of the
basic parameters $Ns$ (which describes selective forces) and
$\frac{s}{\u}$ (which describes the strength of selection relative
to mutation), and it is valid in the regime where both these are
large.  The expression for $q$ above is of order the basic selective
timescale, $\frac{1}{s} \ln [Ns]$ divided by the basic mutation
timescale, $\frac{1}{s} \ln [\frac{s}{\u}]$, which makes sense since
the lead is set by the balance between these two forces.  The two
factors that determine the basic time scales of the multiple
mutation dynamics are \be L\equiv \ln Ns  \ \ \ \ {\rm and} \ \ \ \
\ell\equiv\ln s/\u \ . \ee Although these are both logarithmic in
the population parameters and thus never huge, they can be large
enough to be considered as large parameters.  Many of our more
detailed results are valid in the limit that both $L$ and $\ell$ are
large, with corrections (some of which we include) smaller by powers
of $1/\ell$ or $1/L$.

We will show below that our result for $v$ is consistent with the
fundamental theorem of natural selection, which states that when
mutational effects can be ignored the speed of evolution is equal to
the variance in fitness within the population.  Viewed in this
light, our result for the speed of evolution is not in itself novel:
the speed is just the variance in fitness, as usual.  What our
analysis does is to obtain what this variance is.  In many aspects
of quantitative genetics, the variance of a quantitative trait (such
as fitness here) is taken as some external parameter.  When the
variance has accumulated during a period when it was neutral, and is
only starting to be selected on, this may be appropriate.  But
beyond that, it is surely not. Our analysis deals with the case when
variance is accumulating while being selected on.  That is, when
variance in fitness is increasing due to mutations while at the same
time it is being acted on by selection.  Then, even if the
adaptation speed is only indirectly related to new mutations, it is
essentially dependent on them: without mutations the variance will
rapidly collapse to zero. We analyze here how a balance between the
forces of mutation and selection develops to set a steady-state
variance, and how large that variance is.

However, neither our heuristic analysis above, nor our more careful
work described below ever explicitly involves the fitness variance.
Rather, the natural measure of the width of the fitness distribution
is the lead. It is the lead, not the variance or standard deviation,
that can be most productively thought of as a balance between
mutation and selection.  It is true, of course, that the variance is
also increased by mutation and decreased by selection. However, this
is not the clearest way to understand the behavior. The increase in
the variance from mutations is delayed and indirect. The new
mutations that occur at the nose will only increase the variance
after they have grown enough --- and by then the important new
mutations that will keep the variance high later are happening
further out in the nose.  This is not to say that a variance based
--- together, crucially, with higher moments --- approach is
impossible, but it is unwieldy and prone to hard-to-understand
errors when any approximations are made. We discuss the problems
with moment based approaches in Appendix A.

\section{Simplest Model \label{model}}

We now turn away from crude (although powerful) intuitive arguments
towards more rigorous analysis.  We begin this section by defining
the simplest model more precisely. We consider mutation, selection,
and drift within a purely asexual population of constant size $N$.
We assume that a large number of beneficial mutations, each of which
increases the fitness by $s$, are available and define $\u$ to be
the total mutation rate to these mutations. We consider the
situation where the number of beneficial mutations fixed is small
compared to the total number available so that $\u$ does not change
appreciably over the course of the evolution (we relax this
assumption in Appendix C).  We neglect deleterious mutations and
other-strength beneficial mutations (see later sections below for a
discussion of the consequences of these assumptions).  These
simplifications are not essential and do not change the basic
behavior in many situations, Indeed, we will argue that these
assumptions can all be good approximations even when the situation
is more complex, in particular when $N$ or $\u$ are not constant, or
in the presence of deleterious mutations or variable $s$, as we
discuss in detail in subsequent sections. But, more importantly,
these simplest approximations make the analysis clearer.

In addition to the more innocuous simplifications, we make two
essential biological assumptions: that there is {\it no
frequency-dependent selection}, and that there is {\it no
epistasis}, so that the fitness of an individual with $k$ mutations
is $(k-\ell)s$ greater than the fitness of an individual with $\ell$
mutations.  When either of these conditions fails, the evolutionary
dynamics can be very different from our predictions.  Note, however,
that our model can sometimes be a good approximation even in the
presence of epistasis: the simplifying feature is the assumption
that after one or more beneficial mutations have already been
acquired, the {\it distribution} of available  beneficial mutations
in the new genetic background is similar to that in the ``wild-type"
background, but these need not be the same mutations as those
available initially.

\subsection{Key Approximations}

There are two primary difficulties in analyzing the multiple
subpopulations that occur even in the simplest model.  The first is
the stochastic aspects: when a subpopulation with a given fitness is
small, stochastic drift plays a crucial role and must be handled
correctly. The second is the interactions between the
subpopulations: the constraint of fixed total  population size means
that there is effectively a frequency dependence to the growth of a
subpopulation --- albeit a simple one.

To model the stochastic effects, we assume that the basic process of
birth and death is a continuous-time branching process.  All
individuals have the same constant death rate $1$, which ensures
that the average lifetime of an individual is $1$ (i.e. the units of
time are generations), and the lifetimes are exponentially
distributed. Each individual in the population has some number, $y$,
of beneficial mutations. We define $\bar y$ to be the average value
of $y$ across the population (i.e. the average number of beneficial
mutations per individual). An individual with $y$ beneficial
mutations has a birth rate $1 + (y - \bar y) s$.  This ensures that
the average birth rate in the population is $1$, so the population
stays at a constant size $N$. We assume all individuals give rise to
mutant offspring at rate $\u$, independent of their birth rate (i.e.
mutants arise at a constant rate per unit time).  If mutations
instead occur at a constant rate per birth event, our assumption
underestimates the mutation rate for the most-fit individuals.
However, we always assume $(y - \bar y) s \ll 1$ for all individuals
(i.e. the lead, $Q$, is $\ll 1$), so that the two definitions are
almost equivalent.

The  branching process model allows one to calculate simple analytic
expressions for a number of important quantities which are not
readily available in diffusion approximations of the standard
Wright-Fisher model. However, branching process models cannot easily
deal with the nonlinear saturation effects required to maintain a
constant population size.  By ``saturation" effects, we refer to
when a mutant subpopulation has become large enough to influence the
mean fitness of the population, and hence begins to compete with
itself, slowing its growth: this is the essential effect of the
fixed total population size.  To handle the saturation effects, we
make use of a simple observation: stochastic effects are only
important when a subpopulation is rare, while saturation is only
important when a subpopulation is common.  Thus we use the
stochastic branching process model, ignoring saturation effects, to
describe the dynamics of a subpopulation while it is small.
Conversely, when it is large, we ignore random drift and treat it
with the correctly saturating deterministic equations.  Our use of
both deterministic and stochastic analyses requires an appropriate
way of linking the two together.  In this paper, we will describe a
method for doing so.  This method accounts for all of the important
aspects of genetic drift and is simple and intuitive. It should be
of broad applicability to related evolutionary problems.

This approach works as long as the stochastic regime and the
saturation regime are different.  That is, a subpopulation must
become large enough to neglect random drift before it is too large
to ignore saturation. We can treat a subpopulation of size $n$
deterministically so long as $ns \gg 1$. On the other hand,
saturation can be ignored when $n \ll N$. Thus in order to separate
the stochastic and the saturating phases of growth of a
subpopulation, we require $Ns \gg 1$. Throughout this paper, we will
assume this condition holds.

A situation in which there are multiple subpopulations of varying
sizes is illustrated in Fig. 4: this shows the logarithm of a
typical fitness distribution within a steadily evolving population.
Where the subpopulations are small, at the front of the
distribution, stochastic analysis is necessary but nonlinearities
can be ignored. When a subpopulation represents a substantial
fraction of the total, nonlinear saturation is important but
stochasticity is not. As long as $Ns \gg 1$, there is an
intermediate regime where {\it neither} matters.  We can thus use a
nonlinear deterministic analysis in the bulk of the distribution, a
linear stochastic analysis near the front, and match the two in the
intermediate regime in which both are valid.  These approximations
are {\it fully controlled} and any corrections to our results will
be small for $Ns\gg1$.

Our analysis is inapplicable when $Ns \lesssim 1$, i.e. for small
populations or those experiencing very weak selection. However,
unless $s$ is extremely small ($s \sim \u$), a population small
enough that $Ns \lesssim 1$ will usually be too small for clonal
interference or multiple mutation effects to matter. Thus requiring
$Ns \gg 1$ is not a serious limitation to exploring the effects we
are concerned with here.

\subsection{Relationship of our Model to Wright-Fisher Model}

The deterministic limit of our model is identical to that of the
Wright-Fisher model.  However, the stochastic dynamics are slightly
different.  In the Wright-Fisher model, all individuals have a
lifetime of exactly one generation, while in our model individuals
have a random exponentially distributed lifetime with mean one
generation.  In the Wright-Fisher model, the distribution of the
number of offspring per individual is approximately Poisson, while
in our model the number of offspring is geometrically distributed.
Both the mean lifetime and mean number of offspring per individual
are identical in the two models (hence identical deterministic
dynamics), but the different distributions do lead to slight
differences.  In particular, although the probability a beneficial
mutation of size $s$ ($s \ll 1$) will become established is
proportional to $s$ in both models, it is $\approx cs$ with the
coefficient $c=2$ the Wright-Fisher model and $c=1$ in ours. Since
it is likely that the population dynamics in any real population is
not well represented by either of these models, there is no one
``correct'' model (e.g. for populations dividing by binary fission,
as in many experimental studies of evolution, the fixation
probability is closer to $2.8 s$ \citep{285}). Fortunately, in our
analysis of the behavior of large populations, these differences
only cause negligible corrections in the arguments of logarithms
(e.g. replacing $\ln (Ns)$ with $\ln (cNs)$ when $Ns \gg 1$).  For
smaller populations, however, the speed of evolution is proportional
to the probability of establishment and thus does depend on more
details of the model: in particular, the successional mutation
result for the speed is $v \approx cN\u s^2$.

It would in principle be possible to use a diffusion approximation
to the Wright-Fisher model instead of our branching process model.
This would have the advantage of being able to handle saturation and
drift at the same time, and thus cases where $Ns \lesssim 1$. Such a
model could in principle treat all the different subpopulations
stochastically, including all mutations between these populations.
However, this would lead to a complex and difficult to analyze
infinite-dimensional diffusion process.  There is, however, a
controlled approximation --- valid for large $Ns$ --- to the full
diffusion process that is exactly equivalent to ours; as it would
add little, we will not discuss this explicitly here

\section{Analysis \label{analysis}}

This section contains the primary analysis presented in the present
paper: the accumulation of beneficial mutations in the simple model
described above.  We begin by looking at what happens to a single
mutant individual.  We then ask what happens to a mutant population
which is being fed constantly by new mutations.  We next couple this
analysis to the behavior of the rest of the population to gain an
understanding of the evolution of large asexual populations and
obtain our primary results.  Finally, we connect this behavior to
the small population regime.

\subsection{The Fate of a Single Mutant Individual \label{onemut}}

We begin by considering the fate of a single mutant individual. We
assume that in a large clonal population of size $N$, at time $t =
0$ there is a single mutant individual with a beneficial mutation
conferring fitness advantage $s$.  We denote the size of the
subpopulation carrying this beneficial mutation at time $t$ as
$n(t)$ (by assumption, $n(0) = 1$).  We study
the effects of selection and drift on this population by
calculating the probability distribution of future $n(t)$,
$\pr{n;t}$, assuming that no further mutations occur.  This
provides an essential building block for all the subsequent analysis,
and also illustrates the basic approach in a simple context.

Throughout this analysis, we assume that the number of individuals
with the beneficial mutation is small relative to the total
population size, $n \ll N$.  Thus the mutants do not significantly
affect the overall fitness of the population, and hence do not interfere
with one another. Naturally, if the mutant becomes established it
will supplant the wild-type population and this condition will
cease to be true.  By this time, however, the mutant subpopulation
will be large enough that we can switch from the stochastic analysis
described here to a correctly saturating deterministic analysis.

Because the mutant subpopulation is too small to affect the mean
fitness, mutant individuals have a birth rate $1+s$ and death rate
$1$. We define $g(n, n_0, t)$ to be the probability of having $n$
descendants at time $t$, starting from $n_0$ descendants at $t = 0$.
We are interested in calculating $g(n, 1, t)$.  The probability of a
birth or death event in a unit of time $dt$ is $(2+ \sb) dt$, and
this event is a birth with probability $\frac{1+\sb}{2+\sb}$ and a
death with probability $\frac{1}{2+\sb}$.  This means that \eon g(n,
1, t) = \frac{1}{2+\sb} (2+\sb) dt \delta_{n, 0} +
\frac{1+\sb}{2+\sb} (2+\sb) dt g(n, 2, t-dt) + [1 - (2+\sb) dt] g(n,
1, t-dt), \eoff where $\delta_{n, 0} = 1$ if $n=0$ and is $0$
otherwise. Assuming that individual lineages are independent (valid
since $n \ll N$), \eon g(n, 2, t) = \sum_{\alpha = 0}^n g(\alpha, 1,
t) g(n - \alpha, 1, t). \eoff Using this fact and defining the
generating function \eon G(z, t) = \sum_{n = 0}^{\infty} g(n, 1, t)
z^n, \eoff we find \eon \pd{G(z, t)}{t} = 1 + (1 + \sb) [G(z, t)]^2
- (2+\sb) G(z, t), \eoff with initial condition $G(z, 0) = z$
corresponding to the one individual initially.  We solve this
differential equation for $G(z, t)$, finding \eon G(z, t) = \frac{(z
-1)(1 - e^{\sb t}) + z \sb}{(z-1)(1 - (1+ \sb)e^{\sb t}) + z \sb}.
\label{gone} \eoff

We can now determine $\pr{n;t} \equiv g(n, 1, t)$ from $G(z, t)$. A
standard inversion yields \eon \pr{n;t} = \frac{\sb^2 e^{\sb
t}}{\left[ (1 + \sb) e^{\sb t} - 1  \right] \left[ (1+ \sb) e^{\sb
t} - 1 - \sb \right]} \left[ \frac{(1+\sb) e^{\sb t} - 1 - \sb}{(1 +
\sb) e^{\sb t} - 1} \right]^n \eoff valid for $n > 0$, and \eon
\pr{n=0; t} = \frac{e^{\sb t}-1}{(1+\sb) e^{\sb t} - 1}. \eoff We
are interested primarily in understanding the distribution of $n$
given that the mutant population is not destined to go extinct. This
is given approximately by \eon A(n, t) \equiv \pr{n; t | \textrm{not
extinct}} \approx \frac{\sb}{(1+\sb) e^{\sb t} -1 -\sb} \exp \left[
-\frac{\sb n}{(1 + \sb) e^{\sb t} - 1} \right] \ . \eoff Here we
have approximated the geometric factor by a simpler exponential in
$n$ which is valid for $t\gg \frac{1}{s}$, the regime of primary
interest. Note, however, that although the crucial features are more
apparent in the approximate expression, all the results below follow
from the exact equations.

At this stage, the above results merely reproduce classical
analysis, but it is useful to pause to compare them with various
intuitive predictions.  We first compute the average number of
mutant individuals at time $t$, \eon \langle n \rangle = e^{\sb t},
\eoff which confirms our understanding of what it means to have a
beneficial mutation with advantage $\sb$. However, most of the time
the mutation will die out. Conditional on not going extinct, \eon
\ev{n | \textrm{not extinct}} = 1 + \frac{e^{st} - 1}{s} \ , \eoff
which is larger at long times by a factor of $1/s$. At short times,
$t \ll \frac{1}{s}$, this is $\ev{n | \textrm{not extinct}} \approx
1 + t$.  At long times, $t \gg \frac{1}{s}$, the extinction
probability becomes $\frac{1}{1+s} \approx 1 - s$, and $\ev{n |
\textrm{not extinct}} = \frac{1}{s} e^{st}$. Note that short times
corresponds to $n \ll \frac{1}{s}$, while long times mean $n \gg
\frac{1}{s}$. (Note also that none of these expressions saturate as
$n$ approaches $N$; they are valid for $n \ll N$, as discussed
above.)

It is useful to ignore mutations that are destined to go extinct due
to drift, and focus only on those that are destined to become
established.  We do this for the remainder of this section; all
results are thus implicitly conditional on non-extinction. However,
some care is required. If a mutation occurs at time $t=0$, and
survives drift to become established, it may seem that on average it
will grow as $n(t) = e^{st}$, because it started from one individual
at $t=0$ and grows on average exponentially. However, this is
incorrect.  Given that it survived drift, it is likely to have grown
\emph{faster} than $e^{st}$ in the early stochastic phase of its
growth during which drift is faster than selection \citep{276}.
This is apparent from the expressions above:  for $t \ll
\frac{1}{s}$, $\ev{n | \textrm{not extinct}} \approx 1 + t$ which
is much faster than $\ev{n} = e^{st} \approx 1 + st$. Once the
population is large, and stochastic effects can be neglected, it
naturally grows as $e^{st}$.  However, because it grew faster than
this in the early stochastic phase, it will on average be larger
than if it had grown this fast through its entire history. As is
clear from the expression for the average $n$ at long times,
$\langle n | \textrm{not extinct} \rangle \approx \frac{1}{\sb}
e^{\sb t}$, the behavior can be crudely approximated by assuming
that it started at size $\frac{1}{s}$ (rather than size $1$) at
$t=0$, and then grew exponentially as $e^{st}$ thereafter. This
approximation is of course not valid during the early phase of
growth.  However, as we will later be concerned primarily with
relatively large subpopulations, this is a simple way to take into
account the stochastic effects.  Note that the above also implies
that, given that a mutation is not destined to go extinct due to
drift, it will fix in a time of order $\frac{1}{s} \ln [Ns]$,
\emph{not} $\frac{1}{s} \ln N$, as is sometimes seen in the
literature.  For $s \sim 0.01$, this is a difference of about $500$
generations.  To be more precise, the fixation time is a random
variable with a distribution of width $\frac{1}{s}$ and mean close
to $\frac{1}{s} \ln [Ns]$, rather than the naive $\frac{1}{s} \ln
N$. [For small $s$, this implies that the variation in the fixation
time ($\sim \frac{1}{s}$) is small compared to the difference
between the mean fixation time and the naive result.]

For much of the subsequent analysis, we will be concerned with the
size of a subpopulation only after it is big enough to be
essentially deterministic.  This is because when subpopulations are
rare enough to be stochastic, they are too small to influence the
mean fitness of the population and, as we shall see, are also small
enough that the chance of mutations occurring within these
populations is negligible (the former assumption requires $Ns \gg
1$, while the latter we explore in Appendix G). However, as the
above discussion makes clear, the stochastic phase of growth affects
the later deterministic dynamics.  Thus we are interested in
``summing up'' the stochastic effects in terms of their impact on
later deterministic growth. We want to do better than merely
understanding how stochasticity affects $\langle n \rangle$, as
described above; we want to understand the full effect of
stochasticity in determining the distribution of $n(t)$ in its
deterministic growth phase.

Focusing only on the effects of stochasticity on later deterministic
dynamics allows us to make a key simplification. Once the
subpopulation is large enough to grow deterministically, but still
small enough that saturation can be ignored (i.e. $\frac{1}{s} \ll n
\ll N$), its dynamics can be described by $n= \nu e^{st}$. The value
of $\nu$ is a random variable that depends on how fast the
population grew in its stochastic phase. However, the \emph{only}
effect of this stochasticity on the later deterministic growth is to
create random variation in $\nu$. As almost all this stochasticity
accumulates at short times, at large $t$ (after the population has
become deterministic) we can describe the overall effects of
stochasticity in terms of a probability distribution $\pr{\nu}$.
This is a big simplification, because the full probability
distribution conditioned on non-extinction, $A(n,t)$, depends on
both $n$ and $t$, while for large $t$ $\pr{\nu}$ is
\emph{independent} of $t$, as we show below. This simplification is
possible because at large $t$ the only time-dependence is the
deterministic exponential growth.

We can justify the above heuristic argument rigorously. The
definition of $\nu$ is just a transformation of $n$, $\nu \equiv n
e^{-st}$. This is valid in the early stochastic phase of growth as
well as in the later deterministic phase. However, in the stochastic
phase we do not expect that $\nu$ will be independent of $t$.  As we
have the probability distribution $A(n,t)$, it is straightforward to
transform this to the distribution $\pr{\nu;t}$. When we take the
large-$t$ limit of $\pr{\nu; t}$, it becomes independent of $t$.
This justifies our expectation that at large $t$, we have
$\pr{\nu}$, independent of time.

Rather than using the probability distribution of $\nu$, it will
prove useful to define a related variable $\tau$ by \eon n \equiv
\frac{1}{s} e^{s(t - \tau)}. \eoff The random variable $\tau$ is
simply related to $\nu$: $\tau = -\frac{1}{s} \ln (\nu s)$. Since
$\tau$ is a simple transformation of $n$, we can immediately
calculate $\pr{\tau\in (\tau,\tau+d\tau); t | \textrm{not
extinct}}=\dpr{\tau;t}d\tau$ (with $\dpr{\tau}$ the probability
density as we are treating $\tau$ as a continuous variable) from
$A(n, t)$; again, for the remainder of this section, all
distributions are conditional on non-extinction. We find \eon
\dpr{\tau; t| \textrm{not extinct}} = \frac{s}{(1 + \sb) e^{\sb t} -
1} \exp \left[ \sb (t - \tau) - \frac{e^{\sb (t - \tau)}}{(1+ \sb)
e^{\sb t} - 1} \right]. \eoff As with $\nu$, this describes the
distribution of $n$ both in the deterministic and stochastic phase.
Since $n$ depends on $t$, so does the distribution of $\tau$.
However, as expected from the previous discussion, the distribution
of $\tau$ becomes independent of $t$ for large $t$. We define
$\taunuc$ as $\tau(t \to \infty)$, and find \eon B(\taunuc) \equiv
\dpr{\taunuc | \textrm{not extinct}} = \frac{\sb}{1 + \sb} \exp
\left[ -\sb \taunuc - \frac{e^{\sb \taunuc}}{1 + \sb} \right]. \eoff
The average value (as well as higher moments) of $\taunuc$ can be
easily computed from this distribution.  We have \eon \langle
\taunuc \rangle = \frac{1}{\sb} \ln \left[ \frac{e^{\gamma}}{1 +
\sb} \right] \approx \frac{\gamma}{s}, \eoff where $\gamma$ is
Euler's constant $\gamma = 0.577216$.

The variable $\taunuc$ has an intuitive interpretation:  $\taunuc$
is the time at which $n$ would have reached size $\frac{1}{\sb}$ had
it always grown deterministically at rate $\sb$, as calculated by
looking at $n(t)$ at large $t$ and extrapolating backward. This is
illustrated in Fig. 5a. We can therefore approximate the
destined-to-be-established subpopulation as drifting randomly for a
time $\taunuc$, at which time it reaches size $\frac{1}{\sb}$, and
then grows deterministically thereafter.  With this simplification,
the \emph{only} important stochasticity is the duration of the drift
period.  This is the key simplification which allows us to smoothly
connect the branching process with the nonlinear dynamics once the
subpopulation is no longer rare.  It jibes with our intuitive
expectation that the subpopulation is dominated by drift when rarer
than $\frac{1}{\sb}$, and then behaves deterministically once it
exceeds this size. Note, however, that in addition to telling us
nothing about $n(t)$ before time $\taunuc$, it also gives a slightly
inaccurate picture immediately after $\taunuc$ when $n(t)$ is around
$\frac{1}{s}$. The time $\taunuc$ is \emph{not} in fact the time at
which the subpopulation reaches size $\frac{1}{s}$ (see Fig. 5a).
Rather, it is the time at which $n(t)$ {\it would} have reached size
$\frac{1}{s}$ if we  assumed that it always behaved
deterministically, but it gets the large-$t$ behavior right.  In
fact, some small drift does take place after reaching size
$\frac{1}{\sb}$; our approximation doesn't ignore this drift, but
rather adds up all the drift that takes place through all the time
and rolls it into a change in $\taunuc$.  This can thus be thought
of as the time at which the mutation establishes.  In asking how
quickly beneficial mutations accumulate, this is the most natural
variable.

The caveats above illustrate why it is perfectly consistent to have
$\taunuc < 0$; the distribution $B(\taunuc)$ above shows that this
is not even particularly improbable.  This reflects the fact that,
given that a mutant subpopulation is not going to go extinct, it
is reasonably likely to grow remarkably fast in the early
stochastic phase.  A $\taunuc < 0$ simply indicates that the
mutant subpopulation grew so fast when rare that if we look at the
subpopulation size much later and assume it always grew
exponentially at rate $s$, the subpopulation would have had a size
larger than $\frac{1}{s}$ at $t=0$.

We note that $\langle \taunuc \rangle = \frac{1}{\sb} \ln \left[
\frac{e^{\gamma}}{1 + \sb} \right]$, while $\langle n(t) \rangle =
\frac{1}{s} e^{st}$ for large $t$ (as always, conditional on
non-extinction). This may naively seem inconsistent, since $n(t) =
\frac{1}{s} e^{s(t-\taunuc)}$ for large $t$. However, it merely
reflects the fact that $\langle e^X \rangle \neq e^{\langle X
\rangle}$.  The difference between these two averages is in fact the
essential reason that $\taunuc$ will prove to be such a useful
variable to focus on.  This is because the value of $\langle n(t)
\rangle$ depends much more sensitively on the tails of $\pr{n;t}$
than does $\langle \taunuc \rangle$.

\subsection{Mutants Generated by a Changing Population}

The above analysis of the population size of a clone founded by a
single mutant individual is an important building block. However,
it does not address the full problem.  We must now ask how the
mutants arise in the first place.  In the simplest case, we might
imagine a wild-type population of size $N$, starting with $0$
mutants at time $t=0$. This population generates mutants at rate
$N \u$. Each mutant follows the dynamics given in the above
section, beginning at the time it was created, but now we have
multiple such initial mutants which are created at random times.

Generally, the relevant process is even more complex. Starting from
a wild-type population, a single-mutant subpopulation is generated,
experiences a stochastic period, and then begins to grow
deterministically.  Then double-mutants are created by mutation
within the single-mutant population while it is still growing (i.e.
before it fixes).  The rate at which these double-mutants are
generated increases with time because the single-mutant
subpopulation is growing.  Later, the double-mutants may themselves
generate mutants before they fix (and possibly before the
single-mutants fix), and so on.

We therefore must tackle a more general problem:  the distribution
of the population size $n (t)$ of a mutant clone which starts with
$0$ individuals and is ``fed'' by mutants from a less-fit clone of
(growing) size f(t).  If this less-fit clone is small enough that
its growth is stochastic, calculating the probability distribution
of the mutant population is extremely complex. Fortunately, most
non-viral organisms live in parameter regimes where a subpopulation
will never generate mutants destined to establish while it is still
so small that it must be treated stochastically; we discuss this in
Appendix G. Thus we take $f(t)$ to be some \emph{deterministic}
function describing the growth of the subpopulation from which
mutants arise.  Later we will set the origin of time in $f(t)$
stochastically, to reflect the stochasticity in the establishment of
this feeding population.

Note that we no longer need to condition on the mutant
subpopulation not being destined to go extinct.  Since this
subpopulation is being continuously fed with new mutations,
eventually one of these mutations will survive drift.  Thus at
long times the mutant subpopulation will never be extinct.

Unlike in the previous section, the growth rate of the stochastic
mutant population $n (t)$ is not necessarily $1+s$. Rather, for a
population with a total of $y$ mutations, the growth rate is $1 + (y
- \bar y) s$, where $ys$ is the fitness of the subpopulation $n(t)$
and $\bar y s$ is the mean fitness of the population. For
convenience, we write this as $1 + \p s$. The death rate of this
population is still $1$. Since $\bar y$ increases continuously, $\p$
is time-dependent. Despite this, we approximate $\p$ as a constant.
This is justified because we will only use the stochastic
description of $n(t)$ during the brief period during which it is
rare, and in this time $\p$ does not change significantly.  We
discuss this approximation in Appendix H.

We define $\eta(t - t_k)$ to be the number of descendants at time
$t$ of a single mutant which occurred at time $t_k$.  That is, given
that a mutation occurs from the ``feeding'' population to the
population $n$ at time $t_k$, $\eta (t - t_k)$ is the number of
descendants of this mutation at a later time $t$. Note that $\eta$
is the random variable whose generating function is given by $G(z, t
- t_k)$ above (\eq{gone}), but with $s$ replaced by $\p s$. We have
\eon n (t) = \sum_{k = 1}^M \eta (t - T_k), \eoff where $M$ is the
random number of individual mutations that have occurred and $T_k$
are the random times at which they occurred.

The number of mutations and their timings are an inhomogeneous
Poisson process, fed by the population $f(t)$. We therefore have
\eon \pr{M = m} = \exp \left[ - \int_{-\infty}^t \u f(t') dt'
\right] \frac{\left[ \int_{-\infty}^t \u f(t') dt' \right]^m}{m!}.
\eoff Note the lower limit of integration here represents the
earliest time that mutations are allowed to occur; we have chosen
this to be infinitely early. Although this includes $f(t)$ which are
zero for $t<0$, we discuss this choice of cutoff more generally in
Appendix E.  The timings of the mutations $T_k$, conditional on $M =
m$, are the ordered statistics of $m$ independent identically
distributed samples drawn from the distribution \eon F(x) = \left\{
\begin{array}{ll} \frac{\lambda(x)}{\lambda(t)} & x \leq t
\\ 1 & x > t \end{array} \right. \qquad \lambda(x) = \int_{-\infty}^x \u f(y)
dy. \eoff  This means that the joint distribution of the $T_k$
conditional on $m$ is given by \eon \dpr{t_1, \ldots t_m | M(t) =
m} = m! \prod_{i = 1}^m \frac{f(t_i)}{\int_{-\infty}^t f(x) dx}.
\eoff

The generating function for the distribution of the number of
mutant individuals, $n(t)$, is given by $H(z, t) = \langle
z^{n(t)} \rangle $. Conditioning on the distributions of $M$ and
the $T_k$ given above, and using the fact that $G(z, t-t_k) =
\langle z^{\eta(t - t_k)} \rangle$, we find that \eon H(z, t) =
\exp \left[ \u \int_{-\infty}^t [G(z, t-t') - 1 ] f(t') dt'
\right]. \label{heq} \eoff  To understand the full probability
distribution of $n(t)$, we simply have to plug in the appropriate
form $f(t)$ and then invert this generating function.

\subsection{An Exponentially Growing Population Feeding Another}

In large populations, there will typically be various multiple
mutants present, as described in the introduction and illustrated
in Fig. 3. We can now apply the results of the previous section to
this situation. As before, we define the the most-fit
subpopulation that is large enough to treat deterministically to
have fitness $(q-1)s$ above the mean fitness (note that $q$ is not
necessarily an integer). This subpopulation, $n_{q-1}$, grows
exponentially at rate $(q-1)s$. We define the origin of time such
that $n_{q-1}(t)$ is given by \eon n_{q-1}(t) = \frac{1}{q s}
e^{(q-1) s t}. \eoff Note that, analogous to the previous section,
we are approximating $q$ as constant --- we discuss this further
below.  The reason for defining the origin of time such that
$n_{q-1}(t) = \frac{1}{q s}$ at $t = 0$ will become clear below.
We now want to understand the stochastic dynamics of the
subpopulation a fitness $q s$ above the mean (denote this
population size by $n_q(t)$).  The subpopulation $n_{q-1}$ feeds
mutations to $n_q$; we therefore have $f(t) = n_{q-1}(t)$ in the
notation of the previous section.

This problem involves one exponentially growing population,
$n_{q-1}$, feeding another, $n_q$.  In analyzing it, we first step
back from our specific situation to study the general case of an
exponentially growing population with with size $N_1 = \nu_1 e^{R_1
t}$ feeding mutants at rate $\u$ to a stochastic population $N_2$
which on average grows exponentially with rate $R_2$.  We later will
substitute $\nu = \frac{1}{qs}$, $R_1 = (q-1)s$, and $R_2 = qs$. We
begin by plugging $f(t) = \nu_1 e^{R_1 t}$ into \eq{heq}, using the
obvious generalization of $G(z, t)$ to a population that grows at
rate $R_2$.  This gives us $H(z,t)$, the generating function of the
probability distribution of $N_2$. It is convenient at this point to
pass from generating functions to Laplace transforms by defining the
transform variable $\zeta = 1-z$. For our purposes we can assume
that $\zeta$ is small: this introduces errors into $\pr{N_2; t}$
when $N_2 \sim 1$, but we will never use $\pr{N_2; t}$ in this
regime. We find \eon H(\zeta, t) = \exp \left[ - \u \nu R_2
\int_{-\infty}^t \frac{\zeta e^{R_2 (t - v) }e^{R_1 v} dv}{\zeta
(e^{R_2 (t - v)}(1+R_2) - 1 ) + (1 - \zeta) R_2} \right]. \eoff
Substituting $u = \zeta e^{R_2 (t - v)}$, we find \eon \label{22eqn}
H(\zeta, t) = \exp \left[ - \u \nu \left[ \zeta e^{R_2 t}
\right]^{R_1/R_2} \int_{\zeta}^{\infty} \frac{u^{-R_1/R_2} du}{u +
R_2 u - \zeta +R_2 - \zeta R_2} \right] . \eoff Assuming that
$\zeta$ is small, the integral in this expression is independent of
$\zeta$ and is given by $(\frac{1}{R_2})^{R_1/R_2} \pi \csc \left[
\pi (1 - R_1/R_2) \right]$. We find \eon H(\zeta, t) = \exp \left[ -
\frac{\pi \u \nu \left[ \zeta e^{R_2 t}
\right]^{R_1/R_2}}{R_2^{R_1/R_2} \sin \left[ \pi(1-R_1/R_2) \right]}
\right]. \eoff

We can now substitute our values of $\nu_1$, $R_1$, and $R_2$ to
find that in our case \eon H(\zeta, t) = \exp \left[ - \frac{\pi \u
\zetaexp^{1-1/q}}{q s (q s)^{1 - 1/q} \sin(\pi/q)} \right]. \eoff
This is the standard form for the Laplace transform of a one-sided
Levy distribution, a well-studied special function. An integral
representation of this is the inverse Laplace transform of $H$, \eon
P(n_q, t) \equiv \pr{n_q;t} = \frac{1}{2 \pi i} \int e^{n_q \zeta}
H(\zeta, t) d \zeta, \eoff where the integral is over the imaginary
axis. For large $n_q$ this can be integrated to give $P(n_q) \sim
\frac{1}{n_q^{2 - 1/q}}$. [Note this distribution has infinite
$\ev{n_q}$, an unimportant and unbiological artifact of our choice
of cutoff in the integral for $H(z, t)$; this is discussed in
Appendix E below.]

To understand this distribution $P(n_q, t)$, we define a variable
$\tau_q$ similar to that described in the section above on the fate
of a single mutant.  We first define \eon n_q(t) \equiv \frac{1}{s}
e^{q s (t-\tau)}. \eoff As before, $\tau$ is time dependent, but for
$t \to \infty$ the distribution of $\tau$ is independent of $t$. We
define $\tau_q \equiv \tau(t \to \infty)$. As before, $\tau_q$ is
the time at which the subpopulation $n_q(t)$ \emph{would} have
reached size $\frac{1}{q s}$ had it always grown deterministically
at rate $q s$, as calculated by looking at the size $n_q(t)$ at
large $t$ and extrapolating backwards.  Unlike $\taunuc$, the value
of $\tau_q$ includes both the time for the mutation (or mutations)
to arise in the first place as well as time for their initial
stochastic growth. This is illustrated in Fig. 5b.

As in the section on the fate of a single mutant, we can think of
the mutant subpopulation as drifting randomly for a time $\tau_q$,
at which point it reaches size $\frac{1}{q s}$ and thereafter grows
deterministically.  We therefore sometimes refer to $\tau_q$ as the
``establishment'' time. As before, this is somewhat inaccurate in
describing the dynamics right around $\tau_q$ (or before) when the
population is around or below a size of $\frac{1}{q s}$.  Again
$\tau_q$ is \emph{not} actually the time the population reaches size
$\frac{1}{q s}$.  Now this is because both because future random
drift and future feeding mutations, after the population reaches
size $\frac{1}{qs}$, are included in the estimate of $\tau_q$.
However, for the purposes of understanding the dynamics of the
mutant population once it becomes large compared to $\frac{1}{q s}$,
it is valid to think of $\tau_q$ as the time it takes the population
to reach size $\frac{1}{q s}$.

We often wish to use moments of $\tau_q$.  These are
straightforward to calculate in principle, but somewhat tricky in
practice. We first note that because of the definition \eon n_q(t)
\equiv \frac{1}{q s} e^{q s(t - \tau)}, \eoff we have \eon \tau =
t + \frac{1}{s} \ln \left( \frac{1}{s} \right) - \frac{1}{s} \ln
[n_q(t)]. \eoff  We can therefore calculate $\ev{\tau}$ by
computing $\ev{\ln n_q}$ and plugging into this expression. Higher
moments of $\tau_q$ are easily computed by similar expressions;
these depend also on higher moments of $\ln n_q$.  We can
calculate $\ev{\ln^m n_qt}$ by noting that $\ev{\ln^m n_q} =
\lim_{\mu \to 0} \frac{\partial^m}{\partial \mu^m} \ev{n_q^\mu}$.
Using the integral representation of $P(n_q, t)$, we have \eon
\ev{n^{\mu}} = \frac{1}{2 \pi i} \int_{0}^{\infty} \int n^{\mu}
e^{n \zeta} e^{-a \zeta^{1-1/q}} d \zeta dn, \eoff where the
$\zeta$ integral is over the imaginary axis and we have defined
\eon a \equiv \frac{\u \pi }{s (q-1) \sin(\pi/q)} \frac{e^{(q-1) s
t}}{(q s)^{1 - 1/q}} . \eoff We integrate this to find \eon
\ev{n^\mu} = \frac{q}{q-1} \frac{\sin (\pi \mu)}{\sin \left(
\frac{\pi \mu q}{q-1} \right) } \frac{\Gamma(1 + \mu)}{\Gamma(1 +
\frac{\mu q}{q-1})} a^{\frac{\mu q}{q-1}}, \eoff where $\Gamma(x)$
is the Gamma function.

We can now calculate derivatives of this with respect to $\mu$ to
get $\ev{\ln^m n_q}$ and hence the moments of $\tau$.  For large
$t$, as expected, $\tau$ becomes independent of $t$.  For the mean
of $\tau_q$, we find \eon \ev{\tau_q} = \frac{1}{(q-1) s} \ln \left[
\frac{s}{\u} \frac{(q-1) \sin(\pi /q)}{\pi e^{\gamma/q}} \right],
\eoff where $\gamma = 0.5772$ is Euler's constant.  The variance of
$\tau_q$ is given by \eon \textrm{Var}(\tau_q) = \frac{\pi^2}{6}
\left[ \frac{1}{[(q-1)s]^2} - \frac{1}{[q s]^2} \right]. \eoff
Higher moments are also simple to compute if desired (and
demonstrate that there is substantial skew in the distribution of
$\tau_q$, as $\tau_q$ substantially smaller than $\ev{\tau_q}$ can
occasionally occur, while $\tau_q$ substantially larger than this
almost never do --- this is important in understanding the
fluctuations in the rate of adaptation around its steady state
value, and is discussed in Appendix D).

This calculation of $\ev{\tau_q}$ is somewhat involved because of
the need to use the integral representation of $P(n_q, t)$.  We can
get rough estimates (often useful in other contexts) via a simpler
method.  Namely, we define a ``typical'' population size by defining
$\tilde n_q \equiv \frac{1}{\zeta_1}$, where $\zeta_1$ is defined by
$H(\zeta_1, t) = e^{-1}$.  As is apparent from the definition of a
Laplace transform, $H(\zeta, t) = \int e^{-\zeta n_q} P(n_q, t)$,
for well-behaved distributions this typical value $\tilde n_q$ is
roughly like the median of $n_q$.  We can then get a typical value
$\tilde \tau_q$ from this using the relationship between $\ln n_q$
and $\tau_q$.  Doing this leads to a $\tilde \tau_q$ that is very
close to the $\ev{\tau_q}$ calculated above.

Note that the careful result for $\ev{\tau_q}$ is similar to the
crude calculation in the heuristic analysis section above, which
approximated the time required for a new mutation to arise at the
nose as $\int_{-\infty}^{\tau_q} \u n_{q-1}(t) qs = 1$, roughly the
typical time at which the first mutant destined to establish arises.
Note that this expression is only weakly dependent on the lower
cutoff to the integral, which is good since $n_{q-1}(t)$ is not
given accurately by the deterministic approximation in this regime.
This weak dependence appears for the same reasons in the careful
calculation of $\tau_q$ and is discussed in more detail in Appendix
E. The crude and careful results do differ, however. The careful
result accounts properly for the randomness in the timing of a new
mutation and the fluctuations during its early drift phase. It also
accounts for the fact that not only the first mutant destined to
establish at the nose contributes. Rather, as we will see later, of
order $q$ different mutations contribute significantly to the
establishment of a new most-fit subpopulation at the nose.  For
later considerations it is important to note that this means that
there is significant diversity even among the individuals which have
the same fitness.

\subsection{The Rate of Evolution and Maintenance of Variation at
Large N} We are now in a position to calculate the rate of evolution
and amount of variation maintained in large populations. In the
above calculations, we set $t=0$ to be the time at which the
population $n_{q-1}$ reached size $\frac{1}{qs}$. This corresponds
to the establishment time of this population.  After a (stochastic)
time $\tau_q$, the next more-fit subpopulation, $n_q$, establishes.
For the later deterministic dynamics of $n_q$, we can think of this
as the time when $n_q$ reached size $\frac{1}{qs}$. At this point,
we have reached the identical situation where we started, but with
the nose of the population fitness distribution moved forward by
$s$. In the steady state, the mean fitness of the population must
also have moved forward by $s$ in the average establishment time
$\ev{\tau_q}$. Thus the population at $n_q$ now has fitness only
$(q-1)s$ ahead of the mean.  It has size $\frac{1}{qs}$, but
thereafter grows exponentially only at rate $(q-1)s$, giving a
population size $\frac{1}{qs} e^{(q-1)st}$.

The process now repeats itself --- we can take this establishment
time of the new population ($n_q$, above) as the new $t=0$, and
after that this population  grows as we had described for the
original population $n_{q-1}$.  In fact, it now \emph{is} the
population $n_{q-1}$, since the mean fitness has increased by $s$.
Thus we can see that the mean fitness of the population and the
position of the nose move forward by $s$ in a time $\ev{\tau_q}$.
Thus the average rate of increase in fitness in the population is
\eon v = \frac{s}{\ev{\tau_q}}. \eoff  Note that this discussion
makes clear why, for consistency, we defined  the establishment time
for $n_{q-1}$ to be when this population reached size
$\frac{1}{qs}$, not $\frac{1}{(q-1)s}$. We also note that the
population which we had originally called $n_{q-1}$ is now $n_{q-2}$
and its size is given by $\frac{1}{qs} e^{(q-1)s \tau_q} e^{(q-2) s
t}$.

This change in the growth rate of the population we had originally
called $n_{q-1}$ raises an important point.  We defined $n_{q-1}(t)
\equiv f(t) = \frac{1}{qs} e^{(q-1)st}$, and used this expression in
calculating $P(n_q, t)$, particularly for large $t$. Yet at this
large $t$, our expression for $f(t)$ is not accurate, because the
mean has shifted and the population with original (relative) fitness
$(q-1)s$ is no longer growing exponentially at rate $(q-1)s$.
Fortunately, the mutations that occur after the establishment of
$n_q$ (when the expression $f(t)$ becomes inaccurate) do not greatly
impact its later population size, $n_q(t)$.  In other words, the
mutations that dominate the population $n_q$ happen early while
$n_{q-1}$ is still accurately given by $f(t)$. Yet one must also ask
whether these mutations happen too early when $f(t)$ is also not a
good approximation for $n_{q-1}(t)$ (because the definition of
$\tau_q$, which we used to define $f(t)$, includes mutations and
stochastic behavior that happen later).  Fortunately, the mutations
that matter from $n_{q-1}$ to $n_q$ do occur late enough that
$n_{q-1}$ is accurately described by $f(t)$. This can be checked by
studying the behavior of $\tau(t)$; we discuss this and related
subtle issues in Appendix G.

When $q$ is too small, the approximations above are no longer
justified.  Whenever $q < 2$, the growth rate of the subpopulation
$n_{q-1}$ slows substantially during the period while the important
mutations to $n_q$ are occurring.  That is, $n_{q-1}$ saturates
while $n_q$ is becoming established.  Thus our analysis in this
section is only valid for $q > 2$.  As we will see, this corresponds
to large $N$. We discuss the $q < 2$ case in the next section.
However, it is the large-$N$, $q > 2$ result that we are most
interested in --- this is where there are typically many multiple
mutations at once, and the behavior differs dramatically from the
successional-mutations regime.

Throughout this section, we have asserted a steady state in which
the mean fitness increases at the same rate as new mutations are
established, and have defined the lead in steady state to be $q s$.
Yet we have not discussed the balance between mutation and selection
which sets this steady state. In a very small population, mutations
are few and far between, and fixation times are relatively short.
Thus a single-mutant population at $s$ will establish and fix before
another mutation becomes established. This is thus $q=1$. In a
somewhat larger population, a single-mutant population at fitness
$s$ will again establish and begin to grow. However, since the
population is larger, it takes longer to fix. Before it does so,
another mutation occurs in this single-mutant population, creating a
double-mutant population at $2s$.  If the population is only
moderately large, the single-mutant will replace the wild-type
before a triple-mutant can arise. The process will then repeat; this
is $q=2$.  As $N$ continues to increase, we expect that $q$ will
also rise.

The relationship between $q$ and $N$ can be obtained from
 $\tau_q$.  As we have seen, immediately after
the subpopulation at $q$ becomes established, its size is
$\frac{1}{qs}$.  The subpopulation at $q-1$ has size $\frac{1}{qs}
e^{(q-1)s \tau_q}$, the subpopulation at $q-2$ has size
$\frac{1}{qs} e^{(q-1)s \tau_q} e^{(q-2) s \tau_q}$, and so on. All
of the subpopulations must add up to size $N$; in practice the total
is dominated by one or a few (compared to $q$) subpopulations.
Applying the fixed total populations condition  (and assuming that
all the $\tau_q$ are on average $\ev{\tau_q}$), we find  \eon
\label{qeq} q \approx \frac{2 \ln [N q s]}{\ln \left[ \frac{s}{\u}
\frac{(q-1) \sin (\pi/q)}{\pi e^{\pi/q}} \right]}. \eoff This is a
transcendental equation for $q$, but because of the logarithmic
dependence on $q$ on the right hand side it is easily solved by
iteration. For most purposes, even the zeroth approximation, \eon q
\approx \frac{2 \ln [N s]}{\ln \left[ \frac{s}{\u} \right]},
\label{qeqapprox} \eoff is sufficiently accurate. To get higher
accuracy one can plug this into the right hand side of \eq{qeq}.

As expected, the value of $q$ increases with $N$, and also increases
with $\u$ because when mutations happen more quickly there are more
of them in the population at once. The dependence on $s$ is more
complicated, because increasing $s$ both decreases the fixation time
(leaving less time for additional mutations to occur) and increases
the rate of mutations that establish (because it increases the
establishment probability). Note that $q$ is of order $\frac{1}{qs}
\ln [Nqs]$ (the basic selection timescale) divided by $\frac{1}{qs}
\ln (qs/\u)$ (the basic mutation establishment timescale).  This
makes sense, as $q$ is primarily what is determined by the balance
between these two forces.

With the value of $q$ determined self-consistently above (\eq{qeq}),
the mean fitness shifts by $s$ in exactly the time $\ev{\tau_q}$.
Thus the corresponding distribution of the subpopulations is indeed
a steady state.  By plugging \eq{qeq} into the expression for
$\ev{\tau_q}$ and substituting this into $v =
\frac{s}{\ev{\tau_q}}$, we can obtain the speed of evolution.  Doing
this using the lowest-order result in the iterative expansion for
$q$ (\eq{qeqapprox}), we find that the speed of evolution is roughly
\eon \label{veq} v \approx s^2 \left[ \frac{2 \ln [N s] - \ln \left[
\frac{s}{\u} \right]}{\ln^2 \left[ \frac{s}{\u} \right]} \right],
\eoff valid provided $q$ is reasonably large (basically, when $2 \ln
[Ns]$ is larger than $\ln \frac{s}{\u}$). If a more accurate result
is needed, we can simply carry the iterative expansion for $q$ to
higher order.

The calculations above confirm the intuitive picture and results
described in the heuristic analysis section above. The speed of
evolution is determined by two mostly independent factors. One
factor is the dynamics of the nose --- the feeding process from
$n_{q-1}$ to $n_q$ which sets $\tau_q$. This process depends
directly only on $\u$ and $s$; the only impact of $N$ here is via
its effect on the lead $qs$. The other factor is the dynamics of the
already established populations. This is dominated by selection, and
hence depends directly  only on $N$ and $s$; the only role of
mutation here is its role in setting $q$.

Our result is consistent with the fundamental theorem of natural
selection, which states that the speed of evolution is equal to the
variance of fitness in the population --- provided mutation is
negligible.  To see this, we first note that the bulk of the fitness
distribution is Gaussian.  This is because a population with $\ell$
more mutations than the mean grows as $e^{\ell s t}$, and the mean
shifts by $1$ during every time interval of $\tau_q$. This means
that at the end of an interval, the number of individuals with
$\ell$ mutations more than the mean is determined by its cumulative
growth over all these time intervals: $\exp(- \sum_{k=1}^{\ell} k s
\tau_q) \sim e^{-(s \tau_q/2) (\ell+1/2)^2}$, a Gaussian
distribution. We call the variance of this fitness distribution
$\sigma^2$.  The number of individuals that differ from the mean by
$ks$ is then roughly $N \sigma \exp[-(ks)^2/2 \sigma^2]$, and the
fittest established population
--- with $k \approx q$ --- will have of order $\frac{1}{qs}$
individuals.  We therefore expect $qs \approx \sigma \sqrt{2 \ln
Ns}$.  This means that if the fundamental theorem for natural
selection holds, we expect $v = \sigma^2 \approx \frac{(qs)^2}{2 \ln
(Ns)}$.  And indeed, some algebra verifies that this yields the
expression for $v$ in \eq{veq}.

The fundamental theorem of natural selection should apply whenever
mutation can be neglected compared to selection.  Since this is true
in the bulk (i.e., away from the nose) of the fitness distribution,
the correspondence between our result and the theorem is reassuring.
The speed of evolution is equal to the variance in fitness, as
usual.  Thus the crucial aspect of our calculation of the speed of
evolution can be viewed as an analysis of how much variance in
fitness a population maintains by mutations.  That is, we have
determined how a population maintains variance in fitness when it
continually generates variation due to mutations while at the same
time this variation is being selected on.  As we noted earlier,
however, neither the heuristic arguments nor  the analysis involve
the variance.  To compare with  the fundamental theorem we had to
extract the variance from the analysis  of the full fitness
distribution, but this was not necessary for obtaining the primary
results.  This is because the lead proves to be a more useful
measure of the width of the fitness distribution, because it is the
lead that is directly affected by new mutations at the nose.  The
variance is of course also increased by mutations,  but only as a
consequence of the dynamics of the lead and only after the new
mutant populations have grown to substantial numbers. The key fact
that the distribution is close to gaussian out almost to the nose,
which is many standard deviations above the mean, is indicative of
the little significance of the region near the mean that controls
the variance.

\subsection{Evolution at Moderate N \label{crossover}}

In addition to the evolution at large $N$, we want to understand
the crossover between small-$N$ and large-$N$ behavior. In this
subsection, we explore this crossover.

For very small $N$, the the successional-mutations regime obtains.
In the heuristic analysis section, we noted that mutations take
about $\frac{1}{N \u s}$ generations to establish in this regime,
and then fix in a much shorter time. Thus evolution is
mutation-limited, and we have $v \approx N \u s^2$. It is
instructive to redo this calculation using the machinery we
developed for the large-$N$ case. To do this, we must replace the
exponential form for $f(t)$.  As before, we take the establishment
time of the mutation at $(q-1) s$ to be $t=0$. Of course, here $q =
1$ so $(q-1) s = 0$. In this regime, each mutant fixes soon after
becoming established. For the purposes of the next establishment, we
can therefore approximate the population at $(q-1) s$ by \eon
n_{q-1}(t) \equiv f(t) = N \theta (t), \eoff where $\theta (t) = 1$
for $t > 0$ and $0$ otherwise. We substitute this form of $f(t)$
into $H$ and integrate, and take the inverse Laplace transform of
the result to obtain \eon \dpr{\tau_1} = \frac{s}{\Gamma(N \u)} \exp
\left[ -N \u s \tau_1 - e^{-s \tau_1} \right]. \eoff  This gives
$\ev{\tau_1} \approx \frac{1}{N \u s}$, so the velocity $v \approx N
\u s^2$, as expected.

We now turn to the intermediate regime.  For $N\u$ comparable to
$\frac{1}{\ln \left[ \frac{s}{\u} \right]}$ or larger, the fixation
time is not short compared to the establishment time. Thus we cannot
use $f(t) = N \theta(t)$. At the same time, the establishment time
is not so short compared to the fixation time that saturation in the
feeding population is unimportant (the large-$N$ case we have
focused on thus far). We therefore need to consider the case of a
growing and saturating population feeding another.  We assume that
the single-mutant always fixes before the triple-mutant population
establishes, so that we only have to consider two deterministic and
one stochastic clone in the population (i.e. $q$ between $1$ and
$2$).  The dynamics of the single mutant population a time $t$ after
it establishes are given by \eon f(t) = \frac{N e^{\s t}}{N \s +
e^{\s t}} \eoff Note that $f(t)$ initially grows as $e^{(q-1) \s
t}$, with $q=2$, but later slows to $e^{(q'-1) \s t}$ with $q'=1$
(i.e. it becomes approximately constant).  The crossover occurs over
a time interval of order $1/s$, which is much smaller than the
establishment times and is thus effectively a sharp transition. The
behavior of the feeding population is thus roughly equivalent to
having $q$ between 1 and 2. The stochastic population that it feeds
initially grows at rate $q \s$ with $q=2$. The establishment of this
stochastic population occurs at a time $\tau_2$ when, roughly, \be
\int_0^{\tau_2} \u f(t) dt \equiv \frac{c}{s}, \ee with $c$ of order
unity.  This yields \be \tau_2 \approx \frac{1}{s}\left[\ln Ns +
\ln\left(e^{c/\u N}-1\right)\right] \ . \ee A more careful analysis
(analogous to the earlier calculations of $\tau_q$) that takes into
account the distribution of $\tau_2$ yields a result that is the
same as the above simple argument but with a factor of order unity
inside $\ln Ns$, which is a small correction over the whole range of
validity. While in general $c$ will depend on the detailed birth and
death processes, and the speed of evolution in the successional
mutations regime will be proportional to $c$, for the dynamics we
have analyzed throughout, $c=1$.  We use this below. For $N \u \ll
1$, we obtain \be v \approx\frac{s^2}{\ln Ns + \frac{1}{N \u} },
\label{modn} \ee which crosses smoothly --- and simply! --- over
from the successional mutations behavior  for $Ns \ll \frac{1}{\ln
\left[ \frac{s}{\u} \right]}$ to $v\equiv \frac{s^2}{\ln \left[
\frac{s}{\u} \right]}$, which is just the result we obtain for
$q=2$.  When $N\u$ becomes of order unity,  from the above
expression we have $\tau_2 s \approx\ln Ns + {\cal{O}} (1)$.

For $N\u \gg 1$ the behavior is well into the multiple mutations
regime we analyzed earlier, and the results obtained for general
non-integer $q>2$ apply. The two sets of results match together for
$Ns\approx s/\u$, up to order-unity factors inside logarithms of
$Ns$ and of $s/\u$.  An example of the crossover between the two
regimes is shown in Fig. 6a.

\section{Transient Behavior}

So far, our analysis has assumed that the mutation-selection balance
has already been reached. If a population starts with an arbitrary
distribution of fitnesses, it will gradually approach the steady
state distribution. In this section, we describe this transient
behavior.  We focus on the case where the population is initially
monoclonal.  Other starting fitness distributions can be analyzed
using similar methods. We consider the large-$N$  concurrent
mutations regime (in the successional-mutations regime the
monoclonal population is already essentially in steady state).

Starting from a monoclonal population, we can calculate the dynamics
of the single-mutant subpopulation that arises by using the
small-$N$ results above, since here too the feeding population is
$f(t) = N \theta(t)$. It would now be tempting to assume that this
single-mutant population just grows exponentially at rate $s$ after
first becoming established. We could then immediately import our
previous results for the establishment time of the double-mutant
population $\tau_2$, triple-mutant population $\tau_3$, and so on.
We could then assume that all these populations establish in order
until the $q^{th}$ population, at which point the steady state would
be reached.

Unfortunately, this is wrong, for two reasons.  First, the
single-mutant population grows \emph{faster} than exponentially at
rate $s$ because it is receiving mutations from the still-large
wild-type population.  Because of this, the double-mutant population
establishes more quickly than the steady state $\tau_2$, and then
itself grows faster than exponentially with rate $2s$ because it is
receiving more mutants from the fast-growing single-mutant
population.  This then affects the triple-mutants, and so on.  The
second complication is that the mean fitness does not stay at the
wild-type value until the $q^{th}$ mutation has established, so it
takes more than $q$ establishments to reach steady state.

Rather than attempt to find a closed-form analytical result, we
discuss here an algorithmic solution to the transient dynamics. We
proceed in steps. First, we calculate the lead from the current
fitness distribution. Based on this, we calculate the next
establishment time (interpolating if the lead changes during this
period because of an increase in the mean fitness).  We then
calculate the new fitness distribution and the new lead, and repeat
the process.

When calculating the establishment times, we must remember that the
feeding populations are not necessarily growing as simple
exponentials.  Earlier we used the establishment time $\tau_p$ to
approximate the population size of $n_p$ as $n_p(t) = \frac{1}{ps}
e^{ps(t - \tau_p)}$, a simple exponential.  We noted that this is
inaccurate while $n_p$ is around $\frac{1}{ps}$, because it includes
both future mutations from $n_{p-1}$ to $n_p$ and future
stochasticity. Since we have used this form of $n_p(t)$ to calculate
the establishment time of the next more-fit subpopulation, this
approximation for $n_p(t)$ must be accurate by the time the
mutations which lead to the subsequent establishment occur. In the
steady-state case, this holds, as shown in Appendix G. However, for
the transient dynamics it is not always correct.

This problem is most serious for the single-mutant population, which
we consider now.  The wild-type population has roughly constant size
$N$ during the period when the single-mutant population is rare.
This means that the single-mutant population grows on average as
\eon n_1(t) = \frac{N \u}{s} \left[ e^{st} - 1 \right]. \eoff This
reaches size $\frac{1}{s}$ after a time of order $\frac{1}{N \u s}$
generations.  However, the inferred establishment time (by
extrapolating backwards) is $\tau_1 = - \frac{1}{s} \ln \left[ N \u
\right]$ generations. This is substantially negative because
mutations that occur well after the population reaches size
$\frac{1}{s}$ contribute significantly to $n_1$.  The approximation
we used before would be to take $n_1 = \frac{1}{s} e^{s(t - \tau_1)}
= \frac{N \u}{s} e^{st}$ in calculating the establishment time of
the double-mutant population $\tau_2$. But using the correct form of
$n_1$, we find that the first double-mutants occur roughly at time
$t = \frac{1}{s} \ln \left[ 1 + \frac{s}{\u} \frac{1}{N \u}
\right]$. Thus when $N \u \ll \frac{s}{\u} $ --- corresponding to
$q<4$ --- double mutants do not occur until our usual approximation
for $n_1$ becomes reasonable.  We can therefore use our previous
calculation of the establishment time $\tau_2$ from the steady-state
analysis above.  All future establishment times (i.e. $\tau_3$ for
the triple-mutants, etc.) can similarly be imported directly from
the steady state calculations.  However, when $ N \u\gtrsim
\frac{s}{\u}$ ($q\ge4$), we must use the correct form of $n_1$ to
calculate $\tau_2$ and $n_2$.  In this case, $n_2$ will also grow
faster than our usual approximation $n_2 = \frac{1}{2s} e^{2s(t -
\tau_2)}$ would predict.  We must therefore repeat this procedure to
consider whether it is reasonable to calculate $\tau_3$ based on our
usual approximation, or whether we need to use the more complex form
for $n_2$. However, this effect is much weaker than for $n_1$; it
only matters if $N\u$ is much larger than in the previous condition.
If it does matter, we must again ask if the more complex form for
$n_3$ will be important in calculating $\tau_4$; this will only
matter if $N\u$ is larger yet. The number of establishments for
which we have to take this subtlety into account therefore depends
on $\frac{1}{2}q=\ln(Ns)/\ln(s/m)$:  the larger the steady state
$q$, the more transient establishments we must consider. In
practice, in comparing with previous experiments we have found that
considering the complex form of $n_1$ in calculating $\tau_2$ is
sometimes necessary, but all future establishments can be calculated
using the steady-state large-$N$ results \citep{n1exp}, because in
these experimental situations $q$ is never much larger than $4$.

A second subtlety in the above algorithmic approach is the way in
which the mean fitness changes; it does not increase in evenly
spaced steps of size $s$ as it would in steady state.  For example,
the double-mutant subpopulation can become established soon after
the single-mutant subpopulation does. Then, as it grows twice as
fast, it will outcompete the single-mutant subpopulation while both
are still rare. We call such an event a ``jump,'' since it will lead
to a jump in the mean fitness by $2s$ when the double-mutants become
the dominant subpopulation. Of course, it is also possible that the
triple-mutants will jump past the double-mutants, or that the
double-mutants will jump the singles, and then the quadruple-mutants
will jump the triples, etc.  These effects can lead to complex
dynamics of the mean fitness during the transient time before the
steady state is established.  However, {\it given} the establishment
times of the various populations, the time dependence of the mean
fitness is straightforward to calculate from the {\it deterministic}
dynamics of the competing subpopulations that are growing
exponentially.

Putting all these effects together, we can construct an algorithmic
solution for the transient dynamics.  We calculate the first
establishment time, and note at what time this new subpopulation
will change the mean fitness.  We then calculate the next
establishment time, and again the implied future effects on mean
fitness (modifying previous such results if jumping events will
occur).   We continue to repeat this process.  When the mean fitness
changes, we note how this changes the lead and adjust the
establishment times appropriately. We iterate this process until the
steady-state lead, $qs$, is reached. Even after that there can be
some lingering effects of the transient, as the rest of the fitness
distribution may not yet have reached the steady-state gaussian
profile. Yet soon thereafter the steady-state behavior is indeed
reached.

Rather than using this algorithmic approach, it is also possible
to use a deterministic approximation for the transient behavior.
Starting from a monoclonal population, the timing of the first few
establishments are given accurately by a deterministic
approximation.  However, this typically cannot give us the full
transient dynamics, because stochastic effects at the nose become
important once the fitness distribution grows to a substantial
width, which usually occurs before the transient regime is over.
This deterministic approach is also less versatile, as is only
valid for some starting distributions.

The transient behavior can be quite important.  During the transient
phase, the accumulation of beneficial mutations proceeds more slowly
than in the steady state, because after the first few
establishments, but before the steady state is reached, the lead
will be $ps$ with establishment interval approximately $\tau_p <
\tau_q$ (since $p< q$). Thus a clonal population will accumulate
beneficial mutations slowly at first, before the rate of
accumulation gradually increases to its steady-state rate. This
slower transient phase lasts a substantial time --- longer than it
takes to accumulate $q$ mutations once the steady state has been
established, again because $\tau_p < \tau_q$ for $p < q$ (and, as
noted above, in fact it can take more than $q$ establishments to
reach the steady state).

\section{Deleterious Mutations \label{del}}

Our simplest model neglects deleterious mutations. But deleterious
mutations can alter the dependence of $v$ on the mutation rate (and
on $N$), because increasing $\u$ typically comes at the cost of also
increasing the deleterious mutation rate. This has proved an
important consideration in clonal interference analyses \cite{331,
294}.  In this section, we consider qualitatively and
semi-quantitatively various effects of different sized deleterious
mutations in the simple model in which all the beneficial mutations
have the same $s$.  The effects of deleterious mutations of size $s$
in this model have been studied by \citet{313}.  Here we discuss
briefly the effects of deleterious mutations of various sizes, but
leave detailed analysis for future work.

It is useful to separate the effects of deleterious mutations into
their impact on the dynamics of the bulk of the distribution (and
hence the mean fitness) and their effects on the establishment of
new most-fit clones at the nose.  In the bulk of the distribution,
deleterious mutations come to a deterministic mutation-selection
balance which alters the shape of the fitness distribution and
reduces the mean fitness.  This effect actually \emph{speeds up} the
evolution:  if the deleterious mutations had no effect at the nose,
their impact in reducing the mean fitness would increase the lead
and thus make new establishments at the front occur \emph{faster}.
But deleterious mutations at the nose have the opposite effect:
they slow down the growth of the most-fit populations and  decrease
the fitness of some of these individuals, reducing the rate at which
new more-fit individuals establish.

In understanding these effects, it is useful to consider
large-effect and small-effect deleterious mutations separately.
First we consider deleterious mutations whose cost $s_d$ is larger
than $s$.  When a deleterious mutation with $s_d \gtrsim s$ occurs
at the nose, that individual is no longer at the nose. Thus the
deleterious mutations just reduce the effective growth rate at the
nose.  If $U_d^>$ is the mutation rate to deleterious mutations with
$s\gtrsim s_d$, then the growth rates of subpopulations at the nose
are simply reduced by $U_d^>$.  The effect of deleterious mutations
on the mean fitness is also simple, because the mean fitness of the
population is dominated by the largest subpopulation (which is
exponentially larger than all others). Thus in considering the
effect of the deleterious mutations on the mean fitness, we can
focus on their impact in this subpopulation. This remains the
largest subpopulation for roughly $\frac{1}{s}$ generations, which
for these values of $s_d$ is larger than $\frac{1}{s_d}$. Thus it
comes to a deleterious mutation-selection balance while it is
largest, since this balance is obtained in
($\frac{1}{\sqrt{(U_d^>)^2 + s_d^2}}$) generations. This means that
the deleterious mutations reduce the mean fitness by $U_d^>$ (up to
small corrections due to the dynamics and the other subpopulations).
This reduction in the mean fitness effectively increases the lead by
$U_d^>$, which increases the growth rates at the nose by the same
amount.  This cancels the effect of the deleterious mutations at the
nose.  Thus deleterious mutations with $s_d \gtrsim s$ have very
little  net effect on $v$:  they do not change the rate of new
establishments at the nose, up to the small corrections noted above.
This is not surprising --- the deleterious mutants are all doomed,
so roughly speaking their effect is simply to reduce the effective
fitness of all individuals equally, which has no net effect on $v$.
But they do increase the lead $qs$, which changes the shape of the
fitness distribution and can slow down the speed somewhat.

For weakly deleterious mutations with $s_d \ll s$, which occur at
mutation rate $U_d^<$, the effects are  more complicated. In this
case, the fact that an individual at the nose has a deleterious
mutation does not make it substantially less likely to be the source
of a new nose-extending mutation. Thus the effective growth rates at
the nose are unaffected by deleterious mutations. However, some
nose-extending mutations will occur in individuals with one or more
deleterious mutations, and hence will not necessarily extend the
nose by $s$. Instead, they will sometimes have an effect $s - s_d$,
or $s - 2 s_d$, or less. We can estimate the strength of this effect
by using a deterministic approximation for the deleterious mutation
accumulation at the nose.  When $\frac{U_d^<}{qs} \ln \left[
\frac{s}{\u} \right] \ll 1$ (or, roughly, when $\frac{U_d^<}{s} \ll
1$), we find that on average, nose-extending mutations are burdened
by a deleterious load of $\frac{U_d^< s_d}{(q-1) s^2} \ln
\left[\frac{s}{\u} \right]$. Thus the effect of the deleterious
mutations at the nose is to reduce the effective $s$ by the amount
$\frac{U_d^< s_d}{(q-1) s} \ln \left[ \frac{s}{\u} \right]$, which
is small compared to $s$. This will tend to slow the evolution. An
analogous calculation applies when $\frac{U_d^<}{s} \gg 1$; here the
deleterious mutations have a larger effect, but still only produce
an average fitness cost at most of order $s_d$. The effect of the
deleterious mutations on the bulk of the distribution is again to
reduce the mean fitness of the population. The amount of this
reduction, however, depends on the accumulation of deleterious
mutations throughout the fitness distribution, not just in the
most-fit subpopulation as before, because $\frac{1}{s} <
\frac{1}{s_d}$. Still, this tends to reduce the mean fitness by an
amount of order $U_d^<$. This again speeds the evolution, and
partially cancels the slowing effect at the nose.  Thus deleterious
mutations with $s_d \ll s$ affect $v$ by increasing the effective
lead by $U_d^<$ and reducing the effective $s$ by roughly
$\frac{U_d^< s_d}{(q-1) s} \ln \left[ \frac{s}{\u} \right]$ (when
$\frac{U_d^<}{s} \ll 1$) or by of order $s_d$ (when
$\frac{U_d^<}{s}$ is larger).  These effects are all small.

Small effect deleterious mutations will also slow down the evolution
via the accumulation of them during the  collective-sweep time,
$qs/v\approx \ln(s/\u)/s$, in which a subpopulation grows from being
the lead population to the dominant population.  We expect this
effect to be largest relative to the effects of these deleterious
mutations on the dominant subpopulations when $1/s_d$ is of order
the collective-sweep time.  This effect reduces the speed by an
amount proportional to $U_d$.

To analyze in more detail the quantitative effects of deleterious
mutations (even in the simplest single-beneficial-$s$ model) is
beyond the scope of this paper.  Note in particular that the
analysis in this section is invalid when the deleterious mutation
rate is large enough that the deterministic approximation for their
behavior at the nose becomes incorrect. In this regime --- on the
border between Muller's ratchet and adaptive evolution --- a more
careful analysis is needed. We leave this discussion, which is
essential for understanding the dependence of the rate of evolution
on the mutation rate when mutation rates become large, for future
work.

\section{Simulations}

Our analysis involves a number of approximations.  While we have
analyzed their validity above and in the appendices, we also used
computer simulations to test our results. In this section, we
describe these simulations and the comparisons to our results.

We started our computer simulations with a clonal population with a
birth and death rate of $1$, and a mutation rate of $\u$. We
arbitrarily defined this population to have fitness $0$.  We divided
time into small increments.  At each increment, we first calculated
the average fitness $\bar y$, and then produced births, deaths, and
mutations with the appropriate probabilities. The birth rate of
individuals at fitness $y$ was set to be $1 + (\bar y - y)$ (with
$\bar y - y$ always small compared to unity), their death rate $1$,
and the mutation rate $\u$. We then repeated this process to
simulate the population dynamics, providing a full stochastic
simulation of the simplest constant-$s$, beneficial-only model
analyzed above. We recorded the mean fitness and lead as a function
of time and, for each set of parameters, measured the average $v$
and $q$ once past the initial transient regime.

We carried out these simulations at a variety of different parameter
values.  The match between simulations and our theoretical results
was good, provided the conditions for the validity of the concurrent
mutations regime obtained.  Examples of these comparisons are shown
in Figs. 6 and 7. In Fig. 6, we show the theoretical predictions for
the average speed of adaptation (using both the lowest order
iterative result for $v$ presented in \eq{veq} and a higher order
iterative expansion) compared to simulation results as a function of
$N$, $\u$, and $s$. In Fig. 7, we show similar comparisons for the
average lead $q$ (again using the lowest order iterative result for
our theoretical predictions). The agreement is good in both cases,
though our theory slightly underestimates both $v$ and $q$. This may
be due to the effects of fluctuations in $\tau_q$ (described in
Appendix D) slightly increasing the mean $v$ and $q$ because of
their non-linear effects, or to other factors arising from
$\ln(s/\u)$ not being sufficiently large for the asymptotic results
to obtain to this accuracy.

\section{Distributions of $s$, and Relationship to Clonal Interference Analyses \label{distn}}

The simple model we have analyzed assumes that all beneficial
mutations confer the same advantage $s$.  But in most natural
situations different beneficial mutations will have different
fitness effects. This does not change the basic dynamics of
adaptation in large asexual populations: many beneficial mutations
still occur before earlier ones have fixed and these can interfere
with each other's fixation (Fig. 1b). And the successful mutant
lineages are likely to have had multiple beneficial mutations before
they fix. But because mutations in different lineages cannot
recombine together, many will be wasted when other lineages
outcompete them.

There are two reasons beneficial mutations are wasted.  We have
focused on the wasting of mutations because they occur in
individuals who are not very fit (i.e. away from the nose) and are
therefore handicapped by their poor genetic background.  But when
beneficial mutations have a variety of different effects, there is
another way they can be wasted: small effect mutations can be
outcompeted by larger mutations that occur in the same or similar
genetic background. We refer to this latter process as ``clonal
interference.''  As before, we use the term ``clonal interference''
to refer to this first effect only (despite earlier broader
definitions), consistent with the focus of recent work on the
subject.  This can only occur when not all mutations have the same
fitness increment, and is thus absent in the simple constant-$s$
model.

Recent work by \citet{1} and others \citep{410, 417, 351, 331,292,
294, 562} has taken the opposite approach to the multiple
constant-$s$ mutations approximation and focused instead on the
effects of  clonal interference, while ignoring multiple mutations.
In this section, we first summarize the conclusions of such
analyses, which assume all mutations occur on the {\it same} genetic
background. We then consider the effects of including both clonal
interference and multiple mutations.   As we will argue, whenever
the former plays a significant role, so does the latter.

The now-conventional clonal interference analysis considers how
small effect mutations can be outcompeted by larger mutations.
Specifically, if a mutation $A$ with fitness $s_A$ becomes
established, one considers the probability that another mutation
$B$, with effect $s_B > s_A$, will also become established before
mutation $A$ has fixed.  If this happens, mutation $B$ can drive $A$
to extinction and mutation $A$ is thus wasted. Of course, it is also
possible that mutation $B$ is subsequently outcompeted by a still
fitter mutation $C$, and so on.  The key approximation is that the
largest mutation which  occurs and is not outcompeted by a still
larger one fixes, becomes the new ``wild-type" --- i.e. the majority
population --- and the process then repeats.  Additional mutations
that might occur in a lineage which already has mutation $A$, $B$,
or $C$ are ignored.  For any fixed population size, there is some
selective advantage, $\sci$, such that sufficiently large mutations,
those with  $s>\sci$, are rare enough that they are unlikely to
occur before some less fit mutation arises and fixes. In the
conventional clonal interference analysis, it is assumed that a
mutation of size around $\sci$ will thereby fix before any others,
and the process  will then repeat. This is equivalent to
successional-mutation behavior with a set of mutations each with the
same strength, $\sci$.  Since $\sci$ increases with the population
size, more mutations are  wasted in larger populations, implying
that $v$ increases less than linearly with $N \u$.

Before discussing the problems with the basic successional-fixation
assumption, we consider how the characteristic $\sci$ depends on $N$
and on the distribution of selective advantages, $\rho(s)ds$.
Because only beneficial  mutations with substantial $s$ matter for
large $N$, the total $U_b$ itself is not important.  It is more
convenient to use the mutation rate per generation for mutations in
a range $ds$ about $s$: \be \mu(s)ds\equiv\  {\rm rate\ of\
mutations\ in\  interval\ } (s,s+ds)\ \equiv e^{-\La(s)}ds \ . \ee
We assume that large-effect beneficial mutations are typically much
less common than small-effect ones, so that $\mu(s)$ is small and
decreases rapidly with $s$.  Since $\mu(s)=\rho(s)/U_b$ is
dimensionless, it is convenient to define \be \La(s)=\ln \left[
\frac{1}{\mu(s)} \right], \ee which thus  increases with $s$. We
will see that $\La(s)$ roughly  plays the role that $\ln(s/U_b)$
does in the single-s case; the equivalent condition to $s\gg\u$ is
that $\La(s)\gg 1$ (at least for the important range of $s$).

The basic \CI analysis is simple: in the time that a mutation of
size $s_A$ will take to fix, $t_{fix}^A = \frac{1}{s_A} \ln \left[ N
s_A \right]$, some mutation of larger size $s$ will have time to
occur and become established as long as the total mutation rate for
mutations larger than $s_A$ is sufficiently large: \be
NU_b^{>s_A}\equiv N\int_{s_A}^\infty s \mu(s) ds >
\frac{1}{t_{fix}^A} . \ee Since $\mu(s)$ decreases rapidly with $s$,
this will not happen when $s_A \gtrsim \sci$, where \be
\La(\sci)\approx \ln(N\sci) \ . \ee That is, $\sci(N)$ is the value
of $s$ at which in the whole population there is of order one
mutation per generation: as $\mu(s)\propto \u$, $\sci$ depends only
on the {\it product} $N\u$, with the functional form determined by
$\rho(s)$.  In the clonal interference analysis approximation, the
speed of evolution is assumed to be the size of these mutations
$\sci$ times the rate at which they become established. This yields
\be v_{CI} \approx  C N \left[ \mu(\sci) \sci \right] \sci^2 \approx
C N \sci^3 e^{-\Lambda(\sci)} \sim \left[ \sci(N) \right]^2, \ee
where $C$ is a factor of order unity  which is not really obtainable
from  \CI analysis, as it depends on the details of further
approximations\footnote{Note that the details of how we define
fixation does not make much difference in the \CI result.  We have
also ignored other factors inside logarithms, since $La \gg 1$, but
this will affect the coefficient $C$}. At this point we should note
that various potential improvements are possible. In particular, it
is not at all clear why the establishment time rather than fixation
time should be used to obtain the accumulation rate of the $\sci$
mutations.   As we shall see below, if the latter rather {\it ad
hoc} assumption is made instead, the \CI analysis gives closer to
the correct results for certain distributions: those with long tails
in $\rho(s)$.

The above clonal interference analysis makes a crucial approximation
which is essentially never valid: that double-mutants can be ignored
even when mutations are common enough that they often interfere.
This is manifest in the assumption that the important mutations only
occur in the majority (``wild-type") population.  The basic problem
is that even if a more-fit mutation $B$ occurs before an earlier but
less-fit mutation $A$ fixes, $A$ may still survive. An individual
with $A$ can get another mutation $D$ such that the $A$-$D$ double
mutant is fitter than $B$.  If this happens, mutation $A$ (along
with $D$) can fix after all. Indeed, such events should be expected:
any population large enough for clonal interference to matter is
also large enough for double mutants to routinely appear even for
$s\sim\sci$. This is because  clonal interference can only affect
the fixation of a mutation of size $s$ when  the mutation rate to
mutations stronger than $s$ is large enough that  $N U_b^{>s}> 1$.
But when this occurs, the total beneficial mutations per generation,
$N U_b \gg \frac{1}{\ln(\sci/U_b)}$. Thus, from our analysis of the
single-$s$ model, whenever clonal interference occurs, multiple
mutations also play a role.

The single-$s$ model, in contrast, is unrealistic because it
explicitly excludes competition between mutations of different
effects. Thus the conclusions from this model and the clonal
interference analysis are each only part of the story. In the
remainder of this section, we  outline the behavior for more general
distributions of beneficial mutations, taking into account both
clonal interference and multiple mutations. Fortunately, as we shall
see,  for many forms of $\mu(s)$, the single-s approximation can
implicitly account surprisingly well for the effects of clonal
interference. Detailed analysis will be published elsewhere.

Let us first consider starting from a clonal population (although
this is an oversimplification which misses important aspects of the
dynamics; see below). Depending on $N$ and $\u$, various different
mutants will arise, as well as double-mutants, etc.  One of these
will be the fittest mutant that is established in the wild-type
population before any other mutation or combination-of-mutations
fixes.  All the other mutations that have already occurred will be
driven to extinction and thus do not matter for the long-term
evolution. For a given $N$, $\u$, and $\rho(s)$, there is a typical
fitness effect (call this $\st$) of the beneficial mutations that
create --- singly or in combination --- this fittest mutant. We call
mutations of roughly this magnitude {\it predominant mutations}, and
define --- crudely at this point ---  $\ut$ as the mutation rate to
these mutations. Clonal-interference-like competition determines the
predominant range of  mutations. Unfortunately, however, we cannot
simply lift the definition of $\st$ from clonal interference theory.
Except at very short times,  the population will not be mono-clonal
but will include various single and multiple mutants with a
distribution of overall fitnesses. This means that $\st$ is
determined by a delicate balance between clonal interference and
multiple mutation effects. Given an $\st$, however, the predominant
mutations accumulate via a process similar to that described by our
analysis of the constant-$s$ model, with population size $N$ and the
effective parameters $s=\st$, and $\u = \ut$.

Why should there be a predominant range of $s$? The basic argument
is simple. Mutations significantly smaller than $\st$ occur
frequently. But, by definition, these mutations are routinely
outcompeted by predominant mutants.  Thus these mutations do not
interfere with the accumulation of the predominant mutants. In
contrast, larger-than-$\st$ mutations do interfere with others when
they occur. But, by definition, these must be rare enough that it is
unlikely that such a mutation will arise in the time it takes a
predominant mutant --- or a combination of predominant mutations
--- to fix (else the larger mutation would be the predominant
mutant). Thus the population will primarily evolve via the
accumulation of mutations with $s$ in some range around $\st$.  Our
previous analysis does not predict $\st$, but given a value of $\st$
it determines how these mutations accumulate (see below for more
details). This is a slight oversimplification, as mutations of both
smaller-effect and larger-effect than $\st$ will play some role.
These considerations affect the appropriate definition of $\st$, and
the range of $s$ around $\st$ that is important.

What we must now address is the crucial fact that $\st$ (and $\ut$)
depend on $N$ and $\u$.  As we increase $N$ or $\u$, more mutations
occur before others fix: this suggests $\st$ will thereby increase.
Clonal interference analyses consider part of this process, and
predict that the analog of $\st$ ($\sci$) increases slowly with both
$N$ and $\u$ (indeed, with their product) \citep{1, 410}. But  these
approximations over-suppress smaller mutations by ignoring multiple
mutations, which are more likely to involve the common smaller
mutations. Thus we expect that $\st$ should increase even more
slowly with $N$ and $\u$ than clonal interference models suggest.
Nevertheless, even a slow increase in $\st$ could be important,
since in the single-$s$ model, $v$ increases with $s^2$ but only
increases slowly with $N$ and $\u$. As we now show, the form of
$\rho(s)$ qualitatively affects the behavior.

In the extreme case in which $\rho(s)$ decreases very slowly with
$s$ ($\rho(s) \sim \frac{1}{s^3}$ or slower),  the largest mutation
that can typically occur and establish in a given time always
dominates the {\it cumulative} evolution up until that time.  Thus a
predominant $\st$ does not even exist and neither our analysis nor
clonal interference describes the dynamics: it is controlled by
successional fixations --- but with no steady-state speed ---  no
matter how large the population. We do not discuss this seemingly
unlikely situation further.

Whenever $\rho(s)$ falls off faster than $\frac{1}{s^3}$, the basic
single-$s$ behavior obtains, with a narrow range of $s$ (roughly a
factor of two or less) around some predominant $\st$, with the
effective mutation rate $\ut$ crudely being that for mutations in
this range. But even though one could then simply plug the
appropriate $\st$ and $\ut$ into our earlier expressions for the
speed, $v$, the single-$s$ {\it forms} for the dependence on $N$ and
$\u$ may not be accurate, because $\st$ and $\ut$ themselves depend
on $N$ and $\u$.  There are two possibilities. The first is $\st$
and $\ut$ depend weakly enough on $N$ and $\u$ that our expressions
are roughly accurate. Another possibility is that the evolution is
dominated by larger and larger mutations as the population size
increase, as found in the clonal interference analysis.  Again
mutations in some restricted range will control the behavior (and
some degree of multiple-mutations will still be involved), but $\st$
will increase markedly with $N$. We shall see that both these
behaviors can occur, depending on the form of the distribution of
mutations $\rho(s)ds$.

\subsection{Predominant $s$ approximation}

A simple approximation that might be expected to be valid {\it if} a
sufficiently narrow range of $s$ dominates is to ignore all the
mutations except those in some narrow range about $s$, compute the
evolution speed, $v(s)$ from the single-$s$ analysis, and then
maximize this over $s$ to obtain the {\it predominant} $s$, $\st$,
and an approximation for the actual speed: \be v\approx \max_s v(s)
= v(\st) \ , \ee which defines $\st$.  We call this approximation
the predominant $s$ approximation, as it ignores the question of how
wide a range of $s$ is important.  We can then make a conservative
check of our assumption that a narrow range of $s$ dominates by
computing how quickly $v(s)$ falls off away from $\st$, because
mutations at other $s$ cannot increase the actual velocity by more
than their $v(s)$.

For concreteness, we consider a class of distributions $\mu(s)$
parameterized by three quantities:  a characteristic selective
advantage, $\sigma$, an (effective) overall mutation rate $\u \sim
\sigma e^{-\ell}$ --- so that $\ell$ is like the $\ln(\u/\sigma)$
that appears in the single-$s$ results --- and a parameter $\th$
that characterizes the shape of the distribution of rare large
mutations.  We thus write \eon \mu(s) = e^{-\ell - (s/\sigma)^\th}.
\eoff For convenience we use the shorthand notation \be L\equiv
\ln(N\sigma) \ . \ee

We will see that the behavior depends qualitatively on whether $\th$
is larger or smaller than $1$. For $\th > 1$, the distribution falls
off faster than exponentially, and we refer to this as a
``short-tailed'' $\mu(s)$.  The exponential case is exactly
marginal.  For $\th < 1$, the distribution falls off more slowly
than exponentially.  We refer to this as the ``long-tailed''
$\mu(s)$ case.

\subsection{Short-tailed $\mu(s)$}

We begin by considering the case of $\th > 1$; that is, a
distribution that falls off at least exponentially.  The behavior is
simplest when the population size is large enough that $2 \L /
\La(s)$ is substantially greater than unity. This is loosely like
$q(s)$, the number of mutations by which the nose is ahead of the
mean fitness, being substantially larger than unity (although $2 \ln
[N \sigma] /\La$ differs from the actual $q$ in important ways). In
this regime we have \eon v(s)\approx s^2\frac{2L}{[\La(s)]^2} \eoff
which is maximum at $s=\st$ given by \eon \st
\frac{d\La}{ds}(\st)=\La(\st) \ . \eoff The predominant $s$
approximation is only valid in this regime for $\th>1$.  It yields
\be \st = \si \left[ \frac{\ell}{\th-1} \right]^{\frac{1}{\th}}, \ee
and thus \be v\approx v(\st) = C_\th \si^2 \frac{2\ln [N
\sigma]}{\ell^{2-2/\th}}, \ee with the coefficient $C_\th=
(\th-1)^{2-2/\th}/\th^2$. Note that $\st$ is roughly independent of
$N$ in this regime, but decreases as the overall beneficial mutation
rate increases (i.e. as $\ell$ decreases). In other words, $\st$
does not depend strongly on $N$, but does decrease as $\u$
increases.  This makes sense: as $\u$ grows, multiple small
mutations become more important compared to single larger mutations.
Because of this, the dependence of $v$ on $N$  is very similar to
our single-$s$ approximation, but the dependence on the mutation
rate is \emph{weaker}.  We can check for consistency of the use of
the large-$q$ single-$s$ results. The value of $q$ is $2 \ln [N
\sigma]/\La(\st)$, which yields \be q\approx \frac{2 \L
(\th-1)}{\ell\th}. \ee This is large when $2 \L /\ell$ is large,
unless $\th-1$ is small --- i.e. the tail is becoming long.

The behavior for $\th>1$ can also be analyzed when $\L$ is not so
large.  As $\L$ decreases, the predominant $s$ decreases
--- i.e. it begins to depend on $N$. The resulting expressions are more
complicated, but can be computed from the more general form for
$v(s$), \eq{veq}, in a similar way.  However, they are of
questionable validity, since only some of the significant $s$ will
be in the multiple mutation regime, while others will be in the
crossover regime of $q \lesssim 2$. As we have seen in a previous
section, this crossover is complicated even for the single-$s$
model; it will be even more so with a distribution of $s$.

This brings us to the issue of how wide a range of $s$ plays a
substantial role in determining $v$ (as well as the steady state
shape of the moving fitness distribution). In the successional
mutations regime, the speed is $v\approx N\int s^2 \mu(s)ds$, so
that $s$ of order $\si$ dominates (as long as $\mu(s)$ falls off
faster than $1/s^3$). That is, a range of $s$ within a factor of two
or so of the typical value $\sigma$ dominates the evolution. In the
multiple mutations regime, the maximization of the single-$s$ speed
$v(s)$ over $s$ gives a predominant $\st$ much larger than $\si$,
but no direct information on the range of $s$ that contribute. A
natural estimate is the range over which $v(s)$ is not lower than
$v(\st)$ by more than, say, a factor of two. This would mean that
the width of the range is comparable to $\st$ itself: that is,
mutations with effects between $\st/2$ and $2 \st$ matter. This
confirms our assertion that the single-$s$ model gives at least a
good qualitative picture of the dynamics even when there is a
short-tailed distribution of the effects of beneficial mutations.
Since all the important mutations are of order $\st$,
``leapfrogging'' (by which, for example, a double-mutant gets a
mutation which makes it more fit than an existing quintuple-mutant)
does not have a large effect on the evolution. We can thus indeed
consider the basic dynamics to be the accumulation of mutations of
roughly size $\st$ according to the single-$s$ description given
above.

However, our calculation of the range of $\st$ that matters calls
into question the predominant $s$ approximation: why should the
actual $v$ be $v(\st)$ rather than, e.g., $v(s)$ averaged (or some
other weighted integration) over $s$?  A more sophisticated
analysis, which will be described elsewhere, shows that for
short-tailed distributions ($\th>1$), both $\st$ and $v$ are given
{\it correctly} by the predominant $s$ approximation in the large
$\L/\ell$ limit --- up to only differing factors inside logarithms
and other small corrections. But the range of $s$ that significantly
affects $v$ is much smaller than that guessed from the predominant
$s$ approximation. This should perhaps not be surprising, as the
predominant $s$ approximation assumes that all $s$ contribute to $v$
as if different-sized mutations did not interfere. But interference
will in fact tend to suppress the contribution to $v$ from $s$ away
from $\st$.  We find that for short-tailed distributions, in fact
only $s-\st$ of order $\st/\sqrt{\ell}$ are important. In terms of
the mutation rate $\ut$ to mutations with $s\approx\st$, this range
has width $\st/\sqrt{\ln(\st/\ut)}$. That this difference does not
invalidate the predominant $s$ approximation result for $v$ can be
understood by considering the weak dependence of $v$ on the mutation
rate in the single-$s$ model. As $v$ depends only logarithmically on
$\u$, replacing $\u$ by an effective $\ut$ that includes either a
substantial range around $\st$ or by one that includes only a narrow
range will only alter factors inside the logarithms and thus have
little effect on the inferred $v$.  Since the fuller analysis finds
that an even narrower range around $\st$ matters, it strengthens our
contention that there is a predominant $s$ (albeit one that depends
on $\u$) and that the full dynamics is very similar to that of the
single-$s$ case analyzed in detail in this paper.  The exception to
this is the intermediate $N$ regime in which the crossover from
successive to multiple mutations occurs and the effective $q$ is
less than two or so: we will not discuss this complicated crossover
regime further here, although it may be relevant in many
experimental situations.

We have seen that the predominant $s$ approximation does well for
the primary quantities of interest, $\st$ and $v$, although it
overestimates the range of $s$ that plays a role. In contrast, the
clonal-interference-only analysis yields the incorrect behavior for
short-tailed distributions. For the model distributions,
$\La(s)=\ell+(s/\si)^\th$, the \CI analysis yields \be \sci \approx
\si(\L-\ell)^{1/\th}. \ee For the short-tail case, this is much
larger than the predominant value, $\st$. Indeed it is qualitatively
wrong:  $\sci$ increases with increasing $\u$, while $\st$
decreases. Using $\sci$ instead of $\st$ leads to incorrect
predictions of $v$; in particular, clonal interference predicts $v$
grows only sublinearly with $\ln N$. This problem stems from the
fact that clonal interference analyses have the wrong basic picture
of the dynamics.  The evolution is not in fact dominated by the rare
very large mutations that occur only once per generation in the full
population, as the \CI approximation implicitly assumes. Rather, the
evolution is actually controlled by multiple mutations of smaller
(though still larger than average) fitness that occur frequently
even in the much smaller sub-populations that exist in the nose of
the fitness distribution of the steady state evolving population.
Because the multiple mutation effects depend on there being
sufficiently large  rates for the predominant mutations, increasing
the overall mutation rate allows multiple smaller mutations to beat
larger ones. Thus increasing $\u$ results in decreasing $\st$ --- in
contrast to the increase of $\sci$ with $\u$.

\subsection{Long-tailed $\mu(s)$}

For distributions that fall off more slowly than a simple
exponential --- i.e. $\th<1$ --- the behavior is rather different.
This is apparent even in the crude predominant-$s$ approximation.
Again, we begin by considering the simpler large $2\L/\ell$ limit.
But with the long-tailed $\mu$, we need the fuller single-$s$
expression: \be v(s)\approx s^2\ \frac{2L-\La(s)}{[\La(s)]^2} \ ,
\ee with $\La(s)=\ell+(s/\sigma)^\th$. In the large $2L/\ell$ limit
the predominant $s$ is found to be \be \st\approx \sigma \left[
\frac{4\L(1 - \th)}{2 - \th} \right]^\frac{1}{\th}, \ee with
corresponding effective mutation rate \be \mu(\st)\propto
\left[\frac{1}{N}\right]^{4(1-\th)/(2-\th)} . \ee This yields \be
v\approx A_\th \si^2 (2\L)^{\frac{2}{\th}-1}, \ee with coefficient
$A_\th =\th(2-2\th)^{2/\th-2}(2-\th)^{1-2/\th}$. In this case, we
see that $v$ grows {\it faster} than linearly with $\ln N$.
Surprisingly, the dependence on the mutation rate in this regime is
negligible:  $\u$ only determines how large $N$ has to be to be in
this regime.  The smaller the mutation rate, the larger the $N$
needed. But in contrast to the short-tail case, here \be q \approx
\frac{2 - \th}{2 - 2\th} \ee is {\it not} large, so that even for
very large $N$, the important multiple mutants still involve only
${\cal O}(1)$ of the predominant mutations. The fact that $q$ never
becomes particularly large for long-tailed $\mu(s)$ is because in
this case $\st$ increases substantially with $N$: in the
short-tailed case, many small mutations contribute, while in the
long-tailed case, fewer larger mutations are involved.  But we must
be careful with the above results for the long-tailed case, as they
are not valid if the inferred $q$ is less than two: below this the
crossover from successional to concurrent mutations behavior will
apply. We need to distinguish two cases.

If $\th>\frac{2}{3}$, $q > 2$ and the above results apply.  The
corresponding effective mutation rate decreases with a power of
$1/N$ less than unity, so that the total mutation supply rate for
the predominant mutations ($N\mu(\st)$) grows with $N$ as
$N^{(3-2\th)/(2-\th)}$.  (Of course, many of these are wasted as
multiple mutants outcompete the single mutants and control the
dynamics, as described by our single-$s$ theory).

If $\th<\frac{2}{3}$, then the above analysis would give $q<2$ and
$N\mu(\st)\ll1$, which indicates a breakdown of the approximations.
In this case $q$ sticks at two, and the dynamics is basically
successional, with the predominant mutants being those for which the
total rate $N\mu(\st) \sim \frac{1}{\st}$. This means that $\st
\approx \si L^{\frac{1}{\th}}$, and we expect \be v\approx \si^2
L^{\frac{2}{\th}-1}.  \ee Note that the coefficient coincides with
the earlier expression at $\th=2/3$. The steady state is at the
upper end of the  crossover between the successional and multiple
mutational behavior as discussed in the section on evolution at
moderate $N$.

For $\th<\frac{2}{3}$, the clonal interference-only approximation
agrees with the predominant $s$ approximation, as the total mutation
rate to the predominant mutants is of order unity so that
$\st\approx \sci$.  In contrast, for the intermediate case with
$\frac{2}{3}<\th<1$, \CI analysis yields $\sci\approx \si
\L^{1/\th}$.  This is still the correct behavior, but the numerical
coefficient is wrong: as noted above, the total mutation rate for
the predominant mutants grows as a power of $N$, in contrast to the
\CI approximation in which it assumed to be independent of $N$. For
the speed of evolution, naive application of the clonal interference
analysis gives $v\sim \sci^2 \sim L^{2/\th}$, which is not even the
correct scaling with $\L$. But if, instead, the fixation rate rather
than the establishment rate is used to give an improved (though it
is not {\it a priori} clear why this should improve the result) \CI
estimate of $v$, the correct scaling with $\L$ can be obtained.

At this point, it is not clear how good the predominant $s$
approximation is for the long tailed distributions, nor how wide a
range of $s$ around its predominant value are important.  A more
sophisticated analysis is needed for this, as well as for
understanding the crossover from the successional $N\u\ll1$ to the
large $q$ regime analyzed above.

\subsection{A simple example}

A concrete (albeit artificial) example is useful to illustrate the
points made above.  We consider a simple model with three classes of
mutations, each with a single $s$: weak mutations with a small
$s_s$, intermediate ones with a medium $s_m$ and strong mutations
with a large $s_l$; each class has its own mutation rate.
Specifically, we consider $s_s = 10^{-3}$,  $s_m = 10^{-2}$, and
$s_l = 10^{-1}$, with mutation rates $U_s = 9\e{-6}$, $U_m =
4\e{-6}$, and $U_l = 5\e{-10}$, crudely approximating an exponential
distribution of beneficial mutations (with, in terms of the family
of distributions discussed above: $\th=1$, $\sigma=10\e{-2}$,
$\ell\approx 7$).

For small population sizes, the successional regime obtains and \be
v\approx N [U_s s_s^2+U_m s_m^2+U_l s_l^2] \ee which is dominated by
the medium mutations.  As $N$ increases, we expect multiple
mutations to start to play a role when $NU_m \sim
1/\ln(s_m/U_m)\approx 1/8$, corresponding to crossover out of the
successional fixations regime for $N\sim 3\e{4}$.

To understand the behavior for larger $N$, we first analyze the
three types of mutations separately, similar in spirit to the
predominant $s$ approximation. That is, we consider three
sub-models, each of which have only one of the three types of
mutation.  The corresponding rates of evolution, $v_s$, $v_m$, and
$v_l$ must all be less than $v_{tot}$, that of the full model,
because the full model has more beneficial mutations than any of the
three sub-models. Conversely, we expect $v_{tot} \leq v_s+v_m+v_l$
because, at best, the different mutations can accumulate
independently; in practice, they will tend to interfere (although
multiple mutants with combinations of the different types can matter
and contribute to the actual speed). Each of the three sub-models
has only one type of mutation, so our single-$s$ results can be used
directly to obtain $v_s$, $v_m$, and $v_l$.

For a population of size $N = 10^{5}$ --- just into the multiple
mutations regime --- we find $v_s = 3.5\e{-7}$, $v_m = 1.5\e{-5}$,
and $v_l = 5\e{-7}$. The leads of the corresponding fitness
distributions --- the number of multiple mutants above the mean that
exist at one time --- are $q_s = 2.7$, $q_m = 2.2$, and $q_l = 1$.
Thus the small and medium mutations accumulate primarily as double
and triple mutants, while the large mutations (alone) would be in
the successional-mutations regime. For this moderate size
population, the mutations with effect $s_m$ are the predominant
mutants.  They clearly dominate the full model, since $v_{tot}$ will
be in the very narrow range between $v_m$ and $v_m+v_s+v_l$.
Although the small mutations are common, they do not matter because
even triple-small mutants --- as occur in the small-only model ---
will be routinely outcompeted by single medium-mutations.  The
medium mutations occur frequently in the fixation time of the
triple-small mutants and thus routinely ``leapfrog'' them. The small
mutants never interfere with medium mutations, and those that fix do
so only because they happen to be linked to medium mutants. The
large mutations, in contrast, do interfere with the medium
mutations, but occur so rarely that they are not important for the
overall evolution rate. In this example, a few hundred medium
mutations fix for each large mutation that establishes, so almost
all medium mutations fix without being affected by a large mutation.
Thus the accumulation of mutations is very well approximated by the
process our single-$s$ analysis describes, provided we choose $\st =
s_m$ and $\ut = U_m$.

As the population size is increased, $v_l$ will increase faster than
$v_m$ or $v_s$ because it is not yet in the regime with logarithmic
$N$-dependence.  For $N = 10^{6}$, $v_s = 5\e{-7}$, $v_m = 2\e{-5}$,
and $v_l=5\e{-6}$.  The medium mutations still predominate, but less
strongly than before. By $N=10^{7}$, we have $v_s = 6\e{-7}$, $v_m =
3\e{-5}$, and $v_l = 5\e{-5}$, so the large mutations begin to
dominate. For larger $N$, they will do so even more strongly.  This
shows how $\st$ increases with $N$. With this discrete $\rho(s)$,
$\st$ changes quite rapidly in a small range of $\ln N$, but for a
continuous fitness distribution the increase will be smooth (of
course, continuous distributions present additional complications
involving the proper weighting of mutations near $\st$).

We could also apply clonal interference analysis to this three-class
model.  From these analyses, for a beneficial mutation to fix, it
must establish and then not be interfered with by a more-fit
mutation before it fixes. The probability that a mutation of size
$s$ will be interfered with is $p_I(s) \approx \frac{N \u}{s} \ln N
\int_s^\infty r \rho(r) dr$. Thus the putative distribution of
beneficial mutations that fix will be $\rho_F(s) = K s
e^{-\lambda(s)} \rho(s)$, where $K$ is a normalizing constant. The
average effect of a fixed beneficial mutation --- effectively $\sci$
--- would be the mean, $\ev{s}_F$, of this $\rho_F(s)$. These
mutations arise at average rate $\ev{k}_F =N \u P_{fix}$, where
$P_{fix}$ is the average probability of fixation, $P_{fix} =
\int_0^\infty s e^{-\lambda(s)} \rho(s) ds$. Clonal interference
analysis yields $v = \ev{s}_F \ev{k}_F$.  For our 3-class example,
with $N = 10^{5}$, this gives $v_{tot} \approx 4\e{-5}$, about $3$
times higher than the maximum possible as calculated from $v_{tot} =
v_s + v_m + v_l$. For $N = 10^{6}$, clonal interference predicts
$v_{tot} \approx 4\e{-4}$, about 20 times too high.  The problem is
easy to diagnose.  For both values of $N$, the clonal interference
theory correctly predicts $\ev{s}_F \approx s_m$. However, implicit
in the calculation of $\ev{k}_F$ is the incorrect assumption that
these medium mutations accumulate singly. Conversely, the
predominant mutation approach ( ``max" approximation) is to choose
$s_m$ as the single value of $s$, and then analyze how the
multiple-mutation process sets the rate at which this class of
mutations accumulate.

\section{Discussion}

Beneficial mutations are often assumed to be rare, and adaptation
therefore to be mutation-limited.  This is the basis for the picture
of successional selective sweeps and the conclusion that mutations
arise and fix at a rate proportional to $N \u s$ \citep{566}. This
picture of successional sweeps  underlies the strong selection weak
mutation assumption that is essential to many conclusions in
population genetics and evolutionary theory. This assumption is
likely to be correct for the evolution of some strongly selected
characters in complex multicellular organisms. But most
unicellular organisms and viruses tend to live at much larger
population sizes, and can have larger mutation rates. For such
populations, much of one's intuition from the rare-mutations picture
will often be wrong. This makes it important to go beyond the
successional-mutations regime and to develop an understanding of
evolutionary dynamics when beneficial mutations are common.

This is a very broad subject.  In this paper, we have focused on the
concurrent-mutations regime in which there is strong selection and
strong mutation. By strong mutation, we mean that the total
beneficial mutation production rate $N \u$ is sufficiently large
that the time to establish a mutant population is less that the time
it will take to sweep to fixation.  As the establishment time is
$1/(N\u s)$ and the sweep time is $\frac{1}{s}\ln Ns$, the condition
to be in the concurrent mutations regime is $ N\u \gtrsim
\frac{1}{\ln [Ns]}$, so that multiple beneficial mutations are
present in the population and tend to interfere.  By strong
selection, we mean both $Ns \gg 1$ and $\frac{s}{\u} \gg 1$.  The
former condition is what is commonly meant by strong selection, and
is required to ensure that selection is strong compared to drift
except when subpopulations are rare. The latter constraint makes the
analysis simpler, because it ensures that only one population at a
time needs to be treated stochastically, but is not essential for
the general picture.

The concurrent-mutations regime that we analyze is likely to be
quite common in nature.  Even if there are only ten or so beneficial
point mutations available to a population which has a per base pair
mutation rate of order $10^{-9}$, this gives $\u \sim 10^{-8}$. To
have $N \u \gtrsim \frac{1}{\ln [Ns]}$, we therefore only need
population sizes of order $10^{7}$ ($1/\ln [Ns]$ will typically be
roughly $\frac{1}{10}$ for any reasonable values of $s$ in such
large populations). In other words, if there are even a few
mutations of effect $s \gg 10^{-7}$ available, a population as small
as $10^7$ individuals will experience the multiple concurrent
mutation effects. These sizes are well within normal ranges for many
populations, including, for example, \emph{E. coli} in a single
human gut, cells in an evolving cancer,  pathogens within a single
host, and many others. Moreover, this is a very conservative
estimate. Viral and certain bacterial populations, or mutator
strains in any organism, often have much overall higher mutation
rates. Organisms with more beneficial mutations available will also
have much larger $\u$. In recent experiments in \emph{S. cerevisiae}
adapting to low glucose, we have inferred a beneficial mutation rate
of $\u = 10^{-5.5}$ in non-mutator strains  and an order of
magnitude higher in mutators \citep{n1exp}. Such values are not
atypical \citep{574}.  For these values of $\u$, and $s$ of order a
percent or fraction of a percent, a mutator population of $N \sim
10^{7}$ will have $q \sim 4$, so that quadruple-mutants will be
present and sweep {\it collectively} (for nonmutators, $q \sim 3$).
With these parameters, each factor of about ten increase in $N$ will
increase $q$ by  one. In general, we see that the
concurrent-mutations regime is surely relevant for many microbial
populations.

Within the concurrent-mutations regime, we have explored how a
population accumulates beneficial mutations and maintains variation
in fitness. The fundamental theorem of natural selection states that
the rate of increase in the mean fitness of a population equals the
variance in fitness \citep{567}.  This remains true. Our work
demonstrates how the variance is itself determined:  how fitness
variation accumulates while it is being selected on. The key here is
the balance between selection narrowing the fitness distribution and
mutation broadening it. This is an unusual type of
mutation-selection balance, very different from the deleterious
case.  Only mutations at the nose of the distribution matter. Others
inherit a less good genetic background and do not contribute to the
long term evolution of the population: they are destined to be
outcompeted by new mutations at the nose. The dynamics at the nose,
where subpopulation sizes are small, dominates the behavior. This
means that the natural measure of the width of the distribution is
the lead, not the variance --- in contrast to conventional
treatments. It also means that random drift and finite $N$ effects
are crucial, even for arbitrarily large $N$ as long as there are
more than a few beneficial mutations to be acquired. Thus for any
treatment of evolution in fitness ``landscapes,'' these effects need
to be taken into account whenever the population is not localized
around a fitness peak: in contrast to quasi-species equilibria near
fitness peaks, deterministic approximations give nonsense.

By matching the speed of advance of the nose with the speed of
advance of the bulk of the distribution, we have shown that the lead
depends logarithmically on $N$ and $\u$ according to the formula
\eon q = \frac{2 \ln [N q s]}{\ln \left[ \frac{s}{\u} \right]}.
\eoff  This leads to a speed of evolution that is also logarithmic
in $N$ and $\u$, \eon v \approx s^2 \frac{2 \ln [N s] - \ln \left[
\frac{s}{\u} \right]}{\ln^2 \left[ \frac{s}{\u} \right]}. \eoff

Our work extends and complements earlier work on the
concurrent-mutations regime. \citet{21} and \citet{20} studied a
model like ours, although their initial work did not properly
account for all stochastic effects. Recently they have developed a
moment-based approach which provides results qualitatively similar
to ours in certain regimes \citep{kesslerlevineunpub}. This is a
potentially useful technique, although, as discussed in Appendix A,
it quickly becomes unwieldy as more moments need to be kept, and
numerical analysis is required.

\citet{313} also studied a model similar to ours in the context of
HIV evolution.  Their analysis also involves a separation between
deterministic and stochastic behavior, but treats the stochasticity
at the nose in a different and less explicit manner. To couple this
to the deterministic results, \citet{313} appear to require a
smoothness in the fitness distribution which would obtain only when
it is broad. Thus their analysis is strictly valid only at what we
would call very high speeds: $v \gtrsim s^2$. But, because they
treat only one population stochastically at a time, their analysis
also requires $\frac{\u}{s} \ll 1$, so their results are valid only
at enormous population sizes (and very large $q\gtrsim\ln(s/\u)$).
This regime is likely to be relevant for certain viral populations,
which was their main focus.  Nevertheless, the results of
\citet{313} are similar to ours, in that they involve logarithms of
$Ns$ and $\frac{\u}{s}$ in similar ways (though they do differ
substantially --- in the regimes we have considered their results
lead to errors typically ranging from plus or minus $50 \%$ to $250
\%$). This is unsurprising, since the simple beneficial
mutation-selection balance arguments (in our heuristic analysis
section) apply to the very fast regime of \citet{313} as well, and
lead generally to logarithms of $Ns$ and $\frac{\u}{s}$.  Further
analysis shows, if some algebraic errors are corrected in their
work, and the large $Ns$ and $s/\u$ asymptotics are worked out, that
our result for $v$ can be recovered up to somewhat different factors
inside large logarithms \cite{rouzinepers}.

Various studies have been carried out on clonal interference --- the
other effect that occurs when there are concurrent mutations (i.e.
in the strong selection strong mutation regime) \citep{1, 410, 417,
351, 331, 292, 294}. We have discussed the relationship between this
work and ours and analyzed a model with a distribution of beneficial
mutations which includes both clonal interference and multiple
mutation effects. \citet{562} have also analyzed some of the
interplay between these effects. Clonal interference analysis by
itself makes qualitatively similar predictions about the rate of
accumulation of beneficial mutations as the full theory. Both
predict that $v$ grows much less than linearly in $N$ and $\u$ (as
do the analyses of \citet{kesslerlevineunpub} and \citet{313}),
though the quantitative predictions differ. The major qualitative
differences are in the mechanisms by which the evolution takes
place.  In clonal interference analysis large mutations that occur
in individuals which have roughly the mean fitness --- i.e in the
majority subpopulation --- dominate the evolution. Thus one would
expect to see strong selective sweeps and a population that is
typically either nearly clonal or in the midst of such a sweep
(except occasionally when a smaller mutation becomes transiently
very common before being outcompeted by a larger one.) By contrast,
except when there is a long tail to the distribution of $s$, we have
shown that the evolution is dominated by multiple mutations of
intermediate effect, so the selective sweeps are much less
pronounced and the population always maintains substantial variation
in fitness. And we have shown that even when the distribution of $s$
does have a long tail, some of the quantitative predictions for the
speed of the evolution are different from clonal interference
predictions. We find that the mutations that dominate the evolution
in the concurrent mutations regime have strengths in a narrow range
around some predominant value, $\st$.   The simple single-$s$ model
is thus surprisingly good, provided we use $s = \st$ and $\u$ equal
to the mutation rate to beneficial mutations of this magnitude.

Over the past few years, much experimental evidence has accumulated
which supports the prediction that $v$ grows less than linearly in
$N$ and $\u$ \citep{251, 268, 296, 284, 561}.  This has often been
interpreted as support for the clonal interference picture. However,
the experimental data on the quantitative details of the dependence
of $v$ on $N$ and $\u$ cannot distinguish between clonal
interference analysis and our results. Thus these experiments also
support our general theory.

We have recently (in collaboration with Andrew Murray) conducted
experiments on asexual evolution of yeast in low glucose.   For a
range of different $N$ and $\u$ we measured  the distributions of
fitnesses within the  evolving populations and the dynamics ($v(t)$)
by which the fitness increased \citep{n1exp}.  Since, unlike earlier
work, these experiments measured the widths of the fitness
distributions and the strengths of selective sweeps, we were able to
distinguish between our analysis and clonal interference acting
alone.  The experimental data support the multiple-mutation theory,
with both $v$ and the leads of the fitness distributions depending
on $N$ and $\u$ consistent with our predictions. Clonal interference
analysis, on the other hand, would predict that populations maintain
less variation in fitness, and that this variation would not scale
with $N$ and $\u$ as we predict.  We also measured how the
populations increased in fitness over time, finding smooth increases
suggestive of multiple mutations of intermediate size fixing
together.  This was again consistent with our theory and
inconsistent with clonal interference alone, which would suggest rare
larger mutations dominate the evolution.  Combining all these data,
we found that clonal interference was ruled out unless several
parameters were finely tuned. Thus in the only experimental test
able to distinguish the two effects, multiple mutation effects
explain the data better than clonal interference alone. This
represents only one set of experiments in one organism in one
selective condition, so it is quite possible that in other
circumstances the reverse will be true. Yet, even if clonal
interference is found to better characterize the dynamics in some
situations, we have shown that to understand this properly one needs
to analyze the interplay between this and multiple beneficial
mutations on the same genome.

Despite being consistent with one experimental test, the model we
have analyzed surely has many shortcomings.  We have analyzed one of
the simplest possible situations for positive selection. Violations
of certain simplifying assumptions, such as neglecting deleterious
mutations and assuming a single effect $s$ of beneficial mutations,
may well, as we have argued, have relatively minor effects beyond
modifying the effective parameters $\u$ and $s$ of the model.
Furthermore, the neglect of interactions between effects of
mutations (epistasis) may not invalidate the overall results.  The
key assumption is that the {\it distribution} of the magnitudes of
available beneficial mutations is roughly independent of the genetic
background even though the actual set of these mutations varies.
That is, after each uphill-fitness step is taken, the distribution
of possible next steps is similar, although they may now be in
different ``directions".

However, breakdown of some of our assumptions will surely be
crucial. For example, certain non-multiplicative (epistatic) effects
of beneficial mutations, as well as frequency-dependent selection,
can lead to very different behavior. But our results should serve as
a null model, useful in forming baseline predictions. Departures
from the main results --- especially the scalings with population
size and mutation rates --- indicate the presence of one or more
complicating factors.  Even within the context of our simple model,
however, many important questions remain.

One of these is the expected genetic variation. We have calculated
the expected variation in fitness, but individuals with the same
fitness will often have different sets of beneficial mutations. Thus
the true genetic diversity at the positively selected sites will be
substantially greater than the variation in fitness. Although
sometimes the first new mutant to establish will dominate the lead
population, typically around $q$ different beneficial mutations will
occur and contribute to extending the nose during one establishment.
Subsequent mutations that further extend the nose will occur at
random among these different backgrounds, thus changing (and
typically reducing) this diversity, even as the diversity of the new
mutations is created. Eventually particular beneficial mutations do
sweep, but these sweeps are not necessarily uniform. Instead,
frequencies typically go up and down depending on which backgrounds
future mutations occur in. Understanding this diversity is important
if one is to look for the signature of this type of selection in
sequence data.  It is also important to understand the potential
benefits of sex, as we discuss below.

In addition to the diversity at the positively selected sites, we
would also like to understand the expected patterns of variation at
neutral and deleterious sites.  This will have a very different
character than in neutral evolution or in the successional-mutations
picture of positive selection, and may also help detect
concurrent-mutations evolution in sequence data. The neutral,
deleterious, and beneficial diversity is also important in in
understanding the role of epistasis.  If potential beneficial
mutations have epistatic interactions with other mutations, the
typical variation in the presence of other mutations is crucial.

Another important question is the effect of sex or recombination in
a population in the concurrent-mutations regime. According to the
Fisher-Muller hypothesis, sex should reduce interference effects and
hence prevent the wasting of beneficial mutations. This allows
sexual populations to accumulate beneficial mutations faster than
asexual ones. \citet{309}, \citet{564}, and \citet{297} attempted to
calculate the strength of this effect by comparing the $v$ in an
asexual population to the $v$ in a population with free
recombination. They defined the advantage of sex to be the
difference between these quantities. However, their calculation of
the asexual $v$ assumed that only two beneficial mutations were
possible --- thus ignoring triple and higher mutants, and not
properly accounting for the competing effects of mutations and
selection. With our calculation of the asexual $v$, however, one can
make this comparison.  In the completely free recombination case,
all beneficial mutations behave independently: there is no
interference between them, nor collective behavior among them.  Thus
with free recombination, $v_{fr} = N \u s^2$, as in the
successional-mutations regime. This gives a huge Fisher-Muller
advantage to sex, the difference between $v$ and $v_{fr}$, which is
zero in small populations and grows rapidly as $N$ or $\u$
increases.

However, the above analysis is not directly applicable to the
evolution of sex, since sex and completely free recombination are
certainly not synonymous. Rather, sex may only occur occasionally or
recombination might be infrequent, so that linkage persists for some
time.  An interesting situation is when sex and recombination are
relatively rare.  We would like to understand whether or not a small
amount of sex in an otherwise asexually evolving  population would
be advantageous (and hence be likely to become more common).  To do
so within the simplest model for the asexual evolution, we must
first calculate the true genetic diversity among beneficial
mutations within all the subpopulations at different fitnesses.
Given this, we can then calculate the probability that sex between
any two individuals will produce more-fit offspring.

This is a subtle question, because the average effect of sex on the
variance in fitness or in the tendency to bring together good
mutations more than it breaks them up is largely irrelevant. Rather,
what is important is the rate at which recombination generates (or
eliminates) anomalously fit individuals --- that is, its effect on
the nose.  Sex will tend to break up beneficial mutations at the
nose, and hence tend to destroy some of the most-fit individuals. At
the same time, however, it will occasionally mix two less-fit
individuals in just the right way to create an offspring which is
more fit than the current nose.  It is the competition between these
two effects that determines the advantage of sex. Even if sex on
average tends to increase the variance in fitness, this will not
increase the speed of evolution in the long term if it does not also
extend the nose. Rather, the increased variance from sex will be
balanced by the actions of selection and mutation (in the end, the
mean fitness cannot advance any faster than the nose), and the rate
of adaptation will be largely unchanged. On the other hand, if sex
does extend the nose it will tend to speed up the evolution even if
it has little effect on the variance.  In this case, these
occasional sex-driven expansions of the nose would act like extra
mutations, which modify the mutation-selection balance and cause an
increase in the steady state variance via increasing the lead ---
even though sex has no direct effect on the variance.

In recent years, Otto and Barton have made substantial progress in
understanding the effects of sex, short of completely free
recombination, in the Fisher-Muller picture \citep{568, 276, 234,
263}. This work takes the Hill-Robertson perspective and does not
include the full dynamics of the asexual population that we have
worked out here. As far as we are aware, it is not clear whether the
effect of sex at the nose within our calculated population structure
is the same as the effect of sex in Otto and Barton's analysis.
Future work is needed to unify these perspectives and understand the
effects of sex even within the simplest models.

In this paper, we have explored evolutionary dynamics when
beneficial mutations are common and there are many present
concurrently.  We have laid out an analytical and conceptual
framework for understanding how asexual populations accumulate
beneficial mutations --- the dynamics of adaptation in this
extremely basic situation.  Using this framework, we have
demonstrated that the rate at which a population accumulates
beneficial mutations does increases only slowly with population size
or mutation rate beyond a certain point. Although we have focussed
on the effects of multiple mutations, we have also analyzed the
interplay between this and clonal interference between mutations of
different strengths. The results have implications for comparing
evolution between different populations, and for designing
experiments to investigate various aspects of evolution in the
laboratory.  Statistical tests that can distinguish, based on
sequence data, between various scenarios for ongoing evolution are
needed: our results provide a step in this direction. More
generally, our results provide a framework for starting to address
the effects of sex, of mutators, and of epistatic interactions in
large populations.

\section{Acknowledgments}

We thank John Wakeley, Igor Rouzine, and especially Andrew Murray
for useful discussions. This work has been supported in part by the
NIH via grant P50-GM068763-01, the NSF via DMR-0229243, and the
Merck Foundation.

\section*{Appendix A:  Deterministic and Moment-Based Approaches}

There are a variety of other possible approaches to studying the
problem we have analyzed.  In this Appendix, we briefly discuss two
of these: deterministic approximations and moment-based approaches.
Both of these methods start by considering the distribution of
fitnesses within the population as some function $w(x, t)$, which
describes the number of individuals at fitness $x$ at time $t$. As
long as $s$ is small, $w$ can be treated as continuous: this is
equivalent to the conventional ``diffusion approximation".  The
forces of mutation, selection, and random drift then lead to a
stochastic differential equation which describes the time evolution
of this distribution $w(x, t)$, \eon \pd{w(x,t)}{t} = [x - \bar x(t)
- \u] w(x, t) + \u w(x-s, t) + \sqrt{w(x,t)} \xi(x,t),
\label{eq:stocheqn} \eoff where $\bar x(t)$ is the population mean
fitness and $\xi$ is a gaussian random term but with subtle
correlations needed to ensure that the fluctuations do not change
the total population size $N=\sum_x w(x)$. Studying this equation
can then lead to predictions of the speed of evolution, maintenance
of variation, and other interesting quantities.

The simplest possible approach is to neglect genetic drift and
attempt an ``infinite-$N$'' solution to the problem.  This
deterministic approach is extremely useful in many situations,
including in understanding deleterious mutation-selection balance.
However, when considering beneficial mutations, it is essential to
account for genetic drift and, crucially, the discrete nature of
individuals. Fractional numbers of deleterious mutations, implicit
in the deterministic mathematical analyses that are often
appropriate for large populations, are of little consequence
because they are selected against.  But allowing fractional
numbers of beneficial mutants at the nose yields nonsense because
fractional individuals that are highly fit multiply and take over
the population.  Thus even for very large populations, the
population size, which determines the smallest fraction of the
total population that represents at least one individual, plays a
crucial role.  Infinite-$N$ deterministic approximations are not
even qualitatively correct.

The problems with the simple deterministic approximation to
\eq{eq:stocheqn} are revealed by analyzing the resulting behavior.
This shows that the deterministic solution does not support a steady
state $v$ --- rather, it predicts that the speed of evolution
accelerates without bound. This is clearly unbiological, as it
involves a concomitant exponentially increasing width of the
distribution and thus smaller and smaller numbers in the nose.
Except for very short times (roughly until the nose develops in the
correct analysis), the deterministic approximation is thus
drastically wrong even for very large $N$.  The source of the
problem is that each more-fit population grows faster than the one
before.  Thus early mutants into a new more-fit fitness class at
fitness $x+s$ grow faster than the population at fitness $x$. This
means that even tiny fractions of an individual --- certainly
non-biological! --- will later give rise to a large population even
without further mutations.  Indeed, it is the ``descendants" of
these early fractional mutants  that will later dominate the
population of individuals at fitness $x+s$, despite the fact that
there are more mutants occurring from fitness $x$.  These
descendants then produce fractional mutants to fitness $x+2s$, and
the unrealistic aspects are further exacerbated.

An alternative way to study \eq{eq:stocheqn} is to use a
moment-based approach.  We can can multiply \eq{eq:stocheqn} by $x$
and integrate to find the rate of change of the first moment of the
fitness distribution (the speed of evolution) in terms of the second
moment (the variance).  In the limit that mutation is negligible
compared to selection in the bulk of the fitness distribution,
$d\langle x \rangle/dt\approx {\rm var}(x)$, simply the fundamental
theorem of natural selection.  One can easily work out that the time
derivative of the second moment (the variance) involves the third
moment. The time derivative of the third moment involves the fourth
moment, and so on. This moment hierarchy does not close. Even so,
this approach can yield accurate results for short timescales. The
more moments that are kept, the longer the results will be accurate
for, and if enough are kept the steady state speed of evolution can
be calculated accurately.  The lowest-order version of this is
familiar --- it corresponds to assuming that the variance is given
by its value at $t = 0$ and does not change, and that the speed of
evolution is equal to that.

\citet{kesslerlevineunpub} have carried out a sophisticated analysis
using a moment-based approach; their work contains a more detailed
analysis of the issues involved. Accounting properly for the effects
of mutations, stochasticity, discreteness in population number, and
fixed total population size are very difficult. Thus far, this
analysis involves complex moment equations which unfortunately
provide little intuition and no simple analytic results.

The problems with moment equations are  unsurprising based on our
analysis.  As we have noted, it is the lead $qs$, not the variance
or another moment which is most naturally thought of as being
maintained by the balance between mutation and selection. This lead
is not a moment of the fitness distribution --- it is instead a
measure of its nose, near to which the discreteness in population
number is crucial. The lead thus represents some combination of high
moments of the fitness distribution, with the order of the moments
that matter depending on $N$: to capture the effects of the sharp
nose of the distribution, at least of order $2\ln Ns$  moments are
needed, and such high order moments may be dominated by rare
fluctuations of the lead.  It is hardly surprising that getting at
the dynamics of the lead with a moment expansion is very cumbersome.
Our approach, in contrast, handles the stochastic issues at the nose
in a natural way while simply tracking the effects of selection that
dominate in the bulk of the distribution.

\section*{Appendix B:  Variable $N$ and Effective Population Size \label{changingn}}

We have thus far assumed that the population size is constant. We
now consider what happens when we relax this assumption.

If changes in $N$ are rapid compared to the changes in the mean
fitness, then we can define a constant effective population size
$N_e$.  The definition of $N_e$ can be complicated
--- it is not necessarily the geometric mean of the actual
population sizes.  Rather, $N_e$ is the value of the constant
population size in our model that gives the same stochastic dynamics
as the the changing $N$ situation averaged over a timescale long
compared to the shifts in $N$.  In practice, this means that if our
variable-$N$ population were clonal, $N_e \u s$ must be the
time-averaged rate at which beneficial mutations would establish.
Our theory at constant $N$ is then correct provided we use $N =
N_e$. Serial dilution protocols are one case relevant to many
experimental situations. Here, a population grows exponentially for
$G$ generations, is diluted back to its original size $N_b$, and
then this cycle is repeated.  The effective population size in this
scenario was calculated by \citet{243}, who found $N_e = N_b G \ln
(2)$.

In the opposite regime where the changes in $N$ are much slower than
changes in the mean fitness, the lead and fitness distribution
adjust quickly enough that the correct steady-state behavior for the
current $N$ always obtains.  This means that we can simply replace
the $N$ in our results with the time-dependent $N(t)$.

If the changes in $N$ occur on comparable timescales to the changes
in the mean fitness, the situation is much more complicated.  We
cannot define an effective population size, because the changes in
$N$ are too slow to be ``averaged'' over. On the other hand, the
changes in $N$ are too fast to allow the population to continuously
adjust and stay in steady state. Rather, the population will often
be in a transient regime with a complex dependence on past values of
$N$.  We do not analyze this case.  Though it is an interesting
subject for future work, it is a special situation which is unlikely
to have general importance.

\section*{Appendix C:  Running out of Beneficial Mutations \label{changingu}}

We have taken the beneficial mutation rate $\u$ to be a constant.
However, each beneficial mutation that establishes is likely to
change the total number of beneficial mutations that are available.
Clearly once an individual has a beneficial mutation, that
particular mutation is no longer available.  But it is also possible
that one mutation may open up or close off other possibilities.
Thus the beneficial mutation rate $\u$ may change in complicated
ways.

In many cases, $\u$ will change slowly with each mutation.  Our
theory predicts that the steady-state value of $q$ at a given $\u$
is $q(\u) = \frac{2 \ln [N qs]}{\ln \left[ \frac{(q-1)s}{\u}
\right]}$.  Provided that the change in $q(\u)$ over $q$ mutations
(after which the fitness distribution has moved through its full
width) is small, then the population is always approximately in the
steady state and our theory still holds --- we simply replace $\u$
everywhere with the appropriately varying $\u(t)$.  This condition
holds provided the change in $\u$ from a single mutation is small
enough that $q(\u)$ changes by much less than one.

When $\u$ changes rapidly enough with each mutation that this
condition is violated, the population fitness distribution does not
adjust quickly enough to stay in steady state.  In this case, the
population will often be in a transient regime with a complex
dependence on past values of $\u$.  This situation can be analyzed
with the algorithmic methods described in the section on transient
behavior.

One type of change in $\u$ is of particular interest:  when each
mutation that establishes is no longer available, but does not open
up or close off any other possibilities.  We assume that there are
initially $k$ beneficial mutations, each of which occurs at a rate
$\mu$.  After $i$ such mutations have been established, there are
$\ell = k-i$ left, and the mutation rate is $\u = \ell \mu$.  This
situation has been analyzed in great detail by \citet{313}.  We can
get a sense of the behavior by substituting $\u = \ell \mu$ into our
formula for $q$ to calculate how much $q$ changes after a single
establishment. If this is much less than $1$, our steady-state
theory is a good description of the dynamics; we simply use the
appropriate (changing) value of $\u$. Otherwise, the population will
often be in a more complicated transient regime. This condition
corresponds to \eon q_{\ell} \ln \left[\frac{\ell}{\ell -1} \right]
\ll \ln \left[ \frac{s}{(\ell-1) \mu} \right], \eoff where
$q_{\ell}$ is the value of $q$ corresponding to $\u = \ell \mu$.
Since we have assumed that $\frac{s}{\u} \gg 1$, this condition will
almost always be satisfied, even for very small values of $\ell$
(i.e. when the population has almost reached the fitness ``peak'').
The only potential complication is that if $\frac{s}{\mu} \lesssim
1$, then our assumption $\frac{s}{\u} \gg 1$ may break down for
small values of $\ell$.

\section*{Appendix D:  Fluctuations in $\tau$, Variations in $v$, and Stability
of the Steady State \label{flucts}}

The establishment time $\tau_q$ is a random variable.  Above we
calculated the steady state assuming that each establishment takes
the average establishment time $\langle \tau_q \rangle$.  However,
there are stochastic variations in this establishment time which
lead to fluctuations in the speed of evolution.  These variations
could also affect the average $v$, because the average $v$ is really
determined by the average effect of variable $\tau_q$, not the
effect of the average $\tau_q$ as we have assumed thus far.

The full distribution $P(\tau_q)$ is a special function --- a change
of variable in the one-sided Levy distribution $P(n_q)$. However, we
can calculate arbitrary moments $\ev{\tau_q^m}$. The second moment
is \eon \ev{\tau_q^2} = \frac{1}{[(q-1)s]^2} \left[ \ln^2 \left[
\frac{s}{\u} \frac{(q-1) \sin(\pi/q)}{\pi e^{\gamma/q}} \right] +
\frac{\pi^2}{6} \frac{2q-1}{q^2} \right]. \eoff  From this we can
calculate the variance in $\tau_q$, \eon \textrm{Var} (\tau_q) =
\frac{\pi^2}{6} \left[ \frac{1}{[(q-1)s]^2} - \frac{1}{[qs]^2}
\right]. \eoff  The relative variation in $\tau_q$ is thus \eon
\frac{\sqrt{\textrm{Var} (\tau_q)}}{\ev{\tau_q}} \approx
\frac{\pi}{q \ln \left[ \frac{s}{\u} \right]} \sqrt{\frac{2q-1}{q}}.
\eoff For small $\frac{\u}{s}$, this is small even for $q=2$ and
decreases as $\frac{1}{q}$ for large $q$.  Thus the total
fluctuations in the lead (and the speed of evolution) are small, and
ignoring them in calculating the average $v$ is reasonable.

From these fluctuations in $\tau_q$, we would like to calculate the
expected fluctuations in $v$.  This would explain how much variation
in adaptation we should expect between different populations
experiencing the same conditions (for example, geographically
distinct subpopulations or different experimental lines).
Unfortunately, however, this is a difficult problem.  This is
because successive establishment times are not independent.   A
shorter than average $\tau_q$ immediately increases the lead. This
tends to make subsequent establishments shorter as well. The
opposite is true for longer than average $\tau_q$.  Thus the lead is
unstable to fluctuations in the short term --- increasing the lead
due to a short $\tau_q$ creates a tendency to further increase the
lead, and vice versa. This effect is enhanced because a shorter than
average $\tau_q$ means that the population is less influenced by
subsequent mutations, so its size earlier is slightly bigger than
usual (i.e. $\tau(2 \tau_q)$ is closer to $\tau_q$ than usual).
Again, the opposite is true for longer than average $\tau_q$. This
short-term instability is checked at later times. A subpopulation
with a short $\tau_q$ is more fit relative to the mean than it would
be with an average $\tau_q$. It thus becomes the dominant
subpopulation, increasing the mean fitness, more quickly. When this
happens, the lead is decreased --- roughly $q$ establishments after
the short $\tau_q$. Thus the various $\tau_q$ are correlated in a
complicated way: a short $\tau_q$ tends to favor further short
$\tau_q$, until roughly $q$ establishments later when it favors
longer $\tau_q$, and the opposite is true for longer than average
$\tau_q$.

To understand these complications, it is important to consider more
carefully the form of the distribution of $\tau_q$, especially for
large $q$. Since $\ell\equiv\ln s/\u$ is large, it is convenient to
define \be \tau_q =\frac{1}{(q-1)s}[\ell -\Delta], \ee with $\Delta$
having both average value and stochastic fluctuations of order unity
(and thus small compared to $\ell$).  For small $q$, the
characteristic magnitude of the fluctuations are correctly captured
by the variance.  The behavior for large $q$ is somewhat more
subtle. In this limit, the mean value of $\Delta$ is of order $1/q$,
but its distribution has an interesting form: $\Delta$ is typically
much smaller than $1/q$ and is rarely negative, but with probability
of order $1/q$ it is positive of order unity. The variance of
$\Delta$ is thus of order $1/q$ as can be seen from the above result
for ${\rm Var}(\tau_q)$ but, in contrast to what one might expect,
all higher moments are also of order $1/q$.  The strongly asymmetric
form of the distribution of $\tau_q$ has a simple origin: there is
some chance that an establishment occurs anomalously early, but as
the feeding population is producing mutants at an exponentially
growing rate, it is highly unlikely that the establishment will be
anomalously late.

For large $q$ the form of the distribution of $\tau_q$ has
implications for the distribution of the ``sweep" time, $t_s$, until
new mutants will dominate the population.  This will be $t_s
q\tau_q\approx q\ell/(q-1)s\approx \ell/s$ on average for large $q$.
The variations in $t_s$ will arise from two sources. The first is
the sum of the variations of $q$ successive $\tau_q$'s. From the
above discussion, the sum of $q$ $\Delta$'s will have a distribution
with typical and average value both of order unity. This will give
rise to fractional variations of $t_s$ of order $1/qs$, which is
smaller than the mean $t_s$ by a factor of $1/q\ell$.

But there is another factor that needs to be taken into account: a
short $\tau_q$ will increase the lead and thus make the next
establishment likely to happen somewhat sooner, thereby making
subsequent ones likely to be even earlier. Until the mean population
feels the effects of the series of new mutant subpopulations, the
lead is thus exponentially unstable. But this effect is not large:
the deviation from average, $u(t)$, of the  speed of the lead, grows
proportionally to the increase, $\lambda(t)$, of the lead from $qs$
with \be \frac{du}{d\lambda}\approx \frac{s}{\lambda} \ . \ee Thus
$d\lambda/dt=\lambda s/\ell$, so that for large $q$ in a time
$t_s\approx \ell/s$, an anomalously large lead will only grow
further by a factor of $e$. This means that the effects of the
exponential instability of the lead are only beginning to be felt
before they are counteracted by a sooner than typical advance of the
mean fitness.  The above estimate from a sum of roughly independent
$\Delta$'s thus correctly gives the rough magnitude of the small
variations in $t_s$.  But the correlations between successive
$\tau_q$'s means that the velocity fluctuations are correlated over
times of order $t_s$.

On time scales much larger than $t_s$, the mean fitness $\bar{y}(t)$
will grow, with the mean speed $\bar{v}$ and diffusive fluctuations
around this described by \be \langle
[\bar{y}(t)-\bar{y}(t')]^2\rangle\approx [\bar{v}(t-t')]^2 +
2D|t-t'|, \ee with the diffusion coefficient inferred from the above
to be \be D\sim \frac{s^3}{\ell^3} \ . \ee

\section*{Appendix E:  On the Cutoff in the Integral in H and the Pathologies of
$\ev{n_q(t)}$ \label{cutoff}}

One initially surprising property of the distribution $P(n_q, t)$ is
that it has infinite mean: that is, $\langle n_q \rangle = \infty$.
The infinity arises because we have allowed mutations from $n_{q-1}$
to $n_q$ to occur arbitrarily far back in the past --- even before
the establishment of the $q-1$ population (as described in the main
text, this was implicit in using $-\infty$ as the lower limit of
integration in the expression for $H$).  Naively, it seems that this
is a serious problem, and that the solution is to impose a realistic
cutoff in time before which mutations are disallowed. That is, we
could say that before $t=t_i$ there is a negligible chance of
mutations occurring and therefore set the lower limit of integration
in $H$ to be $t_i$.  This does remove the infinite $\langle n_q
\rangle$. However, it does nothing to address the underlying issue.
Rather than being infinite, we would then have $\langle n_q \rangle$
depending very strongly on $t_i$.  This is biologically
unreasonable, since the population $n_q$ arises from mutations which
tend to occur only after $n_{q-1}$ reaches a relatively large size
(naively, of order $\frac{1}{\u}$). Certainly the important
properties of $n_q(t)$ should be independent of whether we consider
only mutations that occur after $n_{q-1}$ reaches one individual
versus two individuals, for example.  Indeed, since our expression
for $n_{q-1}(t)$ is not valid at these small subpopulation sizes
anyway, for our results to be valid, they had better not depend on
such early times.

The solution to this apparent dilemma lies in the fact that the
average $n_q(t)$ is not an important property of the distribution of
$n_q(t)$.  Rather, $\langle n_q \rangle$ is dominated by events so
rare that they will never actually occur in practice --- namely,
when a mutation occurs in the subpopulation $n_{q-1}$ while
$n_{q-1}$ is extremely small.  The reason for the resulting large
$\langle n_q \rangle$ is that even though mutations are very rare
far back in time when $n_{q-1}$ is small, they have a huge effect on
the future $n_q$ when they do occur and establish.  Since the
subpopulation $n_q$ grows faster than the subpopulation $n_{q-1}$,
the very early mutations dominate over later ones.  This can be seen
explicitly. The probability of a mutation from the population
$n_{q-1}$ at a time $t_0$ is $\frac{\u}{qs} e^{(q-1)st_0}$, and if a
mutation occurs at that time it will on average lead to a lineage
that at later time $t$ is of size $n_q \sim e^{qs(t-t_0)}$. Thus the
contribution to $\langle n_q(t) \rangle$ from mutations at time
$t_0$ is of order $\frac{\u}{qs} e^{-st_0} e^{qst}$.  The dependence
on $t$ is as expected. However, the dependence on $t_0$ is such that
the smaller $t_0$ is (especially at large negative $t_0$), the
larger the contribution to $\langle n_q \rangle$.  This average
$n_q$ is thus dominated by mutations that happened very early.  The
essential point is that although the probability of a mutation
decreases exponentially at rate $(q-1) s$ as we decrease the initial
time $t_0$, its effect on $n_q$ increases exponentially at the
\emph{faster} rate $q s$.

But the lower limit of the mutation times is only important for
determining the very large-$n_q$ form of $P(n_q, t)$. This part of
$P(n_q, t)$ contains extremely small probabilities of extremely
large $n_q$, in such a way  that all integer moments of $n_q$ depend
crucially on this choice. However, this high-$n_q$ part of $P(n_q,
t)$ represents such a small total probability that it would not
occur in any real population. Thus getting $P(n_q, t)$ correct for
this high $n_q$ cannot matter. To get the quantities of interest ---
in particular $\langle \ln n_q(t) \rangle$ --- we can therefore use
any cutoff we choose, and $-\infty$ is a convenient choice.

The problems with $\langle n_q(t) \rangle$ all stem from the fact
that the population grows exponentially once mutations occur. Thus
it is natural to ``factor out'' this deterministic exponential
growth in defining aspects of the distribution $P(n_q, t)$ and then
focus on the distribution of $\ln n_q(t) -qst$. This is what our
definition of $\tau_q$ accomplishes.  The variable $\tau_q$, as we
have seen, has none of the problems of $\ev{n_q(t)}$ and its
distribution is independent of the cutoff we choose (except for a
tiny and irrelevant tail for very anomalously small $\tau_q$).  As
described in the section on the fate of a single mutant, the
essential point here is the difference between $\ev{e^X}$ and
$e^{\ev{X}}$. The former (analogous to $\ev{n_q}$) is very sensitive
to the tails of $P(X)$, while the latter (analogous to
$e^{\ev{\tau_q}}$) is not. And it is the latter that will determine
the mean speed and fluctuations around this.

\section*{Appendix F:  Multiple Stochastic Clones and $\frac{\u}{s}$}

Our analysis rests on a separation between deterministic and
stochastic dynamics, which we used to overcome the limitations of
branching process models. Such a separation is always possible for
$Ns \gg 1$, as noted above, because nonlinear effects are not
important when stochastic effects are, and vice versa. However, we
have made a stronger assumption:  that the separation is possible
right at the nose, so that only the most-fit subpopulation must be
treated stochastically but that all other subpopulations are
deterministic.  This is an important assumption, as a full
stochastic treatment would involve, for example, a double-mutant
subpopulation whose size is a random variable sending mutations
into a triple-mutant subpopulation whose size is also a random
variable, and so on.  These multiply random processes are
difficult to understand analytically.

Fortunately, there is a broad parameter regime in which only the
most-fit subpopulation is small enough to require stochastic
analysis.  Two conditions must be met.  First, the most-fit
subpopulation at the nose cannot generate new mutations that are
destined to fix until it has become large enough that the stochastic
effects are negligible.  Implicit in this condition is the
assumption that the most-fit subpopulation \emph{can} generate
mutations that are destined to go extinct due to drift. This naively
seems reasonable, as mutations destined to go extinct due to drift
should not matter in the long term.  This leads to the second
condition: a population destined to go extinct due to drift cannot
itself generate a mutation that will become established ---
otherwise it does matter after all. Here we consider this latter
condition.  In the next Appendix, we consider the former.

We begin by studying the dynamics of the lineage founded by a single
mutant.  Thus we are concerned with a stochastic subpopulation with
a fitness $\fit$ (or some $p s$) greater than the mean fitness of
the population, evolving by our branching process model starting
from $1$ individual at $t=0$ and with no further mutations.  We
denote the size of this subpopulation at time $t$ by $n(t)$.  We
have already calculated $P(n, t)$, but this quantity offers no
straightforward ways to understand whether mutations can arise while
$n$ is still stochastic.

The expected number of mutations that arise from the mutant
lineage is $\int_0^{\infty} \u n(t) dt$.  Inspired by this, we
define \eon W = \int_0^{\infty} n(t) dt \eoff as the ``weight'' of
the mutant lineage.  If the lineage becomes established, $W$ will
be infinite (the nonlinear saturation effects are not part of the
branching process).  However, if the lineage goes extinct due to
drift, $W$ is the overall integrated population size.  The
expected number of mutations destined to survive drift, $k$, that
arise from this lineage is therefore $k=W \u \fit$.

We can exploit the independence between stochastic lineages (valid
because $Ns \gg 1$) to calculate $W$.  The initial mutant that
founds the lineage will either die (with probability $\frac{1}{2+
\fit}$) or give birth (with probability $\frac{1+\fit}{2+\fit}$).
The time $T$ until this happens is exponentially distributed with
rate $2+\fit$ (i.e. $\dpr{T=t} = (2+\fit) e^{-(2+\fit)t}$).  If it
dies, $W$ is simply $T$. If it gives birth, $W$ is $T$ plus the
$W$ of each of the two offspring. We therefore have \eon \dpr{W=w}
= \frac{1}{2+\fit} \dpr{T=w} + \frac{1+\fit}{2+\fit} \int_v^w
\int_0^w du dv \dpr{T=u} \dpr{W=v} \dpr{W=w-u-v}. \eoff  Converting
to Laplace transforms, we can solve for $W$ to find \eon W =
\frac{2 + z + \fit - \sqrt{z^2 + 2z\fit + \fit^2 +
4z}}{2(1+\fit)}, \eoff where $W(z)$ is the Laplace transform of
$\dpr{W=w}$.  Note that $W(z \to 0) = \frac{1}{1+\fit}$, not $1$.
This is because there is a finite probability (roughly $\fit$)
that the lineage becomes established and thus has infinite weight.
To focus on the lineages that do go extinct, we simply ignore this
weight at infinity.

This form of $W(z)$ is impossible to invert analytically for
general $w$.  However, the small-$z$ behavior controls the
dynamics at large $w$.  For $w \gg 1$ we have that $\dpr{W=w}$
falls off at least as fast as \eon \dpr{W=w} \approx \frac{1}{2
\sqrt{\pi} (1+\fit)} e^{-\fit^2 w/4} w^{-3/2}. \eoff  Values of
$w$ that are larger than roughly $\frac{1}{\fit^2}$ are
exponentially suppressed. Integrating this result, we find that
less than a fraction $\fit$ of the lineages have a weight greater
than $\frac{1}{\fit^2}$, and almost all of these are right at $w =
\frac{1}{\fit^2}$.  This makes intuitive sense.  The largest size
a lineage can reach without establishing is roughly
$\frac{1}{\fit}$. If it does so, it takes roughly $\frac{1}{\fit}$
generations to get to this size and another $\frac{1}{\fit}$
generations to then go extinct. This is because the dynamics are
an approximately neutral process while the lineage size is less
than $\frac{1}{\fit}$ (drift dominates selection in this regime),
so the classical neutral result applies.  During this period its
average size is roughly $\frac{1}{2 \fit}$, so the maximum value
$w$ can take should indeed be about $\frac{2}{\fit} \frac{1}{2
\fit} = \frac{1}{\fit^2}$. The chance of the lineage reaching size
$\frac{1}{\fit}$ is also about $\fit$ (again by analogy to the
classical neutral result) and once there it is about as likely to
establish as to eventually go extinct. So our result that $w$
takes on this maximum value roughly a fraction $\fit$ of the time
also makes intuitive sense.

To assume that mutations destined to establish never arise from a
subpopulation destined to go extinct, we require $k = W \u \fit
\ll \fit$. Note the RHS of this expression is $\fit$ because there
are $\frac{1}{s}$ lineages that go extinct for every one that
establishes, and mutations destined to fix must be much more
likely to arise from lineages that establish. Since the maximum
value of $w$ is roughly $\frac{1}{\fit^2}$ and this occurs a
fraction $s$ of the time, this translates to the condition
$\frac{\u}{\fit} \ll 1$. (Values of $w$ less than
$\frac{1}{\fit^2}$ are more common, but in sum are still less
likely to produce a mutation.) Thus we can ignore mutations from
stochastic lineages destined to go extinct provided \eon
\frac{\u}{s} \ll 1. \eoff

We have not yet considered whether mutations can arise in the
stochastic period of lineages destined to survive.  We address this
question in more detail in Appendix G. However, below a size
$\frac{1}{s}$ the lineages that establish behave similarly to the
lineages that are destined to reach size $\frac{1}{s}$ and then go
extinct, and above this size the surviving lineages quickly become
deterministic.  Thus we expect that whenever mutations never arise
from lineages that go extinct, they will also never arise during the
stochastic period of lineages destined to survive.

\section*{Appendix G:  The $\tau(t)$ Approximation \label{tauoftapprox}}

Our method of linking the deterministic behavior of the bulk of
the population to the stochastic behavior at the nose hinges on
our definition of $\tau_q$.  We defined $\tau_q$ as $\tau(t \to
\infty)$, where $\tau(t)$ is defined by \eon n_q(t) = \frac{1}{qs}
e^{qs(t - \tau)}. \eoff  The variable $\tau(t)$ is just a change
of variable from $n_q(t)$.  From its definition, we see that
$\tau(t)$ is the time at which the subpopulation would have
reached size $\frac{1}{qs}$ had it always grown exponentially at
rate $qs$ until reaching size $n_q(t)$ at time $t$.  Thus
$\tau(t)$ accounts for all the incoming mutations and stochastic
behavior up to time $t$ and allows us to summarize it by saying
$n_q$ reached size $\frac{1}{qs}$ at time $\tau(t)$ and was
deterministic thereafter. The definition of $\tau_q$ as $\tau(t
\to \infty)$ thus summarizes \emph{all} the random behavior and
\emph{all} incoming mutations into a time the subpopulation would
have reached size $\frac{1}{qs}$. Yet this is not actually the
time the subpopulation reached size $\frac{1}{qs}$ (Fig. 5).  It
could, for example, have reached $\frac{1}{qs}$ earlier than this
but by chance grown slower than $e^{qst}$ for a while thereafter.
Despite this, we have assumed that the subpopulation did in fact
reach size $\frac{1}{qs}$ at its establishment time in defining
its size thereafter.  That is, we have written $n_{q-1}(t) =
\frac{1}{qs} e^{(q-1) st}$, defining $t=0$ to be the establishment
time of this population.  And we use this form of $n_{q-1}(t)$ in
calculating how many mutations this subpopulation generates.

In order for this to be reasonable, our form of $n_{q-1}(t)$ must
be accurate once this population becomes large enough that it
starts generating mutants.  This happens roughly $\tau_q$
generations after $n_{q-1}$ became established (by definition, it
takes roughly $\tau_q$ generations for the next mutations to
occur, because $\tau_q$ is dominated by the waiting time for the
first mutation to occur).  Thus for our result to be accurate,
$\tau(2 \tau_q)$ must be approximately $\tau_q$ (to be precise, we
require $\tau(2 \tau_q) - \tau_q \ll \frac{1}{s}$).  That is,
there must not be much stochasticity after the population is large
enough to generate mutations (and additional incoming mutations
must be negligible). Looked at another way, this means that the
population cannot generate mutations while it is stochastic.

To calculate $\tau(2 \tau_q)$, we return to our solution $H(\zeta,
t)$ for the Laplace transform of $P(n_q, t)$.  The time-dependence
of $\tau$ is hidden in \eq{22eqn} --- our assumption that $\zeta$
is small here assumes we are interested only in larger $n_q$ and
is thus equivalent to taking $t \to \infty$.  We can do this
integral more carefully; the result involves hypergeometric
functions. These can be expanded for $\zeta \ll qs$ but nonzero,
corresponding to values of $n_q(t)$ larger than $\frac{1}{qs}$ but
before this subpopulation generates mutations.  We find \eon
H(\zeta, t) = \exp \left[ -\frac{\u}{s} \left( \zeta e^{qst}
\right)^{1 - 1/q} \left[ \frac{\pi}{\sin (\pi/q) (qs)^{1-1/q}} -
\frac{\zeta^{1/q}}{s} \right] \right]. \eoff  Unfortunately, this
form of $H$ is more complex and we cannot exactly compute
$\tau(t)$.  However, we can find typical values of $\tau(t)$ and
$\tau_q$ from this by the same methods as before.  We can also
compare the size of the second term in $H$ (which gives the time
dependence in $\tau(t)$) to the first for values of $\zeta \sim qs
\frac{\u}{s}$, which corresponds to $n_q \sim \frac{1}{qs}
\frac{s}{\u}$, the time this subpopulation begins to generate new
mutations.  Both calculations demonstrate that our approximation
is valid provided that \eon \frac{\u}{qs} \ll 1. \eoff  This
result can be confirmed with a deterministic analysis. About
$\tau_q$ generations after becoming established, a subpopulation
has a size $n = \frac{1}{qs} \frac{(q-1) s}{\u}$. Once it has
reached this size, selection dominates drift and mutations. Thus
subsequent random or mutational events will not significantly
affect $n$, so $\tau(2 \tau_q)$ and $\tau_q = \tau(\infty)$ are
similar.

Thus whenever $\frac{\u}{s} \ll 1$, our method of linking together
stochastic and deterministic dynamics is reasonable.  Populations
never generate mutations while they are stochastic, and hence we
are justified in using a deterministic approximation for all but
the most-fit population.  When this condition fails, we must treat
multiple populations stochastically and the analysis becomes much
more complex.  We could still divide up the population into a
nonlinear deterministic part and a linear stochastic part
(provided only that $Ns \gg 1$), but the stochastic part would
have to include multiple subpopulations.

\section*{Appendix H:  Approximations in the Behavior of $q$ \label{approxq}}

In our analysis to this point, we have assumed that the mean fitness
$\bar y s$ changes abruptly, increasing by $s$ every $\tau_q$
generations.  We used this assumption in calculating $q$ and it is
the reason why we have a constant $q$.  In this Appendix, we discuss
this approximation.

There are two important time scales that determine the relative
sharpness of the changeovers from one dominant population to the
next and, concomitantly, from the lead population growing with rate
$qs$ to rate $(q-1)s$.  Because the second largest population grows
at rate $s$, the time scale for this changeover is $1/s$. But the
time {\it between} such changeovers is $\tau_q$.  The ratio of these
is \be w\equiv\frac{1}{s\tau_q}=\frac{v}{s^2} \ , \ee which is
$1/\ln(s/\u)$ and thus small at the crossover from the successional-
to the multiple- mutations regimes. Indeed as long as $w\ll 1$, the
changeover is relatively sharp on the scale of $\tau_q$ and it is a
good approximation to consider it abrupt, as we have done.

We can make this more precise by computing the actual behavior of
the mean fitness as a function of time. Assume (for convenience)
that at $t = 0$, the mean fitness is at $y = -\frac{1}{2}$, i.e. in
the middle of a changeover. The subpopulations at $y=-1$ and at $y =
0$ are equal in size, and those at other values of $y$ are smaller
by a factor of $e^{-\sum_{k=1}^y ks \tau_q} = e^{-\frac{s \tau_q}{2}
[(y + 1/2)^2 - 1/4]}$. For small $w$, the one or two largest
subpopulations strongly dominate, as these factors are all very
small. This is because the variance of the fitness in the population
in the multiple mutations regime is simply $v$, since the dynamics
of the bulk of the population is controlled by selection.  Thus the
standard deviation is smaller than $s$, making the other
subpopulations far smaller than the dominant one. The parameter $w$
is simply the variance in units of $s^2$.

At future times, the subpopulations all grow (or shrink)
exponentially at a rate $y s$ reduced by the mean fitness (but we
can neglect the mean fitness in this calculation because it affects
all subpopulations equally).  To keep the total population fixed
thus requires that the mean fitness be \eon \bar y(t) =
\frac{\sum_{y = -q}^q y e^{-\frac{s \tau_q}{2} (|y| + 1/2)^2 + y s
t}}{\sum_{y=-q}^q e^{-\frac{s \tau_q}{2} (|y| + 1/2)^2 + y s t}} \ .
\eoff We can perform these Gaussian sums by the Poisson resummation
formula to yield \be \bar y(t) = vst - \frac{1}{2} +
d/d(st)\left\{\ln\left[\sum_{k=-\infty}^{+\infty} e^{2\pi i k w st
-2\pi^2 w k^2}\right]\right\} \ . \ee If $w$ were {\it large}, the
$k=0$ term would dominate, and the $k=\pm 1$ yield relative
variations in the speed \be \frac{v(t)}{\bar{v}}\approx 1-8\pi^2
we^{-2\pi^2w}\cos(2\pi wst) + {\cal{O}} {(e^{-4\pi^2w})} \ \ee and
corresponding variations in $\bar{y}(t)-t/\tau_q$ that are a smaller
by a factor of $1/(2\pi)$.  Thus in practice the parameter that
needs to be large for $\bar{y}(t)$ to increase smoothly is $2\pi^2
w$. Only for $w<0.2$ do the variations in $v$ become more than a
factor of two, and substantial deviations of $\bar{y} (t)$ from
smooth occur only for $w<0.1$.  Above this, our abrupt-transition
approximation is not valid, but despite this our earlier results are
still good; we discuss this below.

The parameter that we have taken to be small throughout is
$1/\ln(s/\u)$. This is the value of $w$ at the crossover from
successional- to multiple-mutations regimes. Strictly speaking, this
means that $w$ is small until $\ln Ns \sim (\ln s/\u)^2$.  For even
larger population sizes, the behavior near the nose changes
somewhat, as discussed in the main text.

When $w$ is small enough that the shifts in $\bar y$ are abrupt, the
dynamics can be worked out more generally than we have done in the
main text. There, we approximated the most-fit deterministic
subpopulation to be growing as $e^{(q-1)st}$ for $\tau_q$
generations, after which $\bar y$ increases by $1$ and the
subpopulation growth slows to $e^{(q-2)st}$, and so on.  This is
strictly valid only for {\it integer} q.  When the naive value for
$q$, $2L/\ell$, is non-integer, the populations shift between growth
rates some fraction of the way between one establishment and the
next.  The effects of this can be taken into account
straightforwardly as long as the shift between growth rates is
indeed abrupt on the scale of $\tau_q$: i.e. that $w$ is small. Here
we ignore factors inside logarithms: to get these one would need to
use the fuller analysis of the feeding and lead population dynamics
used in the text. For our purposes here,  the heuristic derivation
of the establishment times is sufficient.

It is convenient to keep $q$ an integer, with $qs$ the growth rate
of the lead population when it first becomes established.  We then
define a non-integer generalization of $q$ to be $\Q$ with \be
q={\rm GI}(\Q) \ , \ee the greatest integer less that or equal to
$\Q$. It is $\Q$ that is simply related to the population parameters
via \be \Q\approx 2L/\ell \ , \ee i.e. what was previously found for
$q$. The dimensionless speed is found to be \be w\approx
\frac{q(q-1)}{\ell(2q-\Q)} , \ee which is equal to the result in the
text, $w\approx (\Q-1)/\ell$, for integer $\Q$. The difference
between these is small for large $\Q$, with the fractional error of
the simple result (which is an overestimate) largest at
$\Q=q+\frac{1}{2}$, where it is only $1/4q(q-1)$ and thus small even
for the worst case $\Q=2.5, \ q=2$.

In the opposite case where $w$ is not small, the approximation of
abrupt shifts in $\bar y$ is not valid.  In this case, we can make
the opposite approximation that the mean fitness increases at a
uniform rate: from the above discussion, this is valid unless $w$ is
quite small (although strictly speaking this is not true in the
limit that $\ell$ is large with fixed $q$).  In the constant
mean-speed approximation, one obtains \be w\approx
\frac{\Q-1+\sqrt{(\Q-1)^2-1}}{2\ell} , \ee which is an underestimate
that is worst at integer $\Q$: for large $\Q$ the worst fractional
error is $1/4(q-1)^2$.  Since this is small compared to the speed,
the approximation in the main text is reasonable.

We can get an intuitive understanding of why this approximation of
abrupt shifts in $\bar y$ gives reasonable results, even when $\bar
y$ actually increases smoothly.  First we consider the deterministic
dynamics of the bulk of the fitness distribution.  Here the shape of
the distribution (and hence the identity of the most common
subpopulation) depends only on the relative growth rates of the
subpopulations, so assumptions about $\bar y$ are irrelevant. For
the stochastic behavior at the nose, our assumption is more
problematic.  When the mean fitness in fact increases steadily,
rather than jumping by $s$ every time an establishment occurs, our
calculated lead $q$ gives the correct average mean fitness over the
stochastic period. This means we calculate the stochastic dynamics
assuming the correct average mean fitness, but this is slightly
different from the stochastic dynamics given the changing mean
fitness.  Essentially, we have used $q = 3.4$, for example, as an
interpolation for the correct behavior when the lead is just below
$4$ immediately before an establishment, declining gradually to
below $3$ shortly before the next establishment. Rather than
calculate $\tau_{3.4}$ from the stochastic behavior while the lead
shifts correctly, in the main text we have calculated it based on a
constant lead of $3.4$.  As we have seen above, however, the
difference is small.

We conclude with some comments on the stochastic aspects of the
speed of the nose.  These make the above analysis questionable
because of the assumption of deterministic establishments of the
lead populations.  But, as we discuss in Appendix D,the variations
in the establishment times are at worst of order $1/qs$ compared to
the mean $\tau_q$ which is of order $\frac{\ln s/\u}{qs}$ (for large
$q$ they are even smaller than this, as discussed in Appendix D).
Thus the variations in the time intervals between takeovers of the
population by new dominant subpopulations is small compared to the
time intervals themselves.  Hence the deterministic approximation
for the increase of the mean fitness is good, at least as far as its
effects on the dynamics of the lead populations.

\newpage

Figure 1:

For beneficial mutations to be acquired by a population, they must
both arise and fix. (\textbf{a}) A small asexual population in the
successional-mutations (or strong selection weak mutation) regime.
Mutation $A$ arises early on. Provided it survives drift, it fixes
quickly, before another beneficial mutation occurs.  Some time
later, a second mutation $B$ occurs and fixes. Evolution continues
by this sequential fixation process. (\textbf{b}) A larger
population in the concurrent-mutations (strong selection strong
mutation) regime. A mutation $A$ occurs, but before it can fix
another mutation $B$ occurs and the two interfere. Here a second
mutation, $C$, occurs in an individual that already has the first
mutation $A$ and these two begin fixing together, driving the
single-mutants to extinction. This dynamics continues with further
mutations, such as $E$ and $F$, occurring in the
already-double-mutant population.  The key process is how quickly
mutations arise in individuals that already have other mutations.
This picture has elements of both clonal interference and multiple
mutations, illustrated separately in (\textbf{c}) and (\textbf{d}).
(\textbf{c}) The clonal interference effect in large populations: a
weak-effect beneficial mutation $A$ occurs and begins to sweep, but
is outcompeted by a later but more-fit mutation $B$, which in turn
is outcompeted by mutation $C$. $C$ fixes before any larger
mutations can occur; the process can then begin again. Multiple
mutations are ignored here. (\textbf{d}) The multiple mutation
effect: Several mutations, $A$, $B$, and $C$, of identical effect
occur and begin to spread.  Mutant lineage $B$ happens to get a
second beneficial mutation $D$, which helps it sweep, outcompeting
$A$ and $C$. Eventually this lineage gets a third beneficial
mutation $E$. Mutations that occur in less-fit lineages, or those
that do not happen to get additional mutations soon enough (such as
$BDF$), are driven extinct.

\newpage

Figure 2:

The simple positive selection model we study.  A large number of
beneficial mutations are possible, each of which increases the
fitness by the same amount $s$.  Thus the population climbs a long
fitness ``hill.'' All mutations are taken to have the same effect on
fitness, and the supply is assumed to be large enough that they are
not depleted. Epistatic interactions and deleterious mutations are
both ignored.

\newpage

Figure 3:

Schematic of the evolution of large asexual populations.
(\textbf{a}) The population is initially clonal. Beneficial
mutations of effect $s$ create a subpopulation at fitness $s$, which
drifts randomly until after time $\tau_1$ it reaches a size of order
$\frac{1}{s}$, after which it behaves deterministically.
(\textbf{b}) This subpopulation generates mutations at fitness $2s$.
Meanwhile, the mean fitness of the population increases, so the
initial clone begins to decline. (\textbf{c}) A steady state is
established. In the time it takes for new mutations to arise, the
less fit clones die out and the population moves rightward while
maintaining an approximately constant lead from peak to nose, $qs$
(here $q=5$).  The inset shows the leading nose of the population.
Note the logarithmic scale of the populations.

\newpage

Figure 4:

Schematic of a typical fitness distribution on a logarithmic scale.
The total population size is large: $Ns\gg1$. At the front of the
distribution --- the nose --- where only a few individuals are
present, stochastic effects are strong but nonlinear saturation is
not. The reverse is true in the bulk of the distribution.
Stochasticity is strong only when a subpopulation size $n$ is small,
$n \lesssim \frac{1}{s}$, and saturation is strong only when a
subpopulation size is large, $n \sim N$.  Thus there is a wide
intermediate regime where neither matters.  We can therefore use a
nonlinear deterministic model in the bulk of the distribution, a
linear stochastic model at the front, and match the two in the
intermediate regime where both are valid. The bulk of the
distribution is dominated by selection, which gives rise to a
steady-state gaussian shape except near the nose.

\newpage

Figure 5:

(\textbf{a}) The definition of the {\it establishment time}
$\taunuc$.  A single mutant individual is assumed to exist at $t=0$.
It drifts stochastically until it either goes extinct, or eventually
gets large enough that it grows exponentially and its behavior
becomes roughly deterministic. We define $\taunuc$ to be the
inferred time at which the population would have reached size
$\frac{1}{s}$ if one extrapolated backwards from the long-time
deterministic behavior.  Note that $\taunuc$ is {\it not} the time
the population actually reached size $\frac{1}{s}$ (indeed,
$\taunuc$ can be negative). (\textbf{b}) The definition of $\tau_q$:
the time between successive establishments of the lead population
with fitness $qs$ more than the mean. Mutations occur with a rate
that grows exponentially with time. Here, $\tau_q$ is the time the
new lead population would have reached size $\frac{1}{qs}$,
extrapolating backwards from its long-time deterministic behavior.
This includes both the time to generate a mutant destined to
establish and the time for it to drift to substantial frequency (of
order $1/qs$).

\newpage

Figure 6:

Comparisons between simulations and our theoretical predictions for
the mean speed of adaptation $v$ (measured in increase in fitness
per generation, $\times 10^5$).  (\textbf{a}) Speed of adaptation
$v$ versus $\mathrm{log_{10}}[N]$ for $\u = 10^{-5}$ and $s = 0.01$.
Both the large $N$ (\eq{veq}) and the moderate $N$ (\eq{modn})
theoretical results are shown in their regimes of validity, which
are above and below $N\approx 1/\u$ respectively (the crossover
between the two regimes is indicated in the figure). (\textbf{b})
$v$ versus $\mathrm{log_{10}}[\u]$ for $N = 10^{6}$ and $s = 0.01$.
(\textbf{c}) $v$ versus $\mathrm{log_{10}}[s]$ for $N = 10^6$ and
$\u = 10^{-5}$.

\newpage

Figure 7:

Comparisons between simulations and our theoretical predictions for
the mean $q$. (\textbf{a}) $q$ versus $\mathrm{log_{10}}[N]$ for $\u
= 10^{-5}$ and $s = 0.01$. (\textbf{b}) $q$ versus
$\mathrm{log_{10}}[\u]$ for $N = 10^{6}$ and $s = 0.01$.
(\textbf{c}) $q$ versus $\mathrm{log_{10}}[s]$ for $N = 10^6$ and
$\u = 10^{-5}$.

\newpage

\begin{figure}
  \centering
  \includegraphics[width=6.5in]{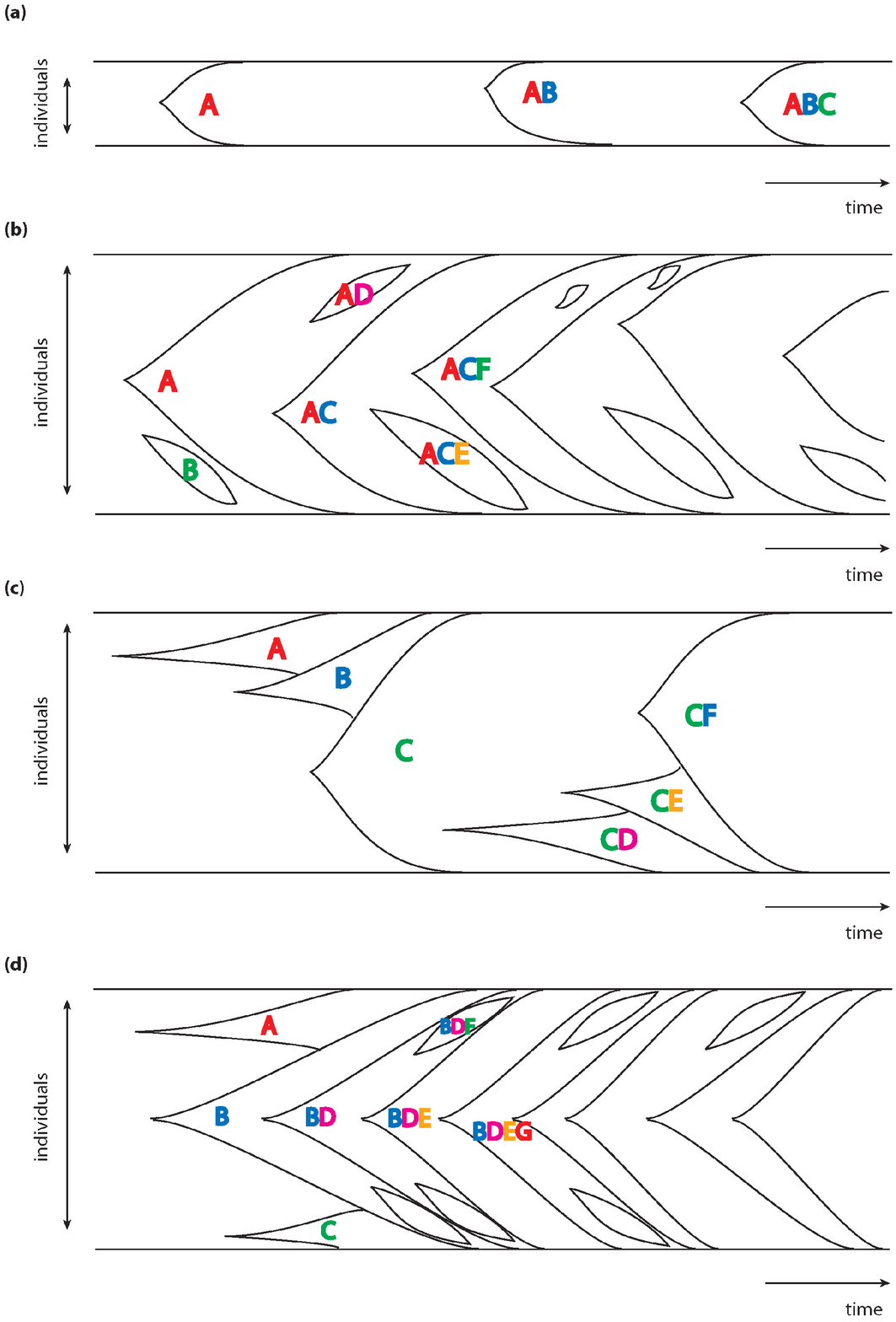}
  \caption{}
\end{figure}

\begin{figure}
  \centering
  \includegraphics[width=6.5in]{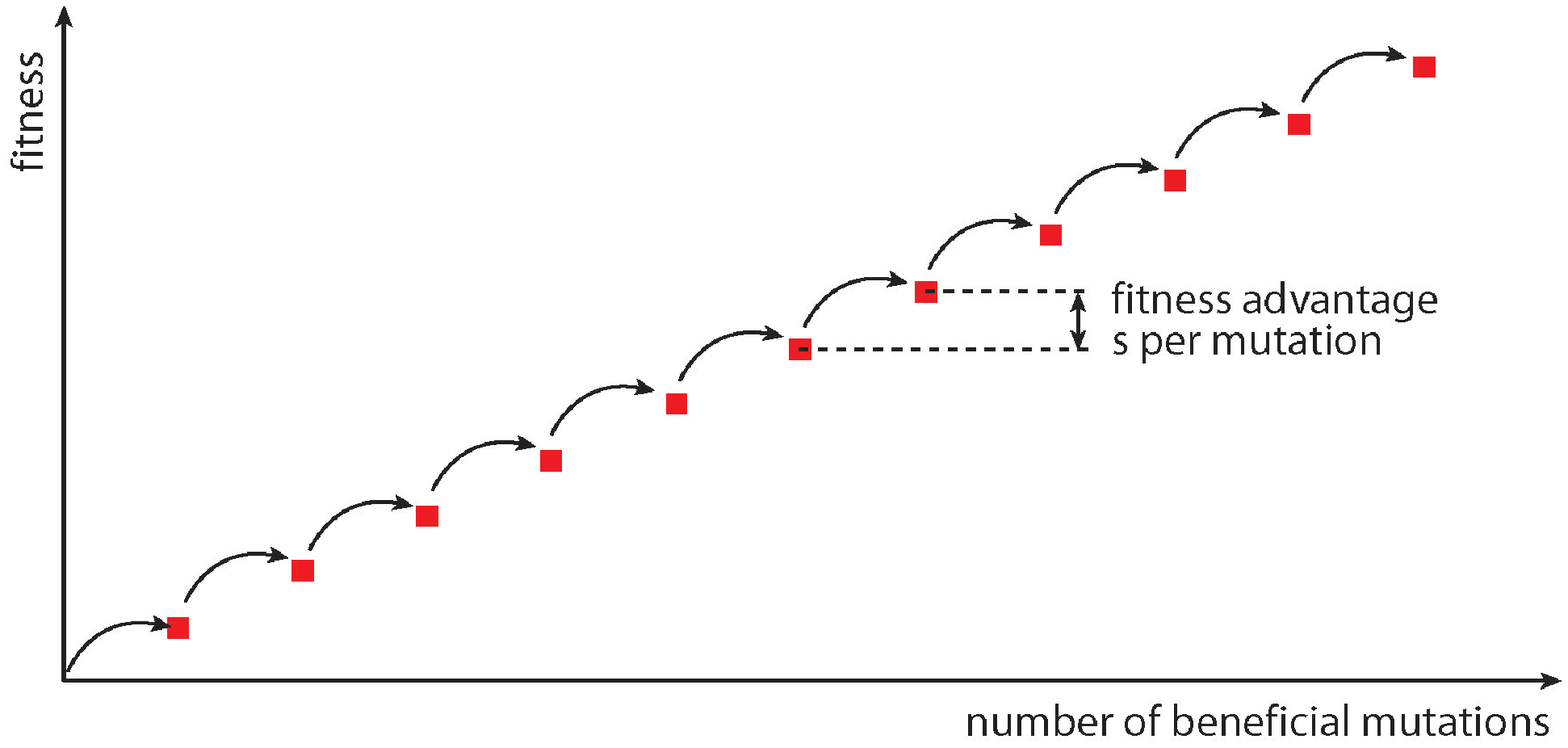}
  \caption{}
\end{figure}

\begin{figure}
  \centering
  \includegraphics[width=6.5in]{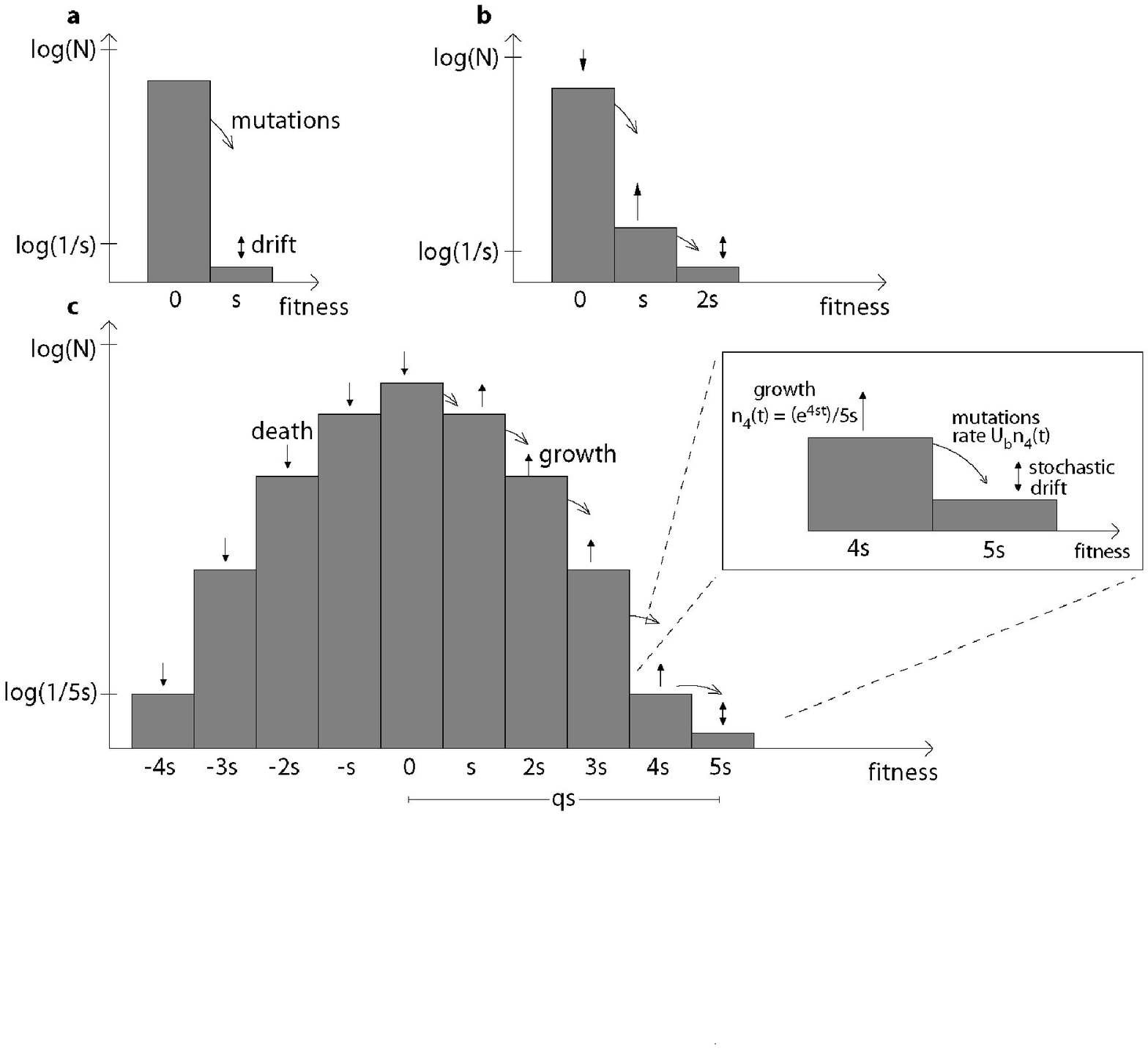}
  \caption{}
\end{figure}

\begin{figure}
  \centering
  \includegraphics[width=6.5in]{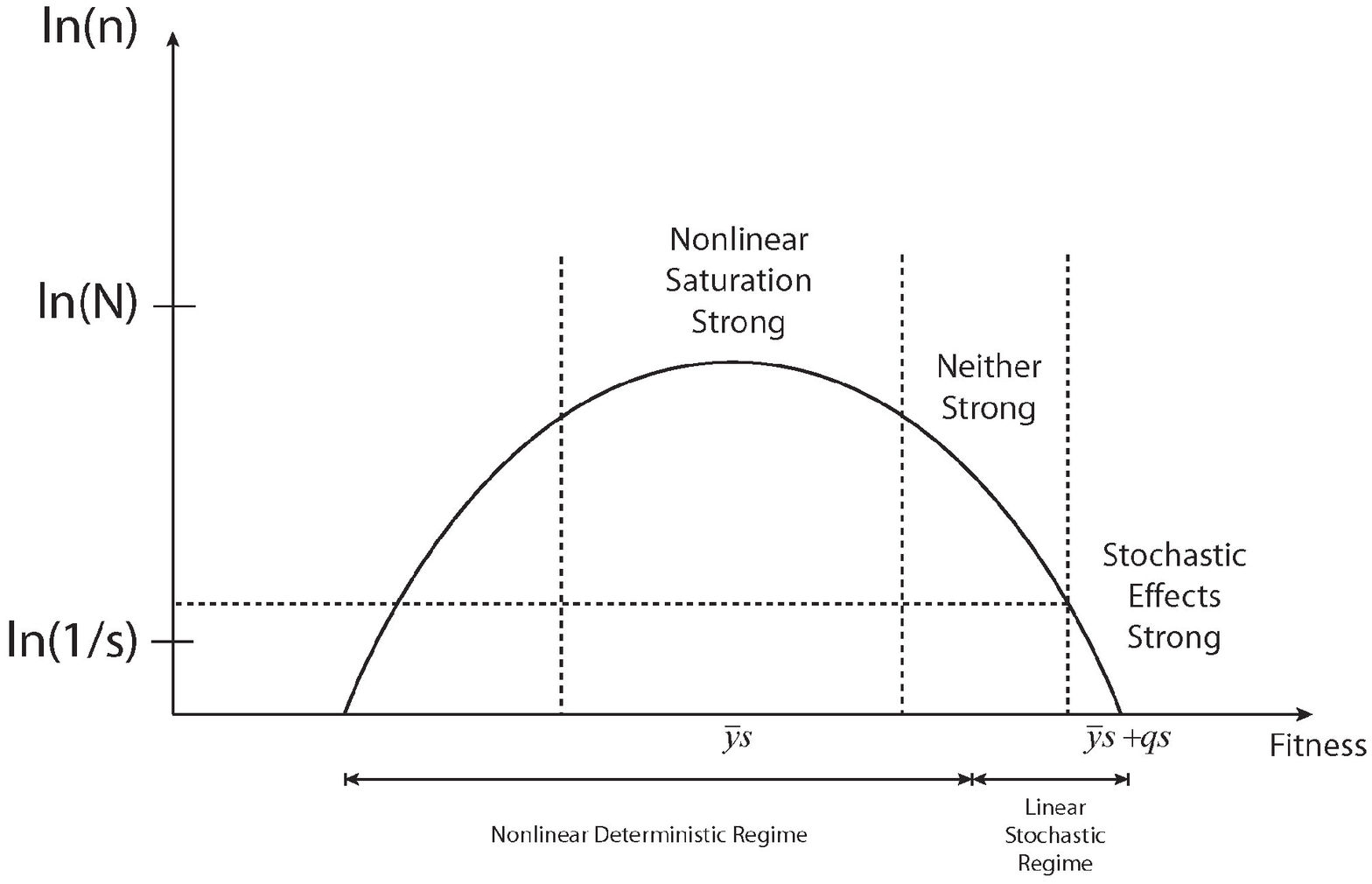}
  \caption{}
\end{figure}

\begin{figure}
  \centering
  \includegraphics[width=6.5in]{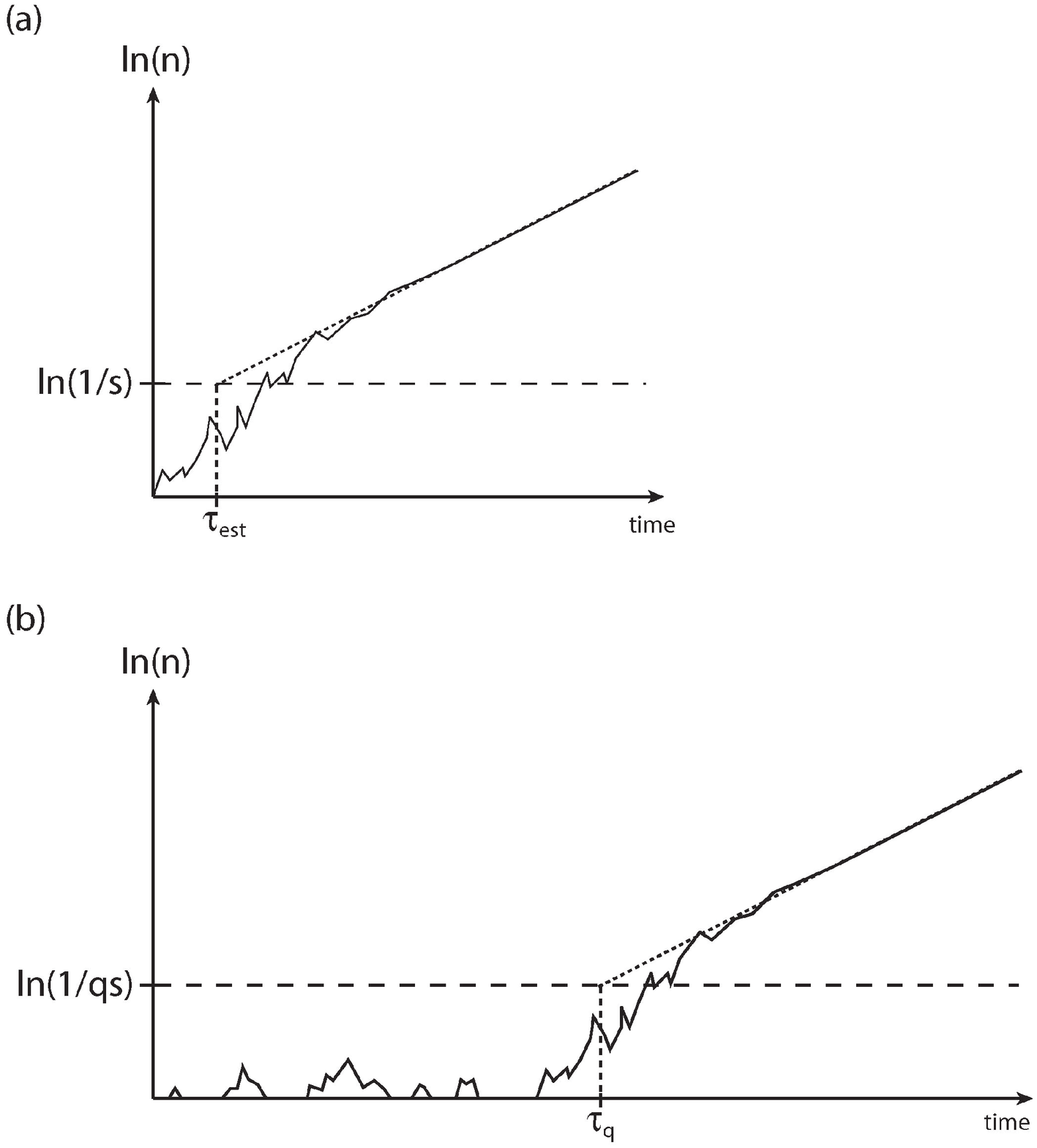}
  \caption{}
\end{figure}

\begin{figure}
  \centering
  \includegraphics[width=6.5in]{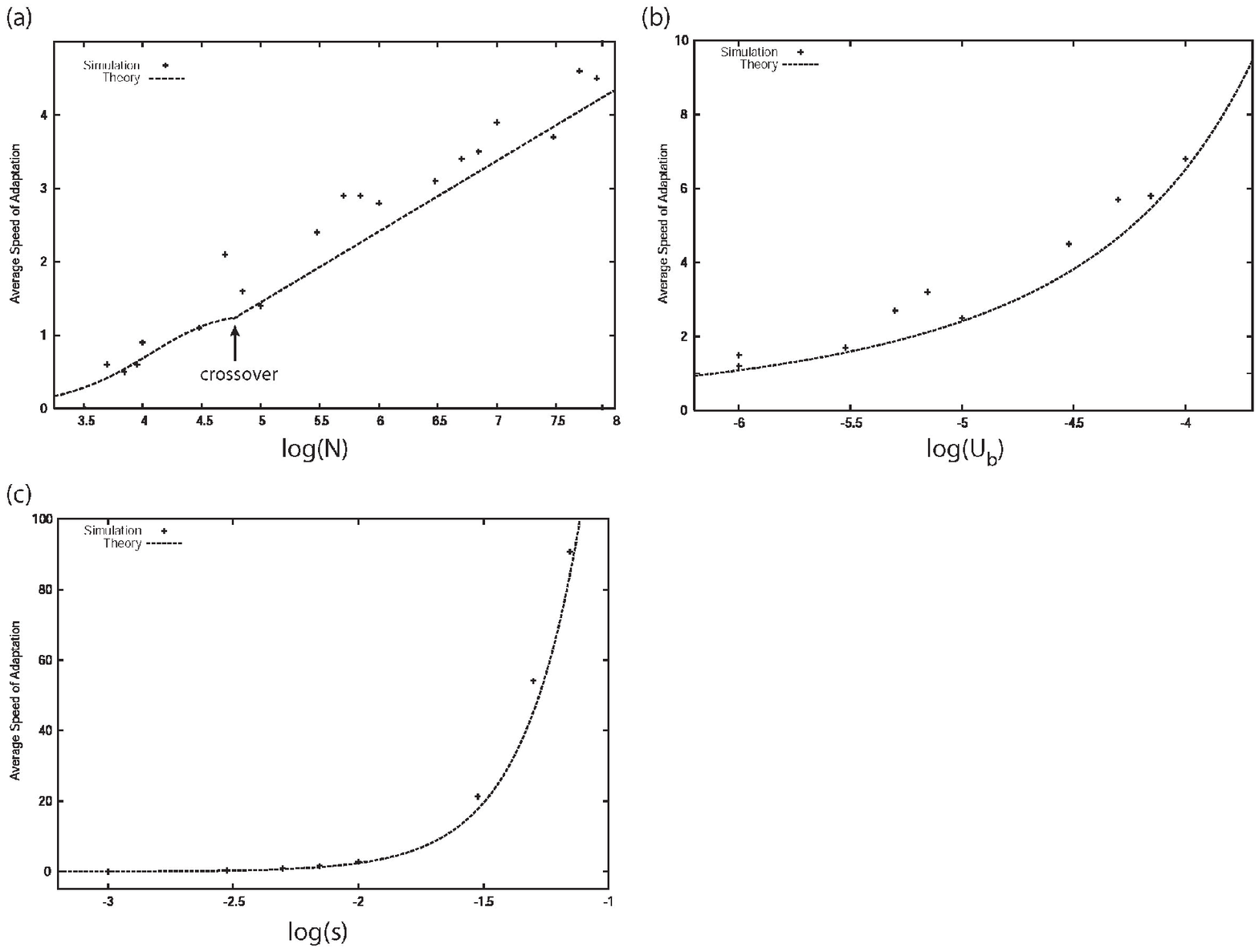}
  \caption{}
\end{figure}

\begin{figure}
  \centering
  \includegraphics[width=6.5in]{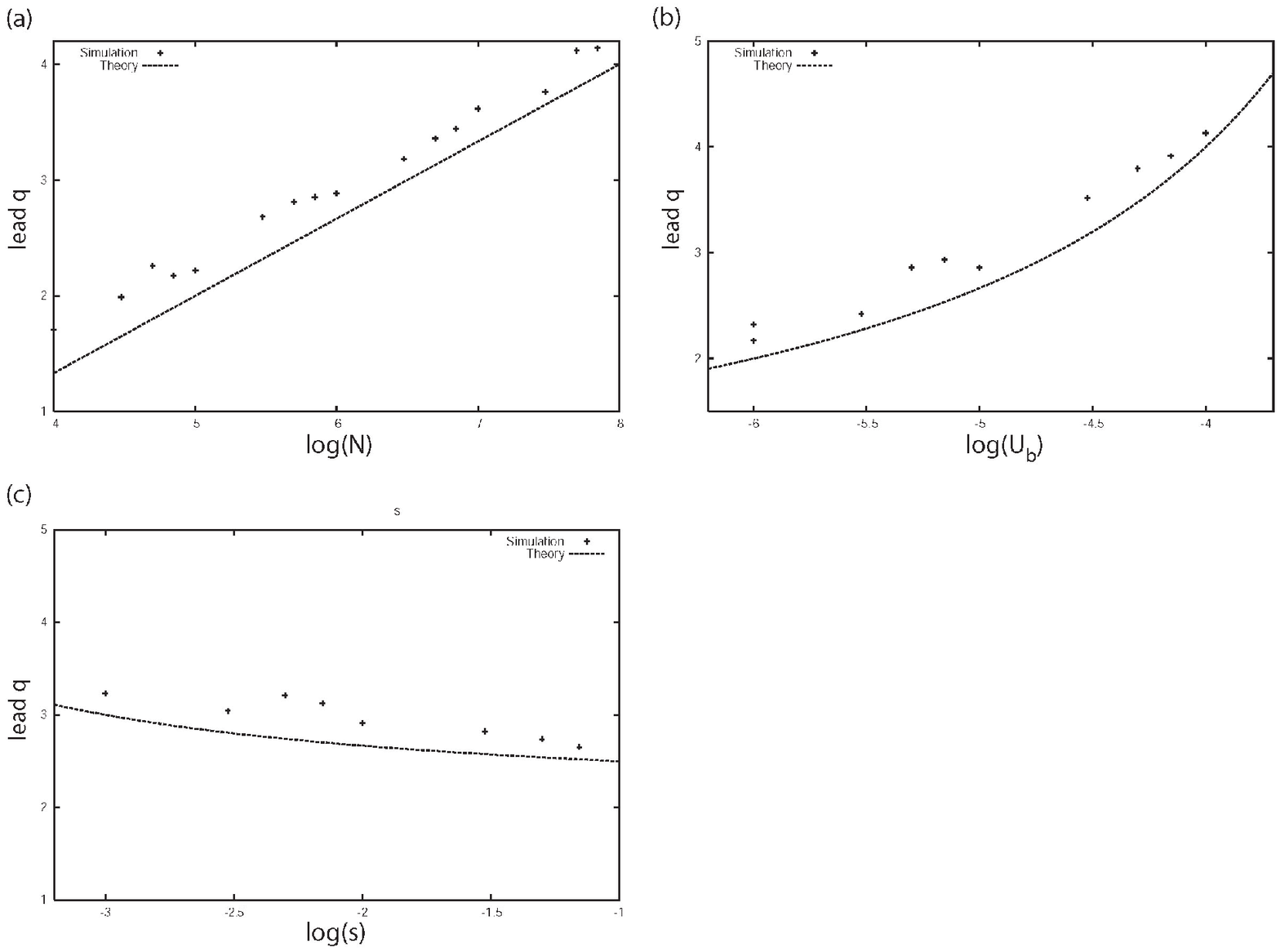}
  \caption{}
\end{figure}

\clearpage

\newpage

\bibliographystyle{genetics}

\end{document}